\documentclass[twocolumn,english,aps]{revtex4-1}
\usepackage[T1]{fontenc}
\usepackage[latin9]{inputenc}
\usepackage{babel}
\usepackage{float}
\usepackage{amsmath}
\usepackage{amssymb}
\usepackage{graphicx}
\usepackage[unicode=true]
 {hyperref}
\usepackage{breakurl}

\makeatletter
\usepackage{amsmath}
\usepackage{titlesec}
\usepackage[below]{placeins}
\titlespacing*{\subsubsection}
{0pt}{1.5\baselineskip}{1\baselineskip}
\titlespacing*{\section}
{0pt}{2\baselineskip}{1\baselineskip}

\makeatother

\begin{document}

\title{Characterization of the quantum phase transition in a two-mode Dicke
model for different cooperation numbers}

\author{L. F. Quezada}

\email{luis.fernando@correo.nucleares.unam.mx}

\affiliation{Instituto de Ciencias Nucleares, Universidad Nacional Aut\'onoma de
M\'exico, Apartado Postal 70-543, 04510 Ciudad de M\'exico, M\'exico.}

\author{E. Nahmad-Achar}

\email{nahmad@nucleares.unam.mx}

\affiliation{Instituto de Ciencias Nucleares, Universidad Nacional Aut\'onoma de
M\'exico, Apartado Postal 70-543, 04510 Ciudad de M\'exico, M\'exico.}

\begin{abstract}
We show how the use of variational states to approximate the ground
state of a system can be employed to study a multi-mode Dicke model.
One of the main contributions of this work is the introduction of
a not very commonly used quantity, the cooperation number, and the
study of its influence on the behavior of the system, paying particular
attention to the quantum phase transitions and the accuracy of the
used approximations. We also show how these phase transitions affect
the dependence of the expectation values of some of the observables
relevant to the system and the entropy of entanglement with respect
to the energy difference between atomic states and the coupling strength
between matter and radiation, thus characterizing the transitions
in different ways.
\end{abstract}
\maketitle

\section*{Introduction}

Quantum phase transitions (QPTs) are informally seen as sudden, drastic
changes in the physical properties of the ground state of a system
at zero temperature due to the variation of some parameter involved
in the modeling Hamiltonian. One model of particular interest for
the study of such phenomena is the Dicke model \cite{key-1}, as it
describes, in a simplified way (electric dipole approximation), the
interaction between matter and electromagnetic radiation. In 1973,
Hepp and Lieb \cite{key-2,key-3}, and Wang and Hioe \cite{key-4}
first theoretically proved the existence of a second-order QPT in
the Dicke model. Wang and Hioe also treated the multi-mode radiation case, where they reduce it to a single-mode case by using an effective coupling constant. To date, this QPT has been experimentally observed
in a Bose-Einstein Condensate coupled to an optical cavity \cite{key-5,key-6}
and it has been shown to be relevant to quantum information and quantum
computing \cite{key-7,key-8,key-9,key-10}.

Even though the formal definition of a QPT requires us to compute
the ground state's energy as a function of any desired parameter in
order to find its transition values, one of the main contributions
of this work is to show how the QPT in the Dicke model influences
the behavior of other quantities relevant to the system, thus characterizing
the transition in different, simpler ways.

\subsubsection*{Quantum Phase Transitions}

The formal definition of the concept of ``quantum phase'' that we
will be using throughout this paper is that of an open region $\mathcal{R}\subseteq\mathbb{R}^{\ell}$
where the ground state's energy $\mathcal{E}_{0}$, as a function
of $\ell$ parameters involved in the modeling Hamiltonian, is analytic.
Thus a QPT is identified by the boundary $\partial\mathcal{R}$ of
the region at which $\frac{\partial^{n}\mathcal{E}_{0}}{\partial x^{n}}$
is discontinuous for some $n$ (known as the order of the transition).

Notice that in the previous definition, for the sake of generality,
we did not consider the thermodynamic limit, as it has been shown
that interesting phenomena regarding QPTs occur even for a finite
number of particles \cite{key-11,key-12}.

\begin{figure*}[t]
\begin{centering}
\includegraphics[scale=0.29]{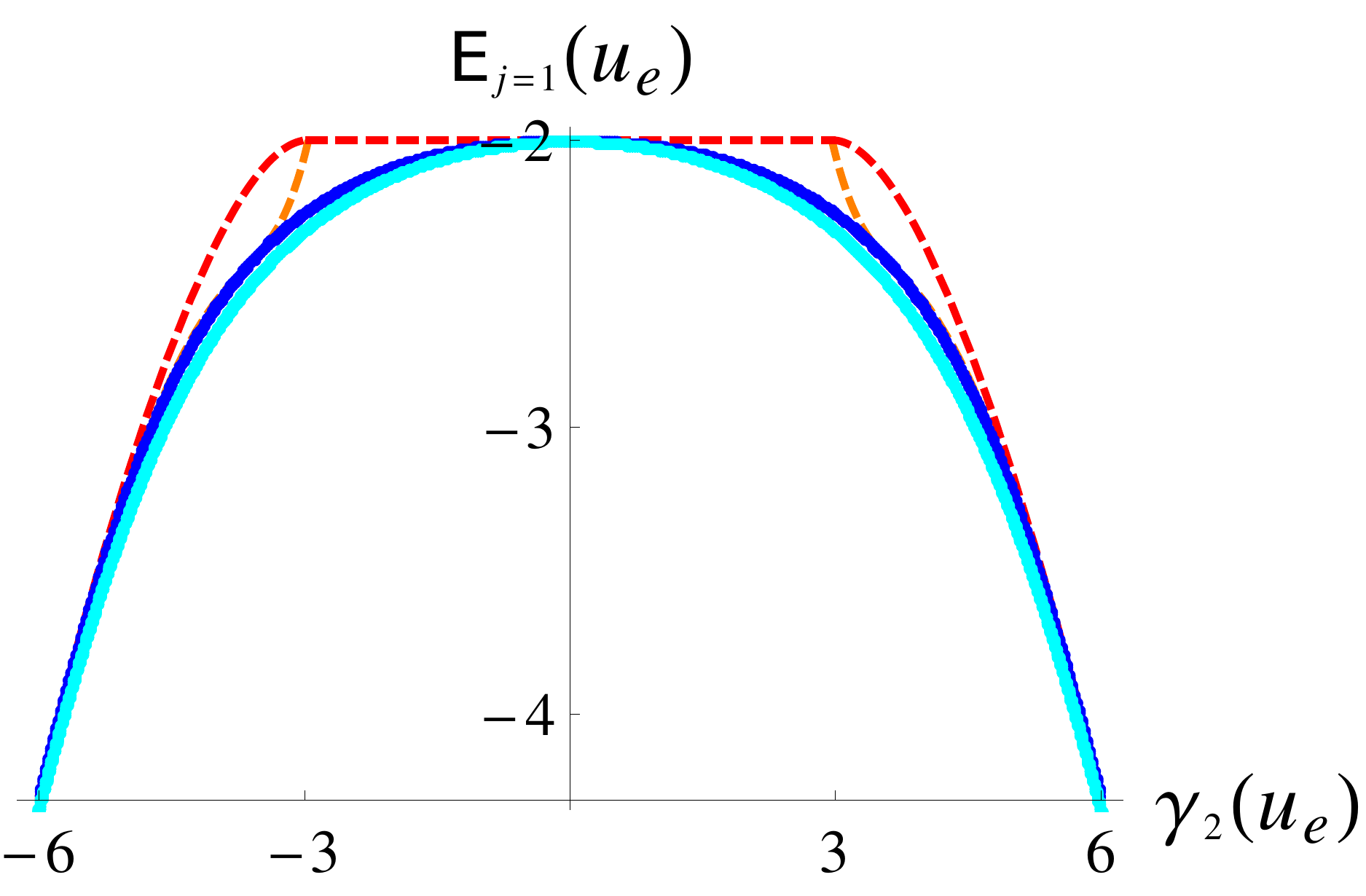}\qquad{}\includegraphics[scale=0.29]{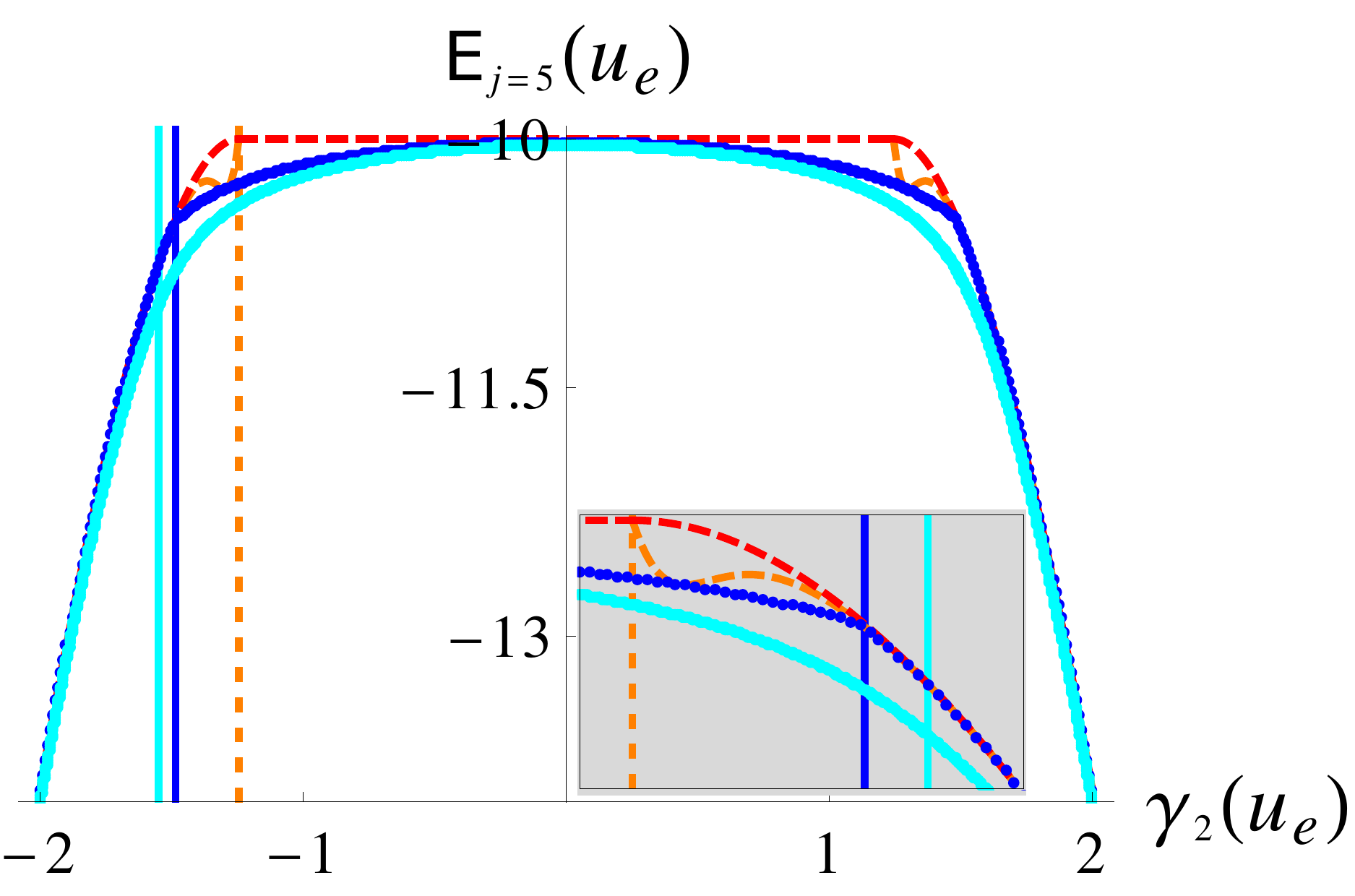}\qquad{}\includegraphics[scale=0.29]{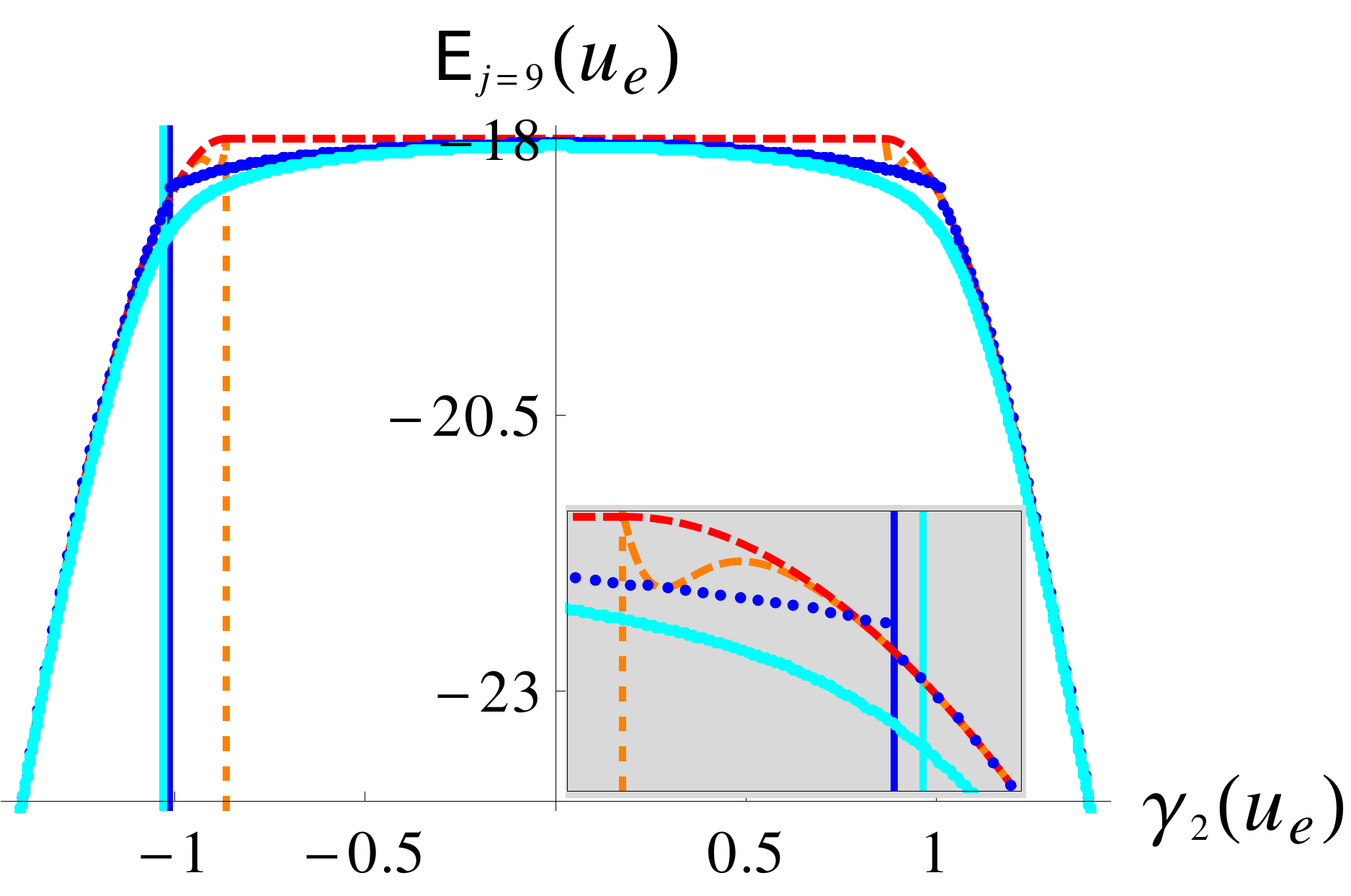}
\par\end{centering}

\caption{Energy of the ground state as a function of $\gamma_{2}$ obtained
using CS (dark gray dashed / red online), SAS with CS's minima (light
gray dashed / orange online), SAS minimized numerically (dark gray
/ blue online) and quantum solution (light gray / cyan online). Vertical
lines show the transition according to the quantum solution via fidelity's
minimum (light gray / cyan online), SAS minimized numerically (dark
gray / blue online) and SAS with CS's minima (light gray dashed /
orange online). Left: j=1, center: j=5, right: j=9. Assuming $k=2$
and using $N=18$, $\omega_{A}=\Omega_{1}=\Omega_{2}=2u_{e}$, $\gamma_{1}=\frac{1}{2}u_{e}$,
where $u_{e}$ stands for any energy unit ($\hbar=1$).\label{fig:1}}
\end{figure*}

\begin{figure*}[t]
\begin{centering}
\includegraphics[scale=0.29]{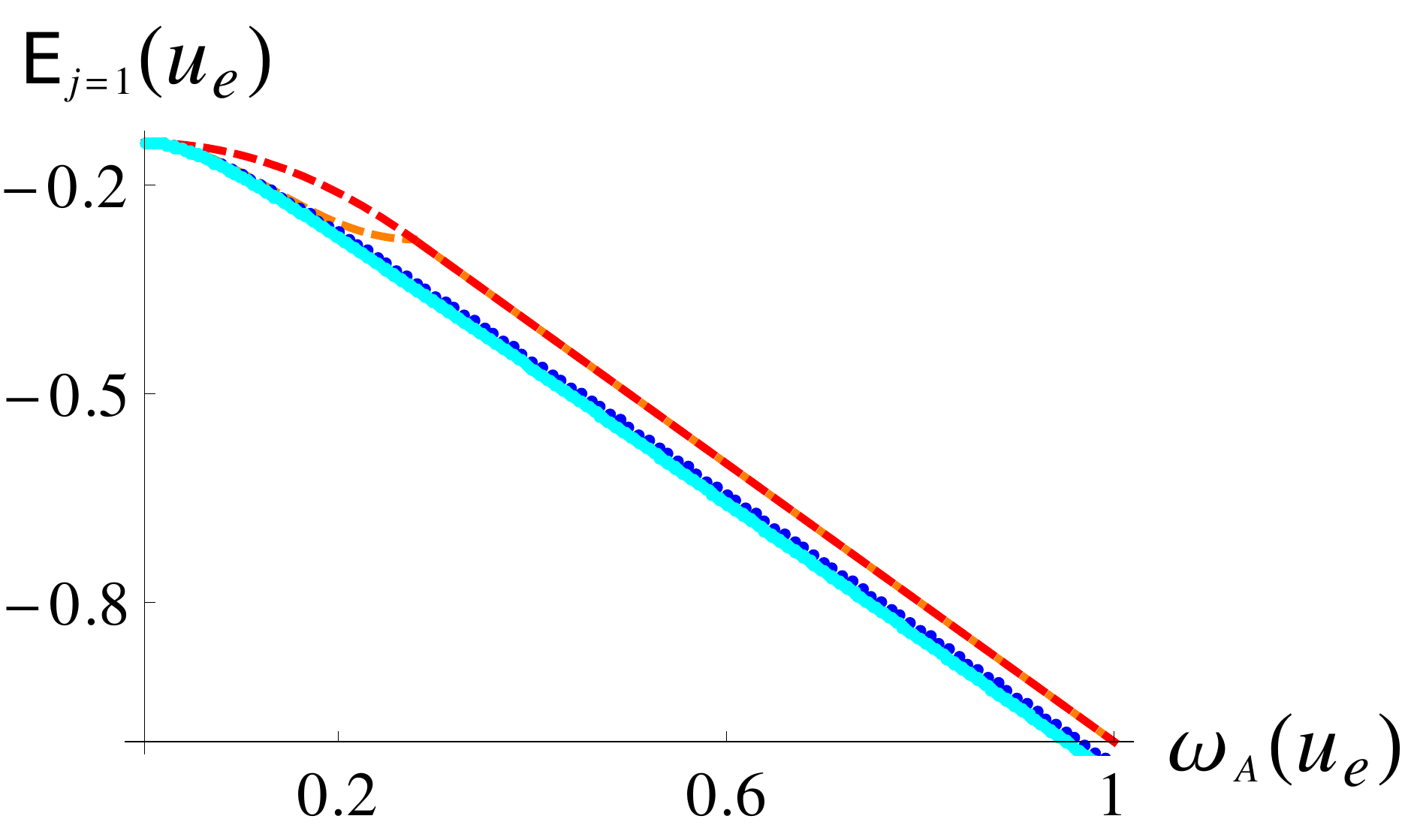}\qquad{}\includegraphics[scale=0.29]{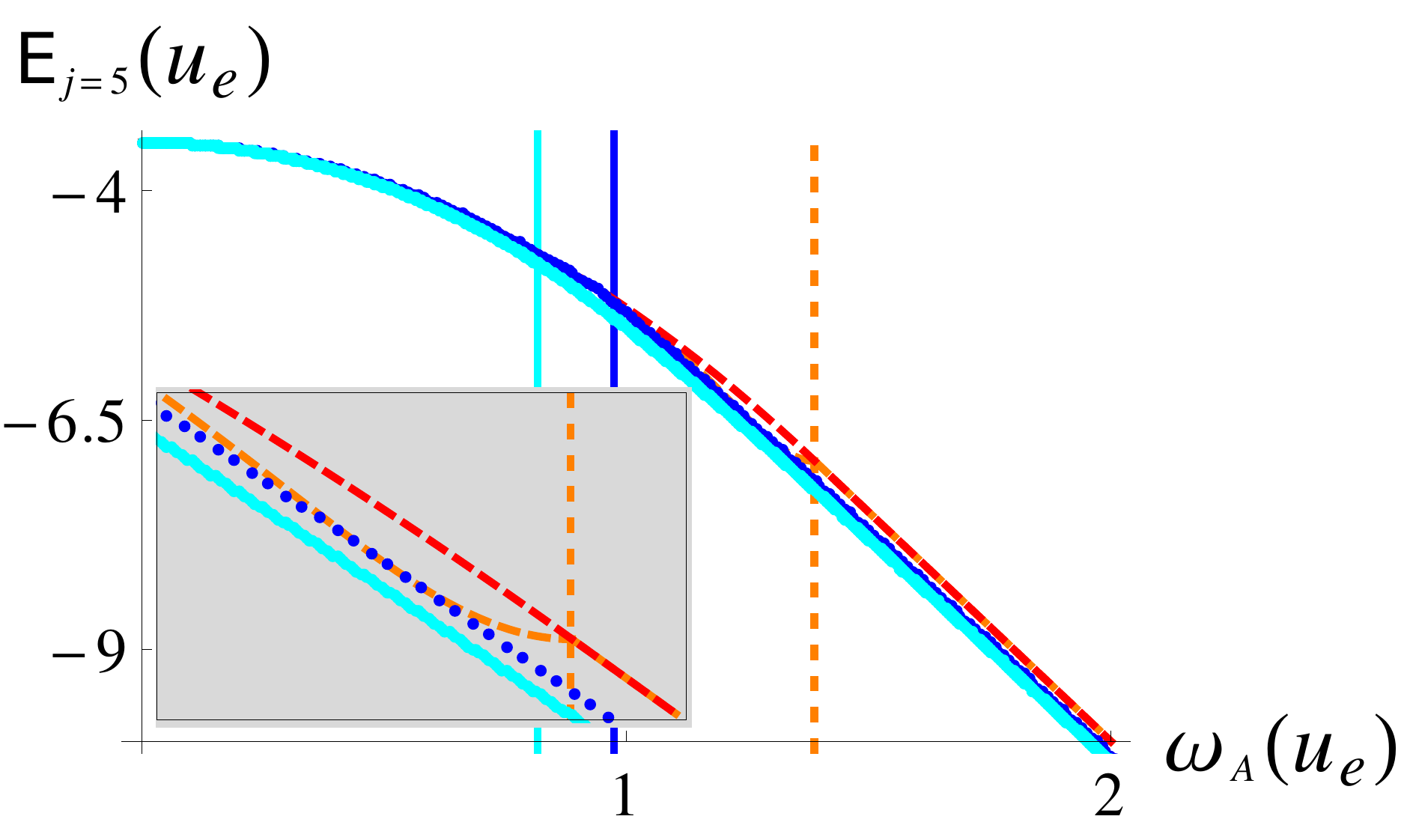}\qquad{}\includegraphics[scale=0.29]{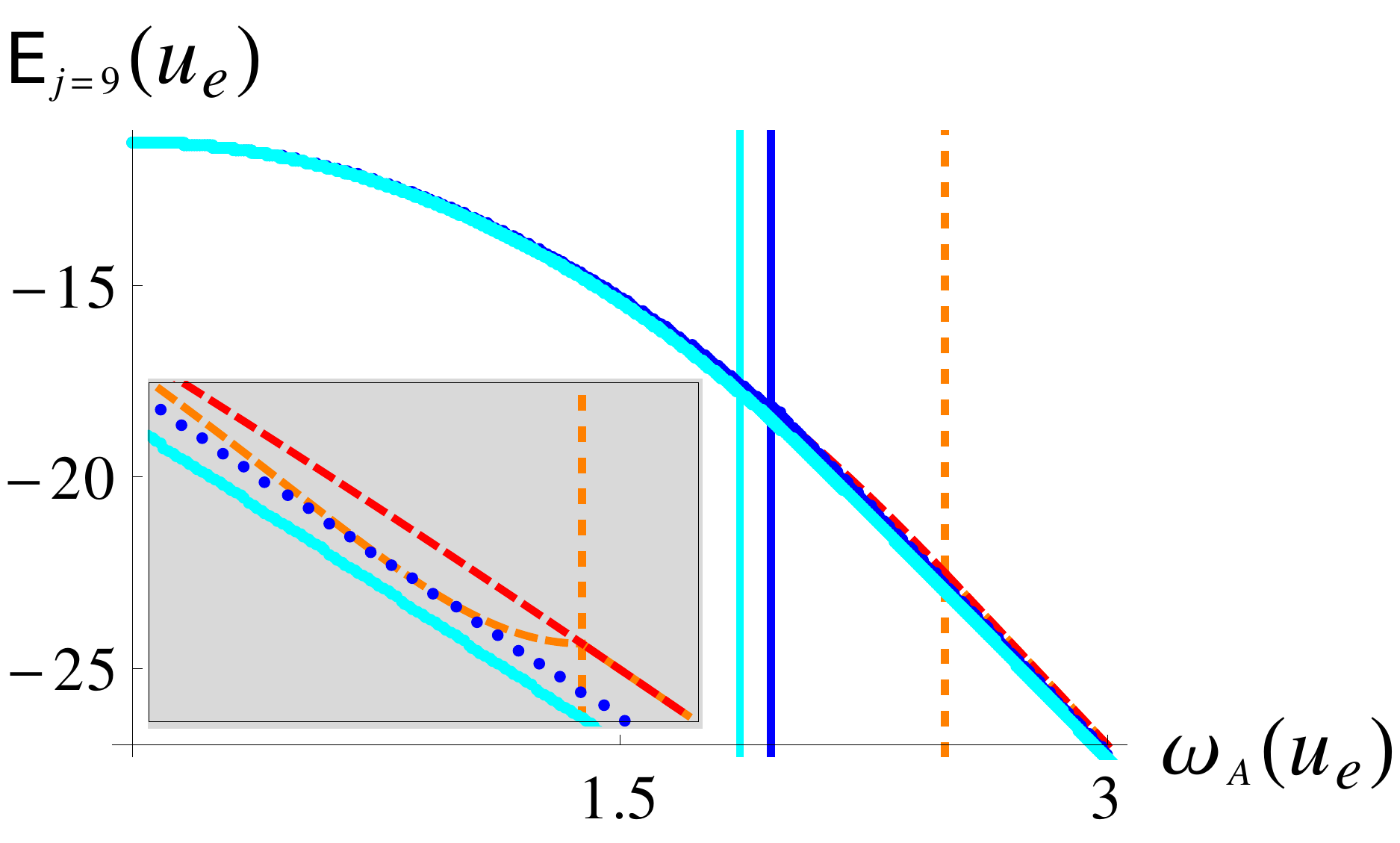}
\par\end{centering}

\caption{Energy of the ground state as a function of $\omega_{A}$ obtained
using CS (dark gray dashed / red online), SAS with CS's minima (light
gray dashed / orange online), SAS minimized numerically (dark gray
/ blue online) and quantum solution (light gray / cyan online). Vertical
lines show the transition according to the quantum solution via fidelity's
minimum (light gray / cyan online), SAS minimized numerically (dark
gray / blue online) and SAS with CS's minima (light gray dashed /
orange online). Left: j=1, center: j=5, right: j=9. Assuming $k=2$
and using $N=18$, $\Omega_{1}=\Omega_{2}=2u_{e}$, $\gamma_{1}=\frac{1}{2}u_{e}$,
$\gamma_{2}=1u_{e}$, where $u_{e}$ stands for any energy unit ($\hbar=1$).\label{fig:2}}
\end{figure*}

\subsubsection*{Modeling Hamiltonian}

The Hamiltonian (Dicke's Hamiltonian) describing the interaction,
in a dipolar approximation, between N two-level identical atoms (same
energy difference between the two levels) and one-mode of an electromagnetic
field in an ideal cavity, has the expression ($\hbar=1$)

\begin{equation}
H_{D}=\omega_{A}J_{z}+\Omega a^{\dagger}a-\frac{\gamma}{\sqrt{N}}\left(J_{-}+J_{+}\right)\left(a+a^{\dagger}\right).
\end{equation}

Here, $\omega_{A}$ is the energy difference between the atomic levels,
$\Omega$ is the frequency of the field's mode, $\gamma$ is the dipolar
coupling constant, $J_{z}$, $J_{-}$, $J_{+}$ are the collective
spin operators and $a$, $a^{\dagger}$ are the annihilation and creation
operators of the harmonic oscillator. The multi-mode Hamiltonian is
obtained summing over the number $k$ of modes \cite{key-13}, and
has the expression

\begin{equation}
H=\omega_{A}J_{z}+\sum_{\imath=1}^{k}\Omega_{\imath}a_{\imath}^{\dagger}a_{\imath}-\frac{1}{\sqrt{N}}\sum_{\imath=1}^{k}\gamma_{\imath}\left(J_{-}+J_{+}\right)\left(a_{\imath}+a_{\imath}^{\dagger}\right).\label{eq:2}
\end{equation}

The $k$ modes of the electromagnetic field are described in terms
of annihilation and creation operators for each mode $a_{\imath}$,
$a_{\imath}^{\dagger}$, acting on the tensor product of $k$ copies
of the Fock space \ensuremath{\bigotimes\limits_{\imath=1}^{k}\mathcal{F}_{\imath}}
and satisfying the commutation relations

\begin{equation}
\left[a_{\imath},a_{\jmath}^{\dagger}\right]=\delta_{\imath\jmath},\qquad\left[a_{\imath},a_{\jmath}\right]=\left[a_{\imath}^{\dagger},a_{\jmath}^{\dagger}\right]=0.\label{eq:3}
\end{equation}

A two-level atom is described using the $\frac{1}{2}$-spin matrices
$S_{z}=\frac{1}{2}\sigma_{z}$, $S_{\pm}=\frac{1}{2}\left(\sigma_{x}\pm i\sigma_{y}\right)$
($\sigma_{x}$, $\sigma_{y}$ and $\sigma_{z}$ being the Pauli matrices),
which act on a two-dimensional complex Hilbert space $\mathbb{C}^{2}$
and satisfy the commutation relations

\begin{equation}
\left[S_{+},S_{-}\right]=2S_{z},\qquad\left[S_{z},S_{\pm}\right]=\pm S_{\pm}.
\end{equation}

When considering a system of N two-level atoms, we use the collective
spin operators $J_{z}$, $J_{-}$, $J_{+}$ defined as

\begin{multline}J_{\diamond}=S_{\diamond}\otimes I_{2}^{\otimes\left(N-1\right)}+I_{2}\otimes S_{\diamond}\otimes I_{2}^{\otimes\left(N-2\right)}\\+\cdots+I_{2}^{\otimes\left(N-2\right)}\otimes S_{\diamond}\otimes I_{2}+I_{2}^{\otimes\left(N-1\right)}\otimes S_{\diamond}\end{multline}where
$I_{2}$ is the identity operator on $\mathbb{C}^{2}$ and $\diamond\in\left\{ z,-,+\right\} $.
These collective spin operators satisfy the commutation relations

\begin{equation}
\left[J_{+},J_{-}\right]=2J_{z},\qquad\left[J_{z},J_{\pm}\right]=\pm J_{\pm}\label{eq:6}
\end{equation}
and act, in principle, on the complex Hilbert space $\left(\mathbb{C}^{2}\right)^{\otimes N}$;
however, working with this space is physically equivalent to studying
a system of N fully distinguishable atoms, which we don't usually
have in the experimental setups used in the study of the QPT in the
Dicke model. To overcome this issue, we must use the common set of
eigenvectors $\left\{ \left|j,m\right\rangle \right\} $ of the two
commuting observables $J_{z}$ and $J^{2}=\frac{1}{2}\left(J_{+}J_{-}+J_{-}J_{+}\right)+J_{z}^{2}$,
where the label $j$ is limited to the values $j\in\left\{ r,r+1,\ldots,\frac{N}{2}\right\} $
($r=0$ for even N and $r=\frac{1}{2}$ for odd N) and the label $m\in\mathbb{Z}$
is constricted by $\left|m\right|\leq j$. These vectors do not form
a basis of $\left(\mathbb{C}^{2}\right)^{\otimes N}$ for $N>2$,
as the dimension of their linear span is

\[
\mathrm{dim}\left\{ \mathrm{span}\left\{ \left\{ \left|j,m\right\rangle \right\} _{\left|m\right|\leq j}^{j=r,\ldots,\frac{N}{2}}\right\} \right\} ={\displaystyle \sum_{j=r}^{\frac{N}{2}}\left(2j+1\right)}\leq2^{N}.
\]

\begin{figure*}[t]
\begin{centering}
\includegraphics[scale=0.29]{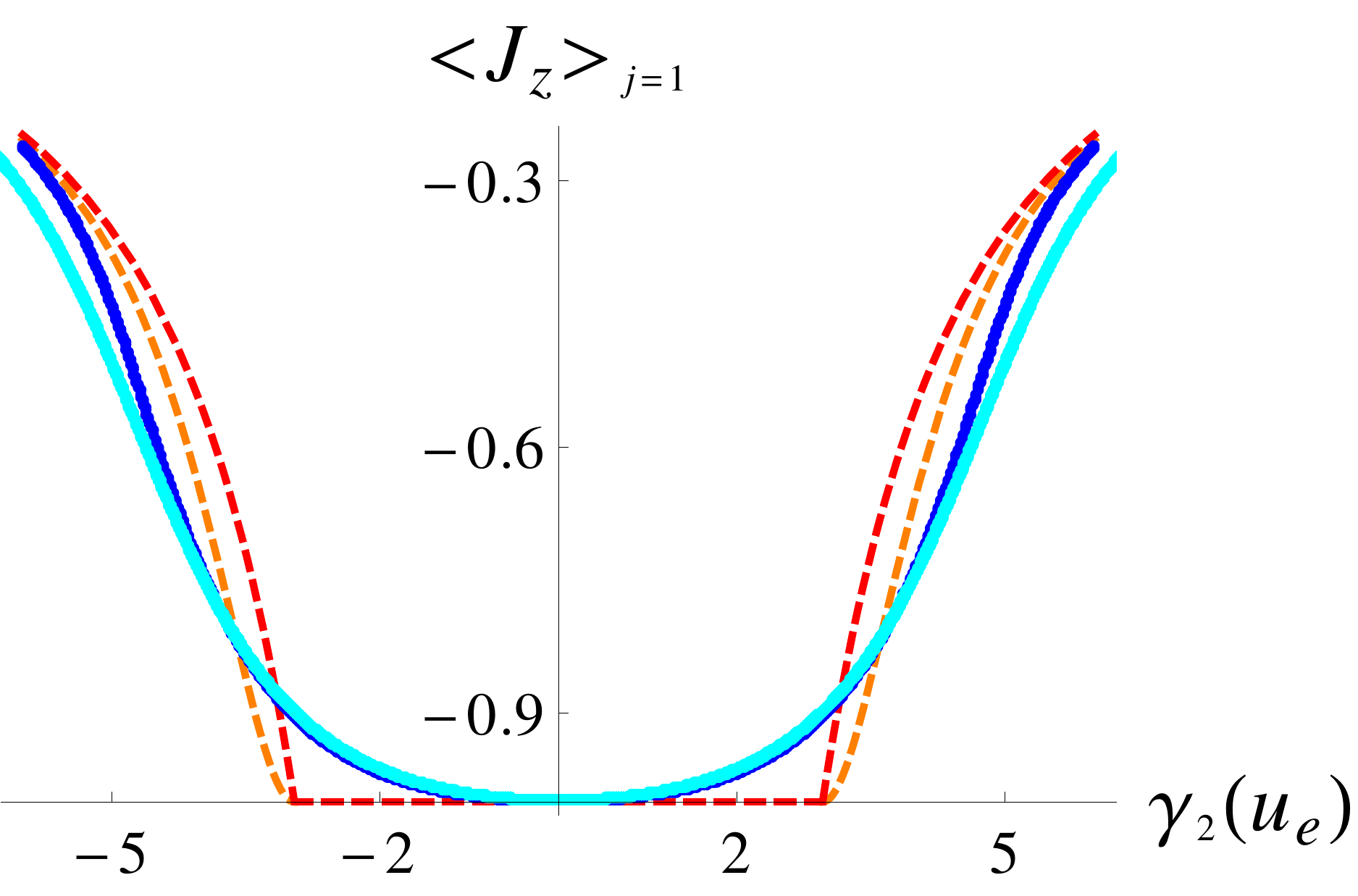}\qquad{}\includegraphics[scale=0.29]{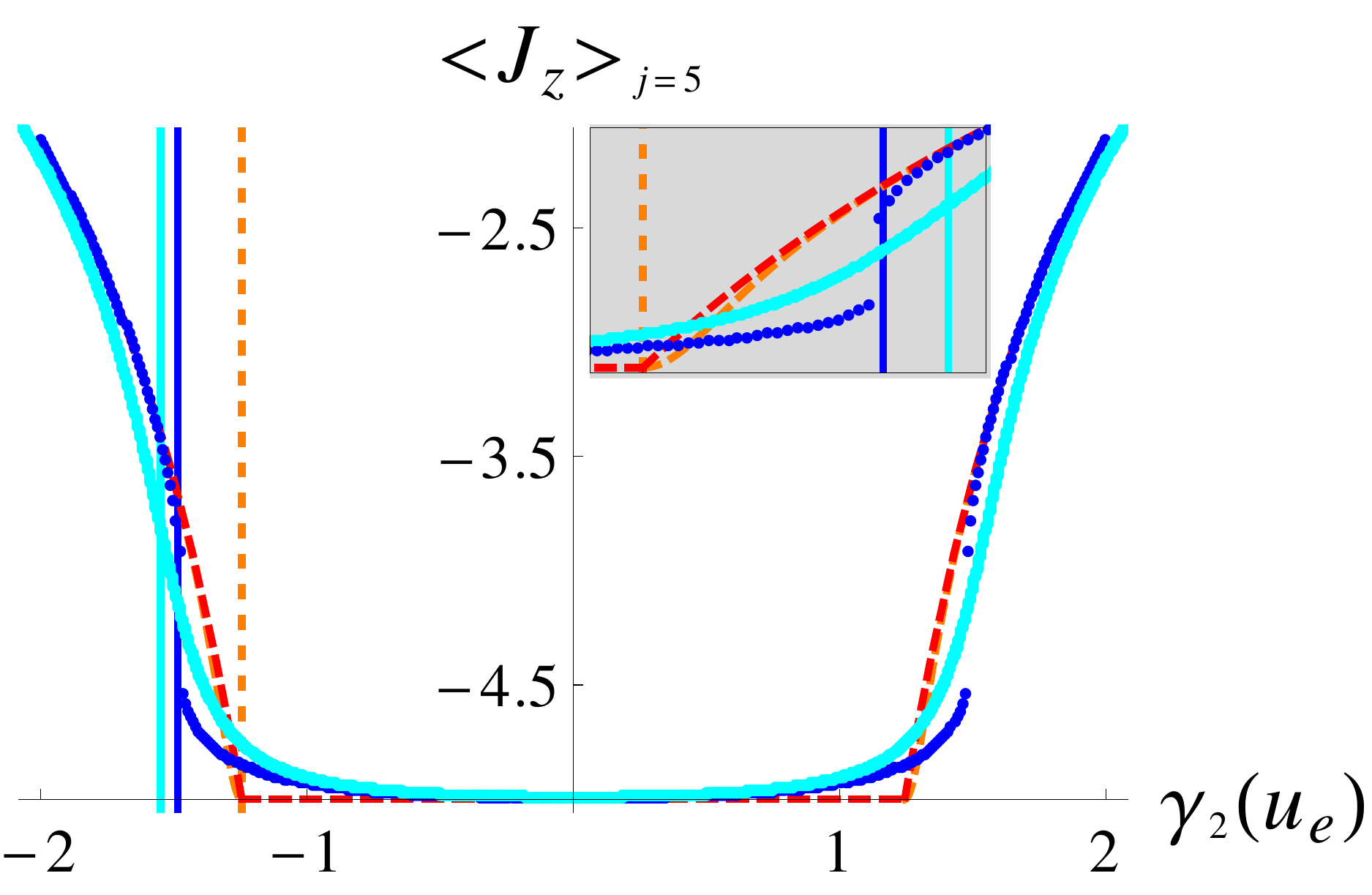}\qquad{}\includegraphics[scale=0.29]{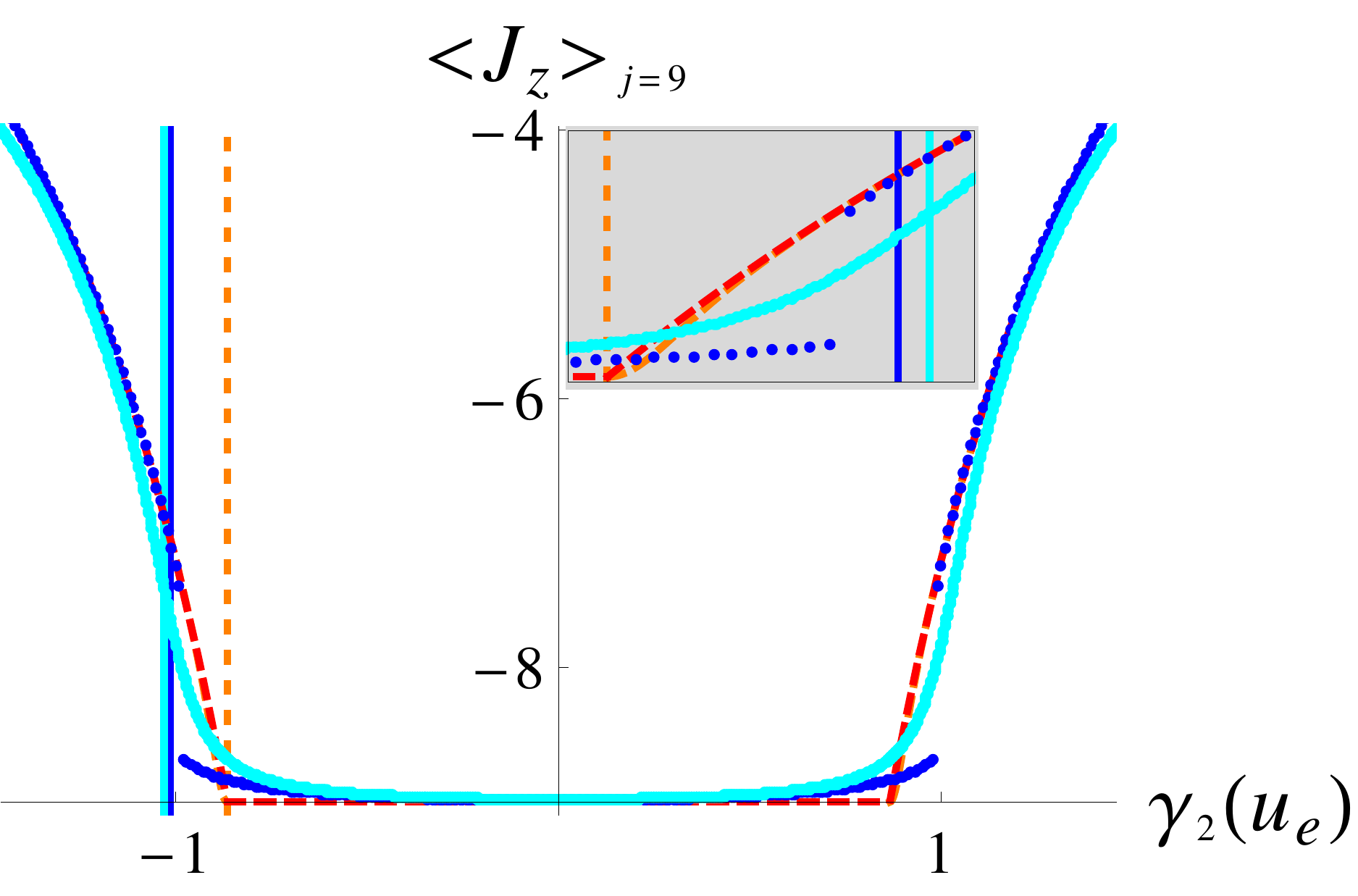}
\par\end{centering}

\caption{Expectation value of $J_{z}$ as a function of $\gamma_{2}$ obtained
using CS (dark gray dashed / red online), SAS with CS's minima (light
gray dashed / orange online), SAS minimized numerically (dark gray
/ blue online) and quantum solution (light gray / cyan online). Vertical
lines show the transition according to the quantum solution via fidelity's
minimum (light gray / cyan online), SAS minimized numerically (dark
gray / blue online) and SAS with CS's minima (light gray dashed /
orange online). Left: j=1, center: j=5, right: j=9. Assuming $k=2$
and using $N=18$, $\omega_{A}=\Omega_{1}=\Omega_{2}=2u_{e}$, $\gamma_{1}=\frac{1}{2}u_{e}$,
where $u_{e}$ stands for any energy unit ($\hbar=1$).\label{fig:3}}
\end{figure*}

\begin{figure*}[t]
\begin{centering}
\includegraphics[scale=0.29]{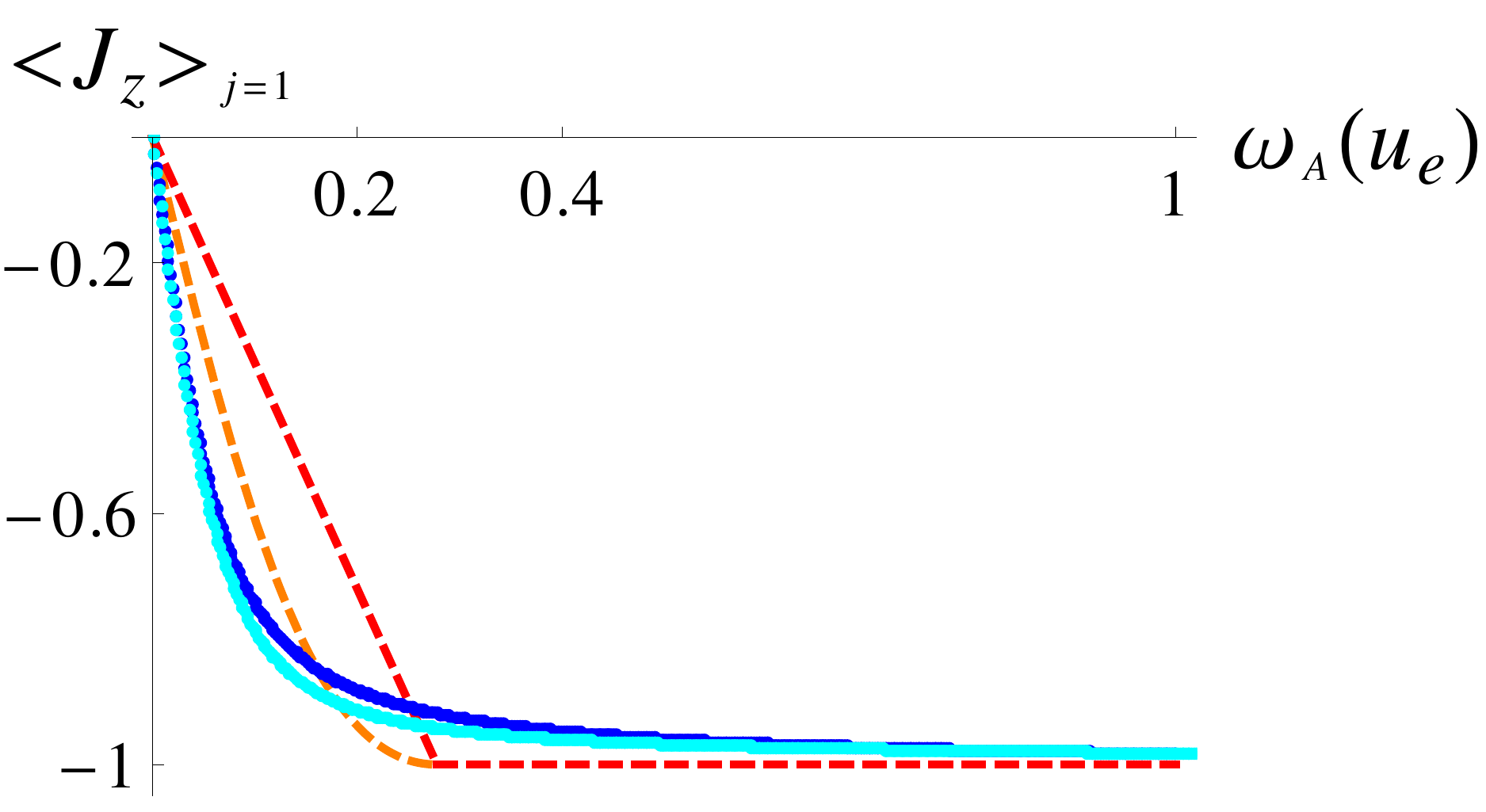}\qquad{}\includegraphics[scale=0.29]{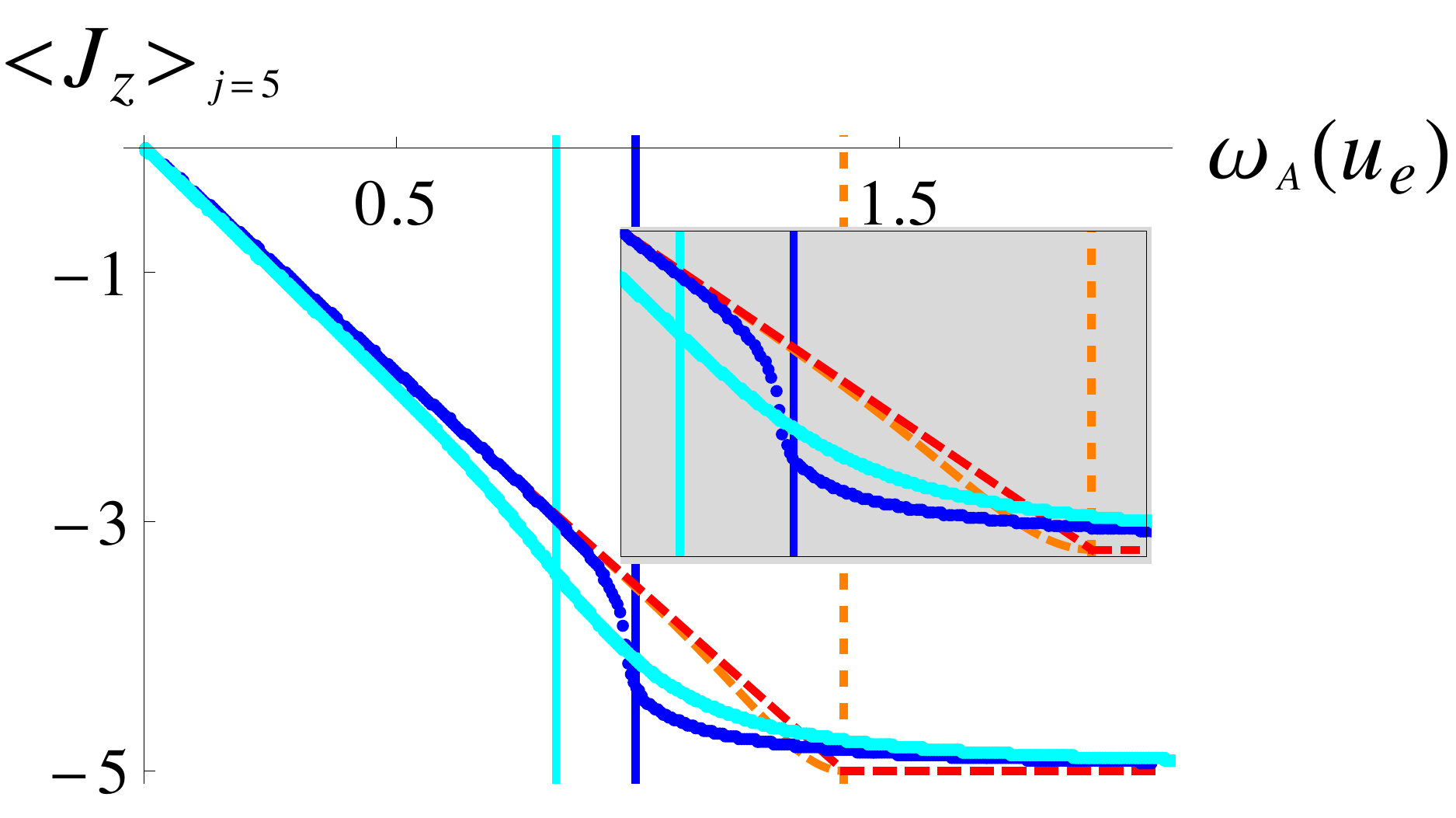}\qquad{}\includegraphics[scale=0.29]{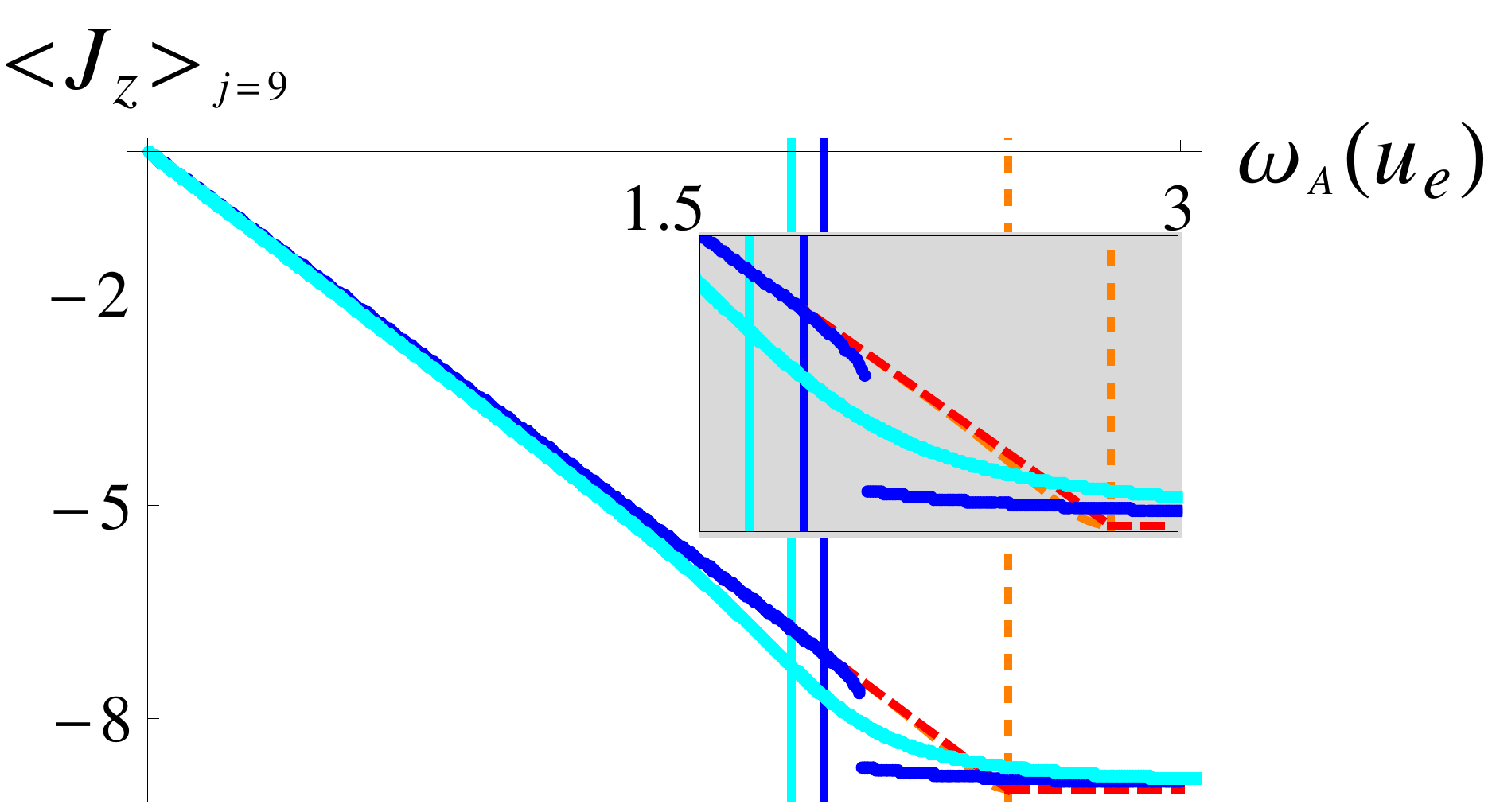}
\par\end{centering}

\caption{Expectation value of $J_{z}$ as a function of $\omega_{A}$ obtained
using CS (dark gray dashed / red online), SAS with CS's minima (light
gray dashed / orange online), SAS minimized numerically (dark gray
/ blue online) and quantum solution (light gray / cyan online). Vertical
lines show the transition according to the quantum solution via fidelity's
minimum (light gray / cyan online), SAS minimized numerically (dark
gray / blue online) and SAS with CS's minima (light gray dashed /
orange online). Left: j=1, center: j=5, right: j=9. Assuming $k=2$
and using $N=18$, $\Omega_{1}=\Omega_{2}=2u_{e}$, $\gamma_{1}=\frac{1}{2}u_{e}$,
$\gamma_{2}=1u_{e}$, where $u_{e}$ stands for any energy unit ($\hbar=1$).\label{fig:4}}
\end{figure*}

We will denote by $\mathcal{H}_{A}$ the subspace of $\left(\mathbb{C}^{2}\right)^{\otimes N}$
generated by the states $\left\{ \left|j,m\right\rangle \right\} _{\left|m\right|\leq j}^{j=r,\ldots,\frac{N}{2}}$.

There are two main results concerning the states $\left\{ \left|j,m\right\rangle \right\} _{\left|m\right|\leq j}^{j=r,\ldots,\frac{N}{2}}$
and the space $\mathcal{H}_{A}$: the first comes from noticing that
$\left[H,J^{2}\right]=0$, which means that the label $j$ of the
eigenvalues of $J^{2}$ remains constant during the system's evolution;
the second is the decomposition $\mathcal{H}_{A}=\ensuremath{\bigoplus\limits _{j=r}^{\frac{N}{2}}\mathcal{H}_{j}}$,
where each $\mathcal{H}_{j}$ is the subspace of dimension $\mathrm{dim}\left\{ \mathcal{H}_{j}\right\} =2j+1$
generated by the states $\left\{ \left|j,m\right\rangle \right\} _{\left|m\right|\leq j}$
with a fixed $j$. In this treatment, in order to study indistinguishable
atoms, we are ignoring the multiplicities $g(j)$ of the irreducible
representations of $SU\left(2\right)$, i.e. the number of times that
each $\mathcal{H}_{j}$ appears in the full decomposition $\left(\mathbb{C}^{2}\right)^{\otimes N}=\ensuremath{\bigoplus\limits _{j=r}^{\frac{N}{2}}g\left(j\right)\mathcal{H}_{j}}$.

To make it clear that the space $\mathcal{H}_{A}$ is the one we must
work with when indistinguishable atoms are considered we should inquire
into the physical interpretation of the labels $j$ and $m$. In order
to give a physical interpretation to the label $j$ we must notice
that the energy of the atomic system is bounded by $\pm j\omega$
independently of the number of atoms $N$ (but with the restriction
$j\leq\frac{N}{2}$), this leads us to interpret the quantity $2j$
as the effective number of atoms in the system and define it as the
\textit{cooperation number}. To make the notion of the cooperation
number more intuitive, Dicke, in his original paper \cite{key-1},
compares a state with $j=0$, which exists only for an even number
of atoms, with a classical system of an even number of oscillators
swinging in pairs oppositely phased. The interpretation of the label
$m$ is clear from the definition of $J_{z}$: $m=\frac{1}{2}\left(n_{e}-n_{g}\right)$,
where $n_{e}$ and $n_{g}$ are the number of atoms in the excited
and ground states, respectively.

In this paper we restrict our analysis to the space $\mathcal{H}_{A}$,
as it allows us to choose $j$ as an initial condition (which will
remain constant) and work in $\mathcal{H}_{j}$, where the atoms are
indistinguishable.

\begin{figure*}[t]
\begin{centering}
\includegraphics[scale=0.29]{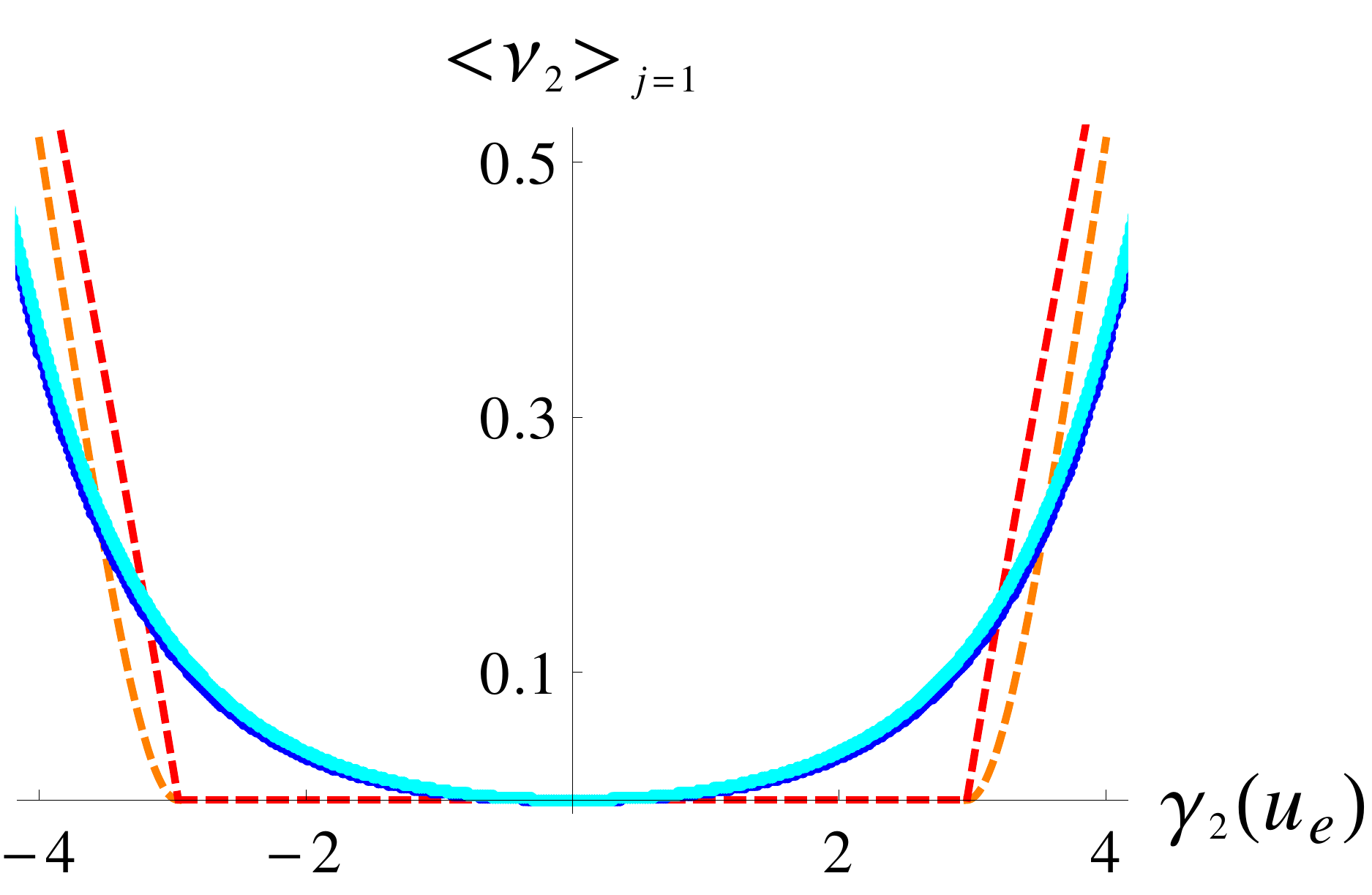}\qquad{}\includegraphics[scale=0.29]{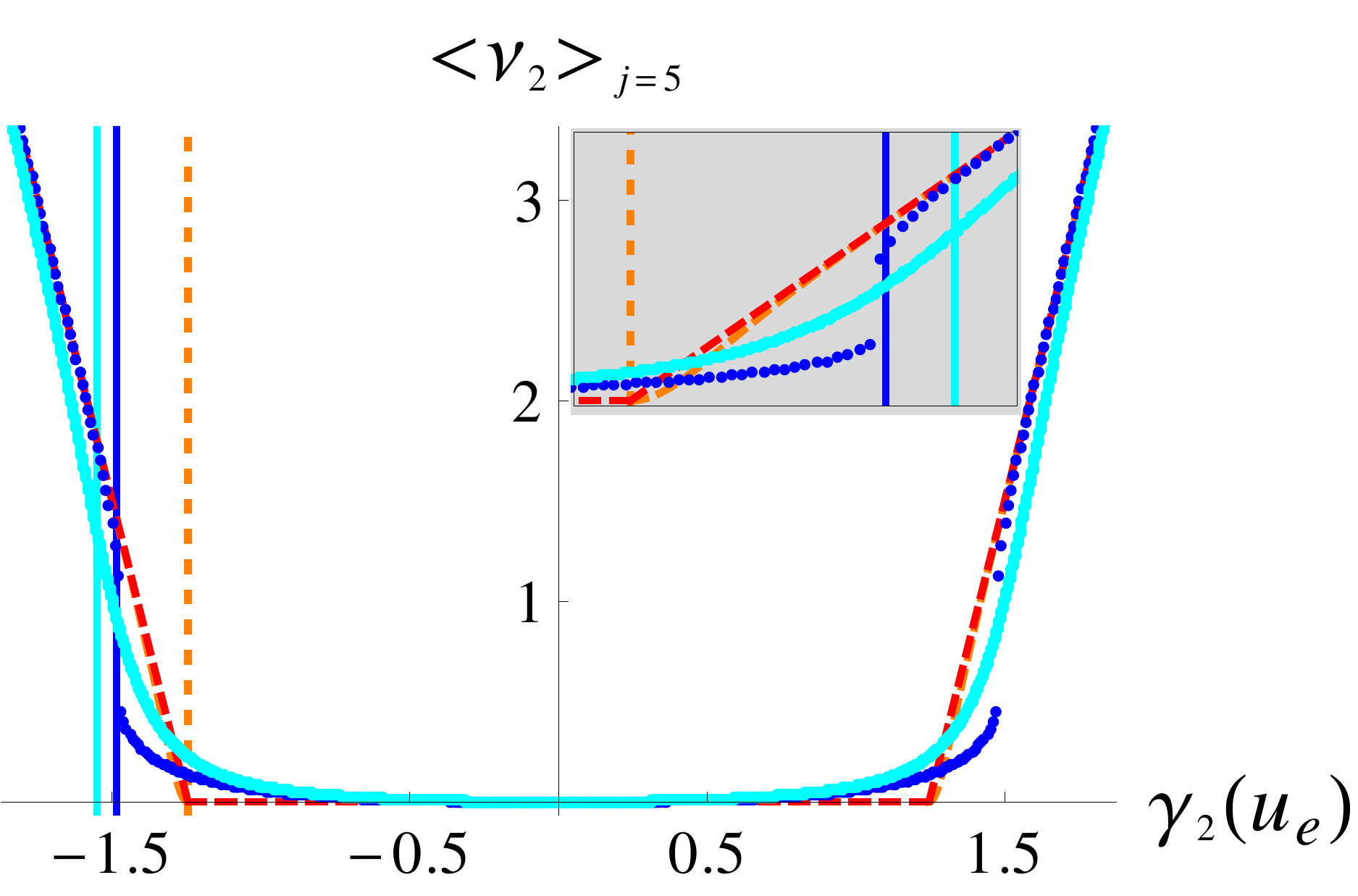}\qquad{}\includegraphics[scale=0.29]{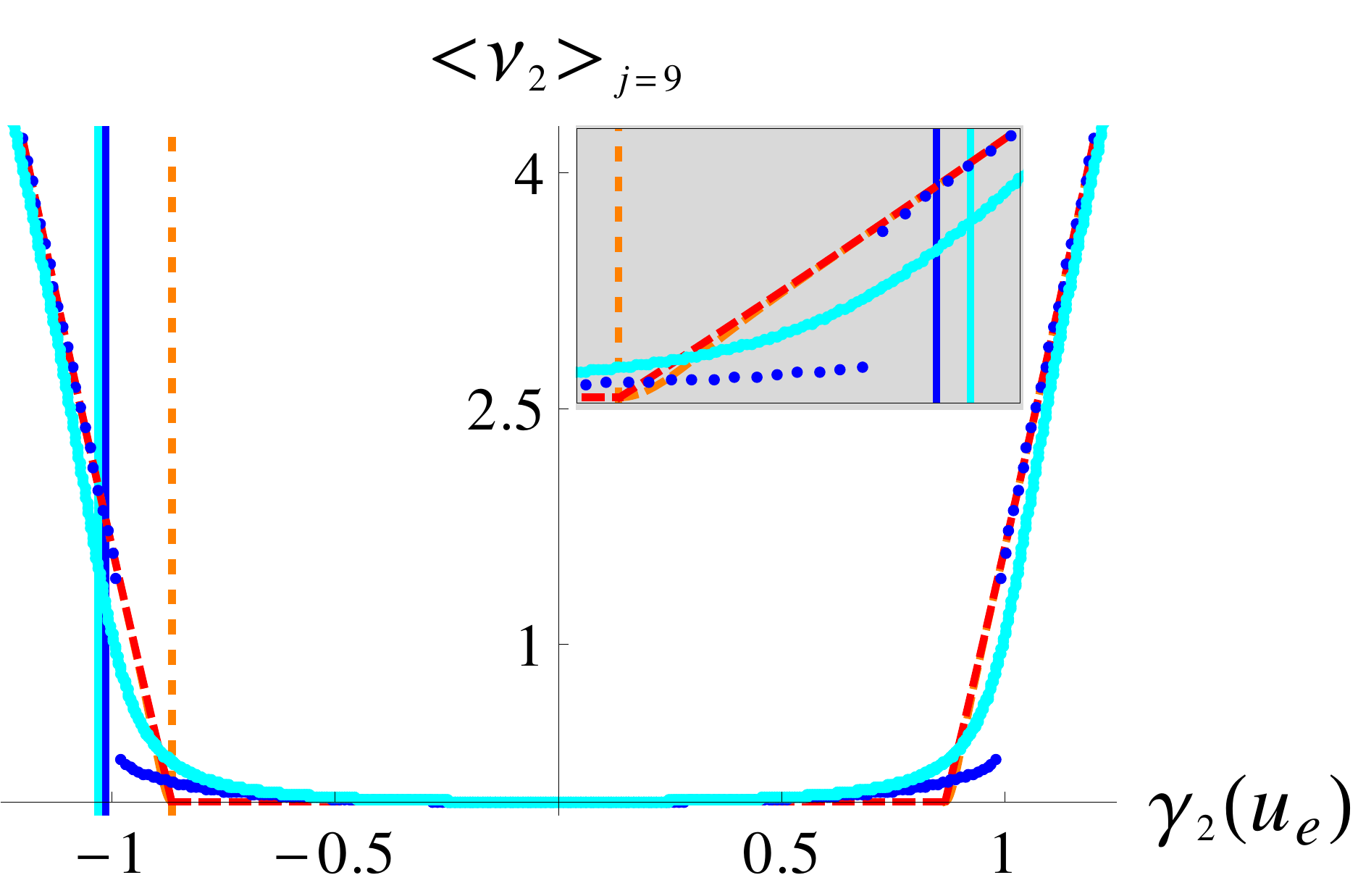}
\par\end{centering}

\caption{Expectation value of $\nu_{2}=a_{2}^{\dagger}a_{2}$ as a function
of $\gamma_{2}$ obtained using CS (dark gray dashed / red online),
SAS with CS's minima (light gray dashed / orange online), SAS minimized
numerically (dark gray / blue online) and quantum solution (light
gray / cyan online). Vertical lines show the transition according
to the quantum solution via fidelity's minimum (light gray / cyan
online), SAS minimized numerically (dark gray / blue online) and SAS
with CS's minima (light gray dashed / orange online). Left: j=1, center:
j=5, right: j=9. Assuming $k=2$ and using $N=18$, $\omega_{A}=\Omega_{1}=\Omega_{2}=2u_{e}$,
$\gamma_{1}=\frac{1}{2}u_{e}$, where $u_{e}$ stands for any energy
unit ($\hbar=1$).\label{fig:5}}
\end{figure*}

\begin{figure*}[t]
\begin{centering}
\includegraphics[scale=0.29]{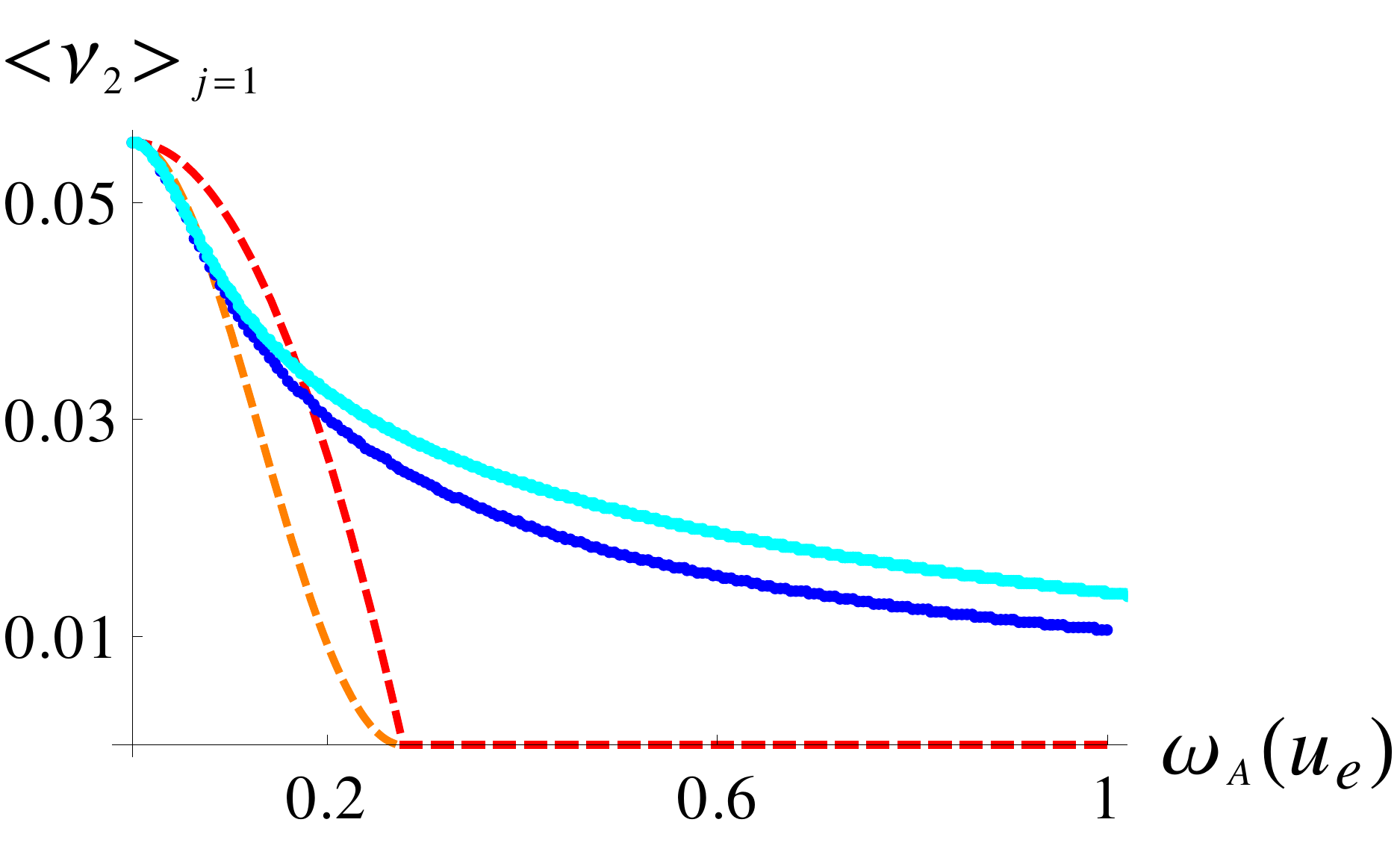}\qquad{}\includegraphics[scale=0.29]{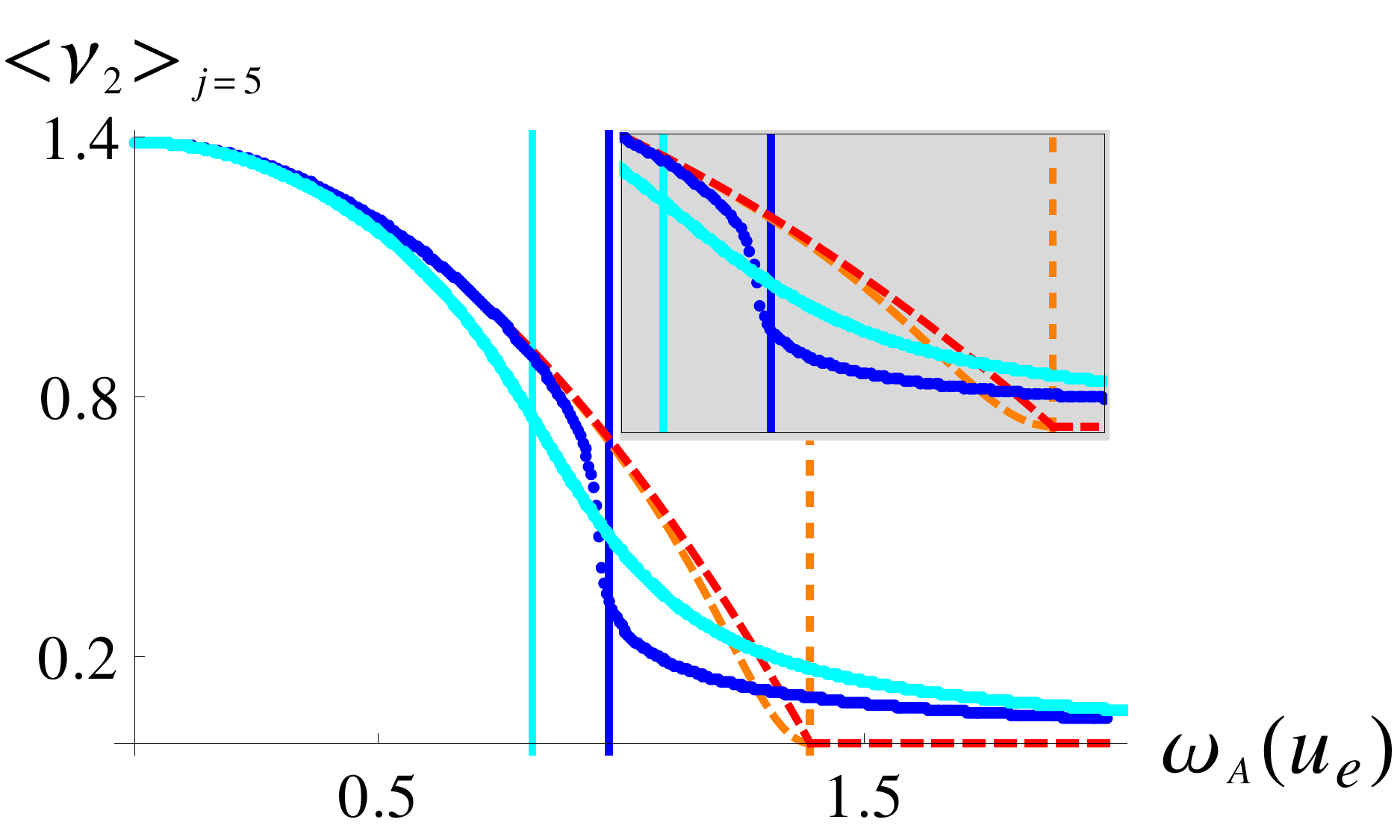}\qquad{}\includegraphics[scale=0.29]{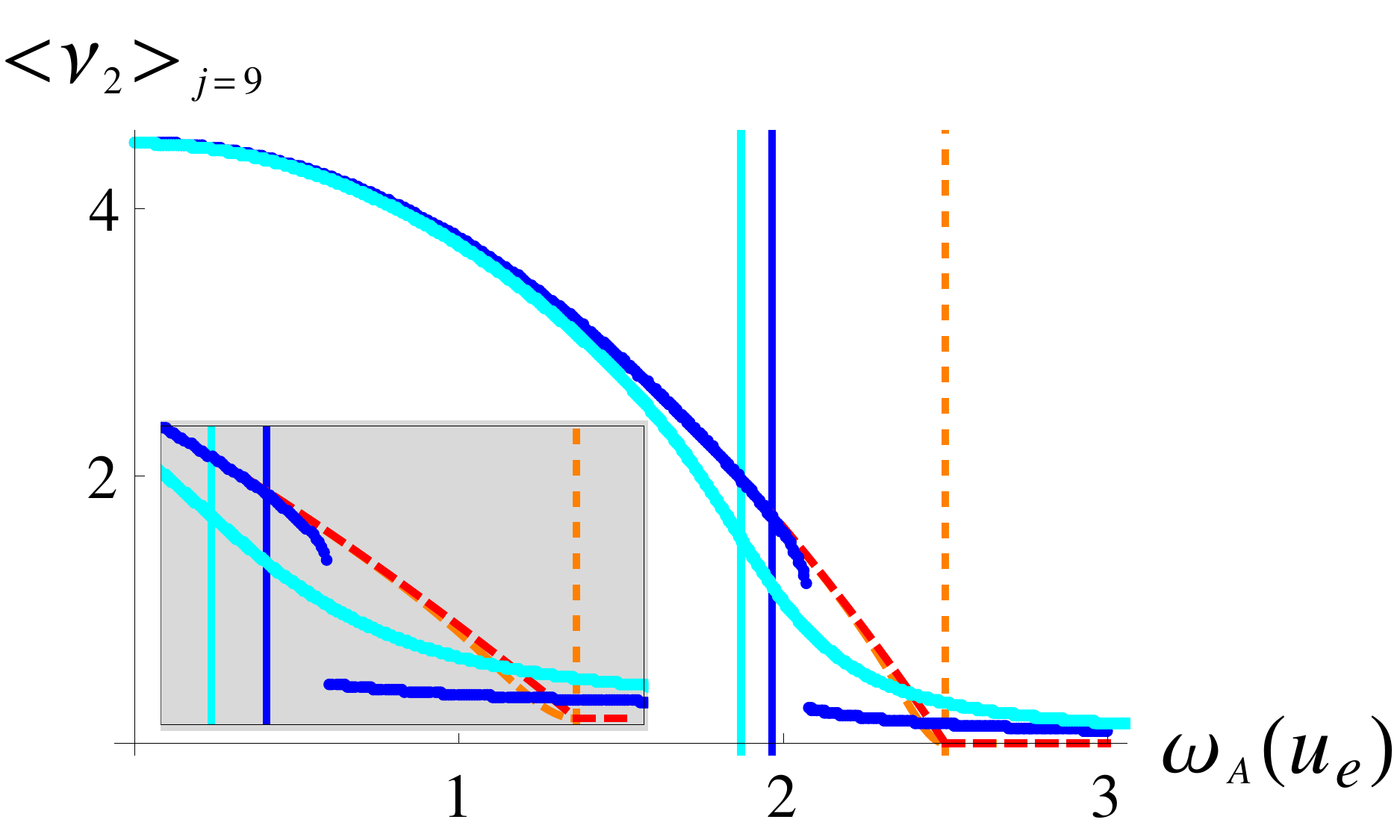}
\par\end{centering}

\caption{Expectation value of $\nu_{2}=a_{2}^{\dagger}a_{2}$ as a function
of $\omega_{A}$ obtained using CS (dark gray dashed / red online),
SAS with CS's minima (light gray dashed / orange online), SAS minimized
numerically (dark gray / blue online) and quantum solution (light
gray / cyan online). Vertical lines show the transition according
to the quantum solution via fidelity's minimum (light gray / cyan
online), SAS minimized numerically (dark gray / blue online) and SAS
with CS's minima (light gray dashed / orange online). Left: j=1, center:
j=5, right: j=9. Assuming $k=2$ and using $N=18$, $\Omega_{1}=\Omega_{2}=2u_{e}$,
$\gamma_{1}=\frac{1}{2}u_{e}$, $\gamma_{2}=1u_{e}$, where $u_{e}$
stands for any energy unit ($\hbar=1$).\label{fig:6}}
\end{figure*}

\section*{Methodology}

There have been various contributions to the study of the phase transition
in the Dicke model (and other two-level models) \cite{key-14,key-15,key-16,key-17,key-18}
and different approaches such as Husimi function analysis \cite{key-19},
entropic uncertainty relations \cite{key-20} and energy surface minimization
\cite{key-21,key-22,key-23,key-24,key-25,key-26}, have been used
for its investigation.

In this work we use the energy surface minimization method, which
consists on minimizing the surface that is obtained by taking the
expectation value of the modeling Hamiltonian with respect to some
trial variational state. The strength of this method lies on the choice
of the trial state, as it is the latter, after minimization, the one
that will be modeling the ground state of the system.

Here we take a variational approach for both matter and radiation
fields, and show how to calculate the QPT of the system modeled by
the Hamiltonian $H$ given in eq. (\ref{eq:2}) via four means:
\begin{enumerate}
\item Using a tensor product of Heisenberg-Weyl HW(1) coherent states for
each mode of the electromagnetic field and SU(2) coherent states for
the atomic field as trial states, and analytically minimizing the
obtained energy surface with respect to its parameters.
\item Using a projection operator on HW(1) coherent states and SU(2) coherent
states to obtain trial states that preserve the parity symmetry of
the Hamiltonian with respect to the total excitation number of the
system (\textit{symmetry adapted states}), and numerically minimize
the obtained energy surface with respect to its parameters.
\item Using symmetry adapted states, as in (2) above, to obtain the energy
surface and ``minimize'' it with the minimizing parameters obtained
in (1) above, thus allowing us to have analytic expressions for the
ground state.
\item Numerically diagonalizing the Hamiltonian, which gives us the exact
quantum solution.
\end{enumerate}
\begin{figure*}[t]
\begin{centering}
\includegraphics[scale=0.29]{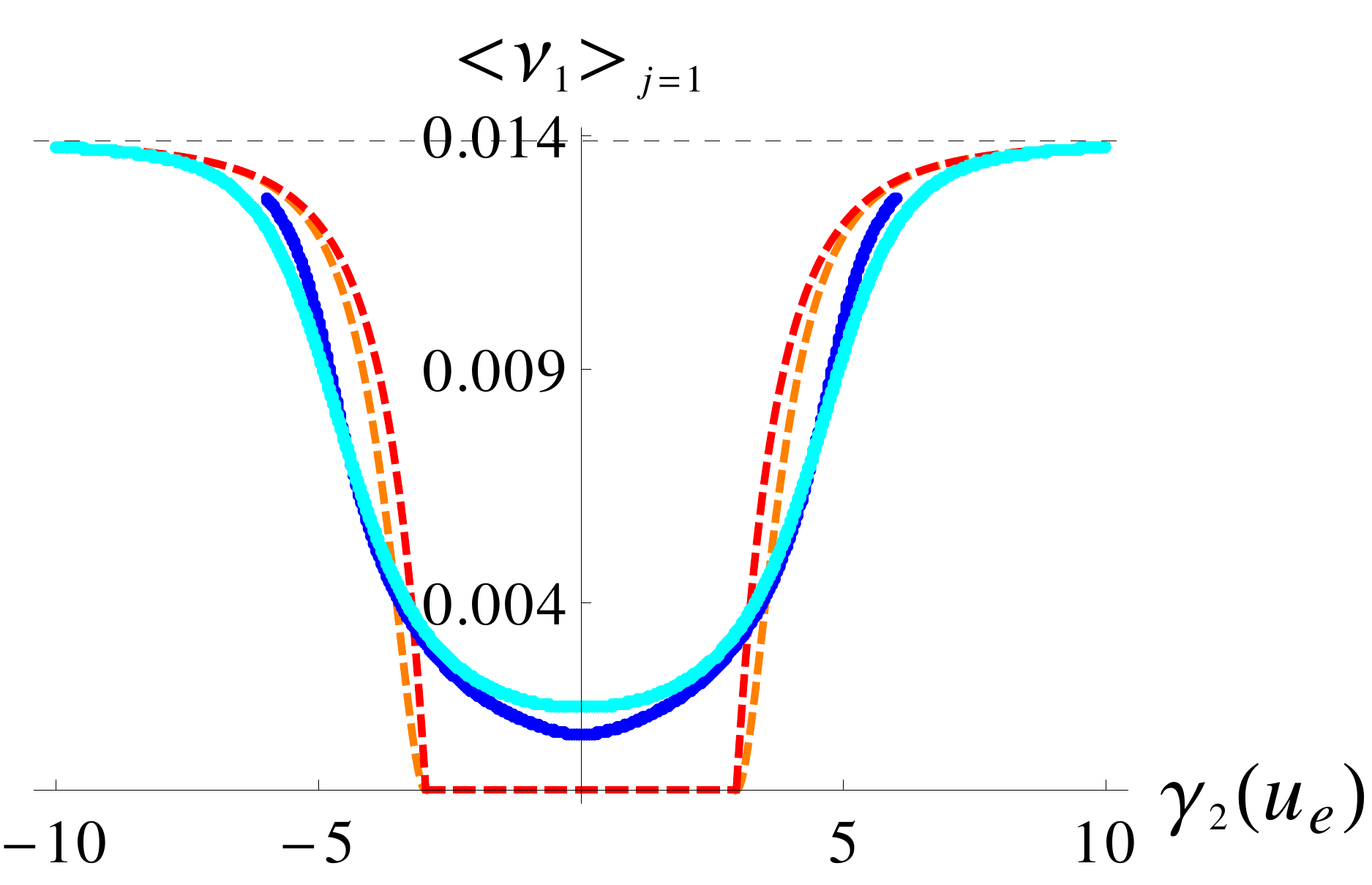}\qquad{}\includegraphics[scale=0.29]{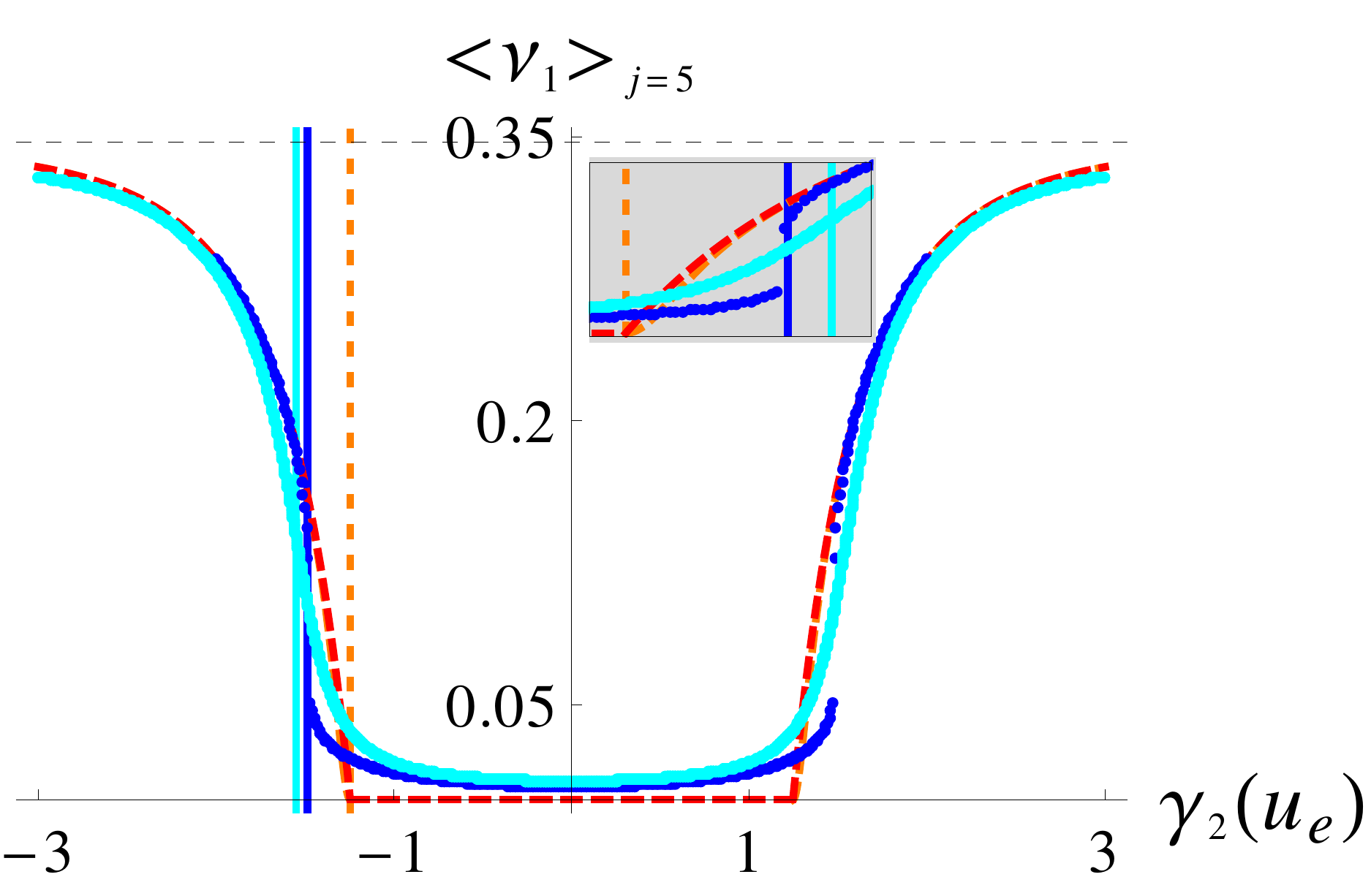}\qquad{}\includegraphics[scale=0.29]{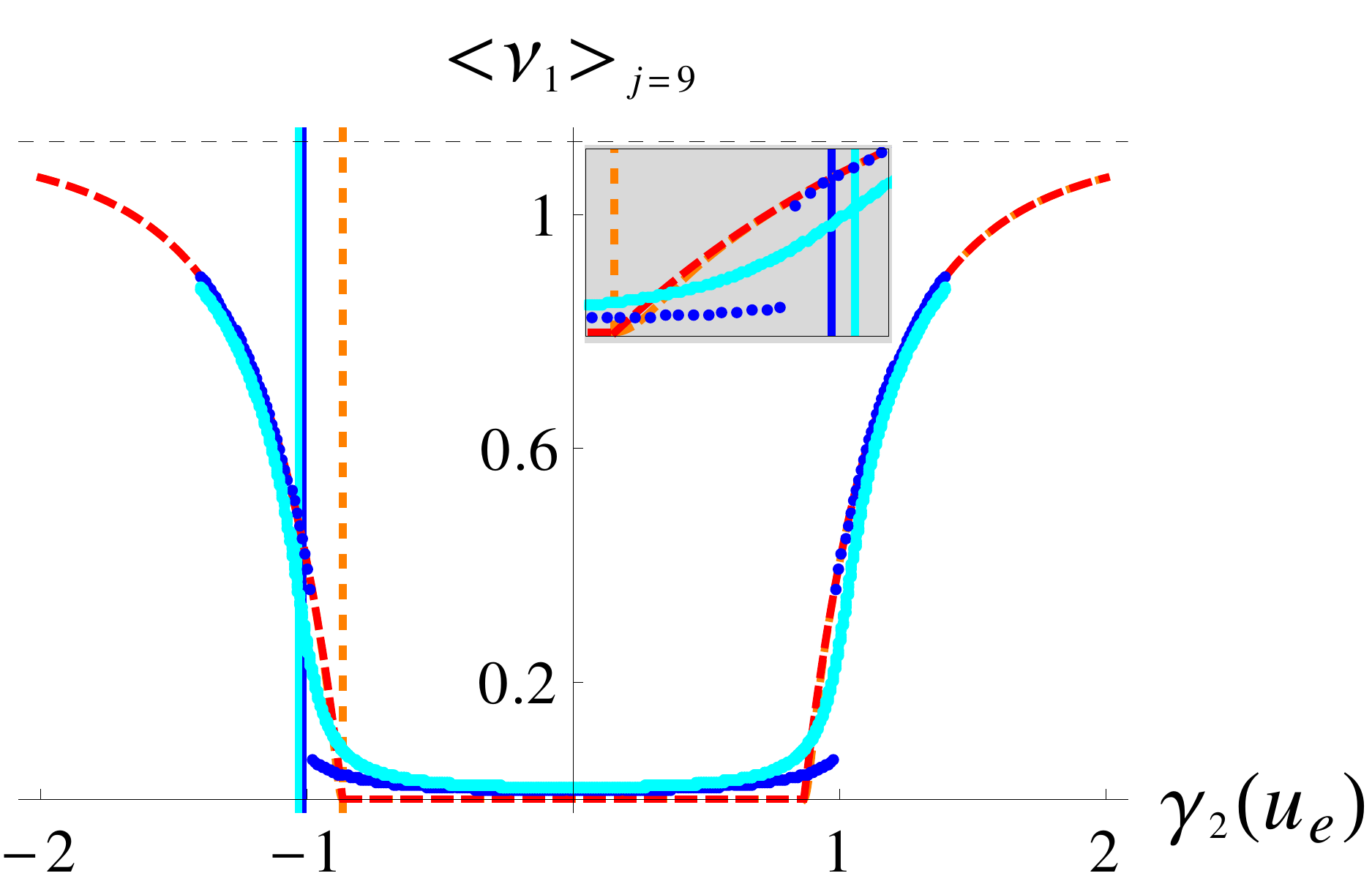}
\par\end{centering}

\caption{Expectation value of $\nu_{1}=a_{1}^{\dagger}a_{1}$ as a function
of $\gamma_{2}$ obtained using CS (dark gray dashed / red online),
SAS with CS's minima (light gray dashed / orange online), SAS minimized
numerically (dark gray / blue online) and quantum solution (light
gray / cyan online). Horizontal dashed line shows the asymptote of
$\left\langle \nu_{1}\right\rangle $ at ${\displaystyle \frac{4j^{2}\gamma_{1}^{2}}{N\Omega_{1}^{2}}}$.
Vertical lines show the transition according to the quantum solution
via fidelity's minimum (light gray / cyan online), SAS minimized numerically
(dark gray / blue online) and SAS with CS's minima (light gray dashed
/ orange online). Left: j=1, center: j=5, right: j=9. Assuming $k=2$
and using $N=18$, $\omega_{A}=\Omega_{1}=\Omega_{2}=2u_{e}$, $\gamma_{1}=\frac{1}{2}u_{e}$,
where $u_{e}$ stands for any energy unit ($\hbar=1$).\label{fig:7}}
\end{figure*}

\subsubsection*{Coherent states (CS)}

For each mode of the electromagnetic field the annihilation and creation
operators $a_{\imath}$ and $a_{\imath}^{\dagger}$, appearing in
the modeling Hamiltonian $H$, satisfy the commutation relations (\ref{eq:3})
of the Lie algebra generators of the Heisenberg-Weyl group HW(1);
hence, a natural choice of a trial state for the radiation field is
a tensor product of $k$ (number of modes) coherent states of HW(1)

\begin{equation}
\left|\bar{\alpha}\right\rangle :=\left|\alpha_{1}\right\rangle \otimes\cdots\otimes\left|\alpha_{k}\right\rangle ,
\end{equation}
where each $\left|\alpha_{\imath}\right\rangle $ is defined as

\begin{equation}
\left|\alpha_{\imath}\right\rangle :=e^{\alpha_{\imath}a_{\imath}^{\dagger}-\alpha_{\imath}^{*}a_{\imath}}\left|0_{\imath}\right\rangle =e^{-\frac{\left|\alpha_{\imath}\right|^{2}}{2}}\sum_{\nu_{\imath}=0}^{\infty}\frac{\alpha_{\imath}^{\nu_{\imath}}}{\sqrt{\nu_{\imath}!}}\left|\nu_{\imath}\right\rangle .
\end{equation}

Furthermore, the commutation relations of the collective spin operators
$J_{-}$, $J_{+}$ and $J_{z}$ (\ref{eq:6}) are the same as the
ones of the Lie algebra generators of the special unitary group SU(2).
Thus, analogously as for the radiation field, we use the coherent
states of SU(2)

\begin{multline}\left|\xi\right\rangle _{j}:=\left|\frac{\upsilon\tan\left|\upsilon\right|}{\left|\upsilon\right|}\right\rangle _{j}:=e^{\upsilon J_{+}-\upsilon^{*}J_{-}}\left|j,0\right\rangle \\=\frac{1}{\left(1+\left|\xi\right|^{2}\right)^{j}}\sum_{m=0}^{2j}\binom{2j}{m}^{\frac{1}{2}}\xi^{m}\left|j,m-j\right\rangle .\end{multline}as
trial states for the matter field.

\subsubsection*{Symmetry adapted states (SAS)}

The modeling Hamiltonian we are considering has a parity symmetry
given by $\left[e^{i\pi\Lambda},H\right]=0$, where $\Lambda=\sqrt{J^{2}+\frac{1}{4}}-\frac{1}{2}+J_{z}+{\displaystyle \sum_{\imath=1}^{k}}a_{\imath}^{\dagger}a_{\imath}$
is the excitation number operator with eigenvalues $\lambda=j+m+{\displaystyle \sum_{\imath=1}^{k}}\nu_{\imath}$.
This symmetry allows us to classify the eigenstates of $H$ in terms
of the parity of the eigenvalues $\lambda$; however, as states with
opposite symmetry are strongly mixed by the CS defined in the previous
section, we should then adapt this symmetry to the CS by projecting
them with the operator $P_{\pm}=\frac{1}{2}\left(I\pm e^{i\pi\Lambda}\right)$,
i.e.

\begin{multline}\left|\bar{\alpha},\xi_{j}\right\rangle _{\pm}:=\mathcal{N}_{\pm}P_{\pm}\left|\bar{\alpha}\right\rangle \otimes\left|\xi\right\rangle _{j} \\=\mathcal{N}_{\pm}\left(\left|\bar{\alpha}\right\rangle \otimes\left|\xi\right\rangle _{j}\pm\left|-\bar{\alpha}\right\rangle \otimes\left|-\xi\right\rangle _{j}\right),\end{multline}with
$\mathcal{N}_{\pm}=\left(2\pm2E\left(-\cos\theta\right)^{2j}\right)^{-\frac{1}{2}}$
the normalization factors for the even (+) and odd (-) states (where
${\displaystyle E=\exp\left\{ -2{\displaystyle \sum_{\imath=1}^{k}}\left|\alpha_{\imath}^{2}\right|\right\} }$).

As we are interested in the ground state of the system, which has
an even parity, we only focus on the state $\left|\bar{\alpha},\xi_{j}\right\rangle _{+}$.

\subsubsection*{Entropy of entanglement ($S_{\varepsilon}$)}

Entropy of entanglement is defined for a bipartite system as the Von
Neumann entropy of either of its reduced states, that is, if $\rho$
is the density matrix of a system in a Hilbert space $\mathcal{H}=\mathcal{H}_{1}\otimes\mathcal{H}_{2}$,
its entropy of entanglement is defined as

\begin{equation}
S_{\varepsilon}:=-Tr\left\{ \rho_{1}\log\rho_{1}\right\} =-Tr\left\{ \rho_{2}\log\rho_{2}\right\} ,
\end{equation}
where $\rho_{1}=Tr_{2}\left\{ \rho\right\} $ and $\rho_{2}=Tr_{1}\left\{ \rho\right\} $.

Our Hamiltonian $H$ models a bipartite system formed by matter and
radiation subsystems, which means that their entropy of entanglement
can be used to see the influence of the QPT on its behavior; this
we do below.

\subsubsection*{Fidelity between neighboring states ($F$)}

Fidelity is a measure of the \textquotedbl{}distance\textquotedbl{}
between two quantum states; given $\left|\phi\right\rangle $ and
$\left|\varphi\right\rangle $ it is defined as

\begin{equation}
F(\phi,\varphi):=\left|\left\langle \phi|\varphi\right\rangle \right|^{2}.
\end{equation}

Across a QPT the ground state of a system suffers a sudden, drastic
change, thus it is natural to expect a drop in the fidelity between
neighboring states near the transition. This drop has been, in fact,
already shown to happen \cite{key-15,key-25} for the case $2j=N$.
We study it here also, and its behavior with the cooperation number.

\begin{figure*}[t]
\begin{centering}
\includegraphics[scale=0.29]{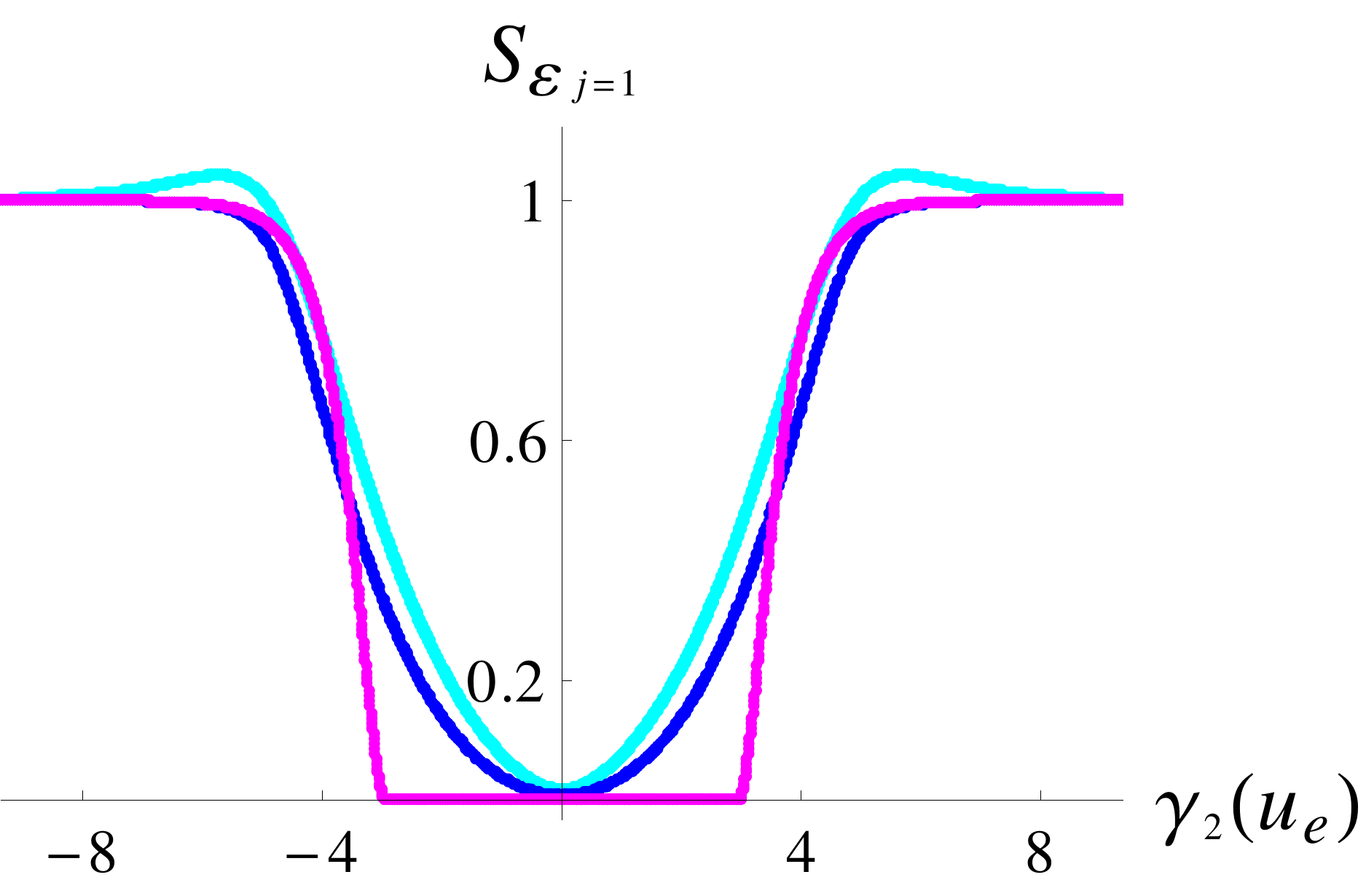}\qquad{}\includegraphics[scale=0.29]{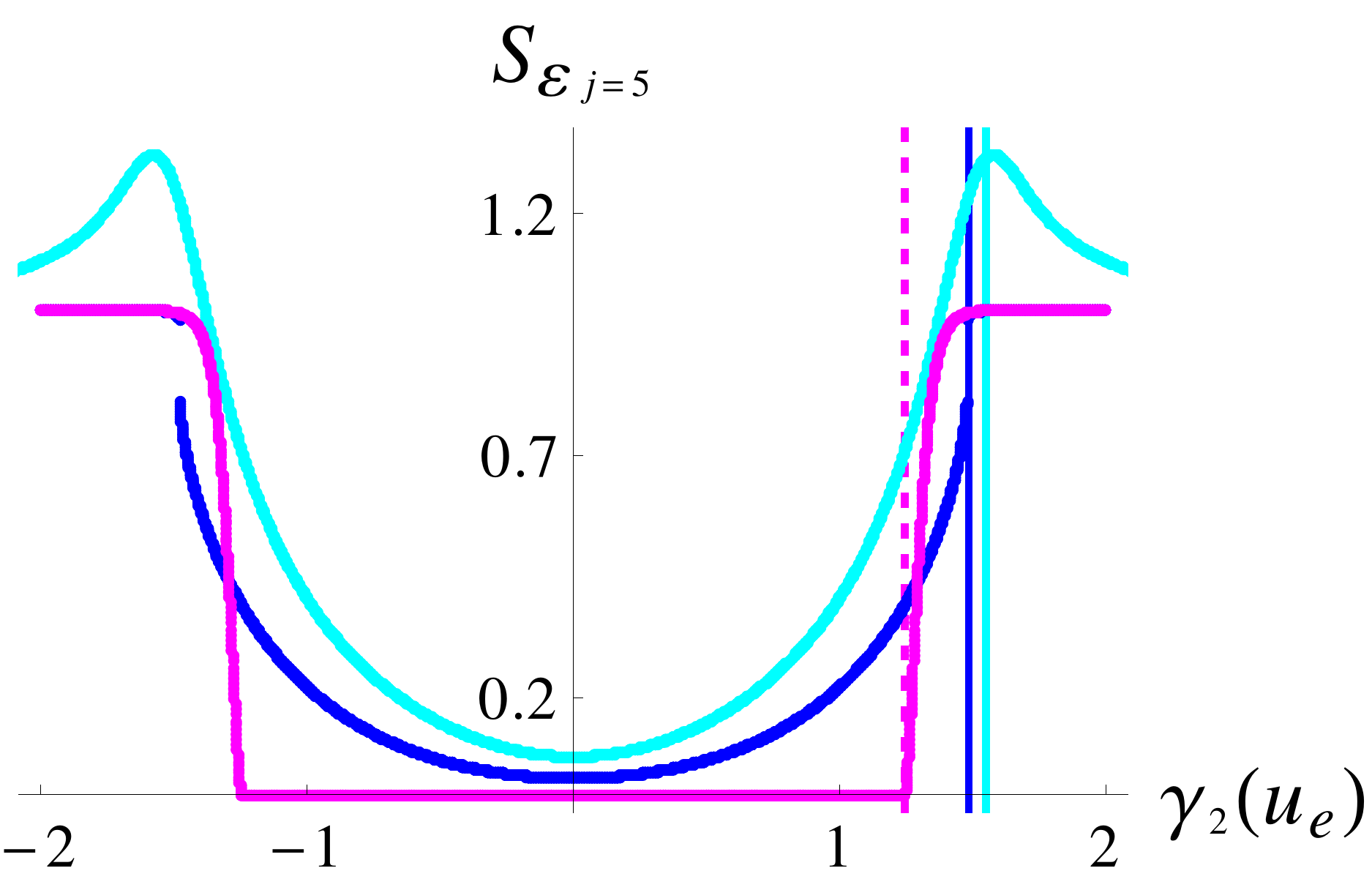}\qquad{}\includegraphics[scale=0.29]{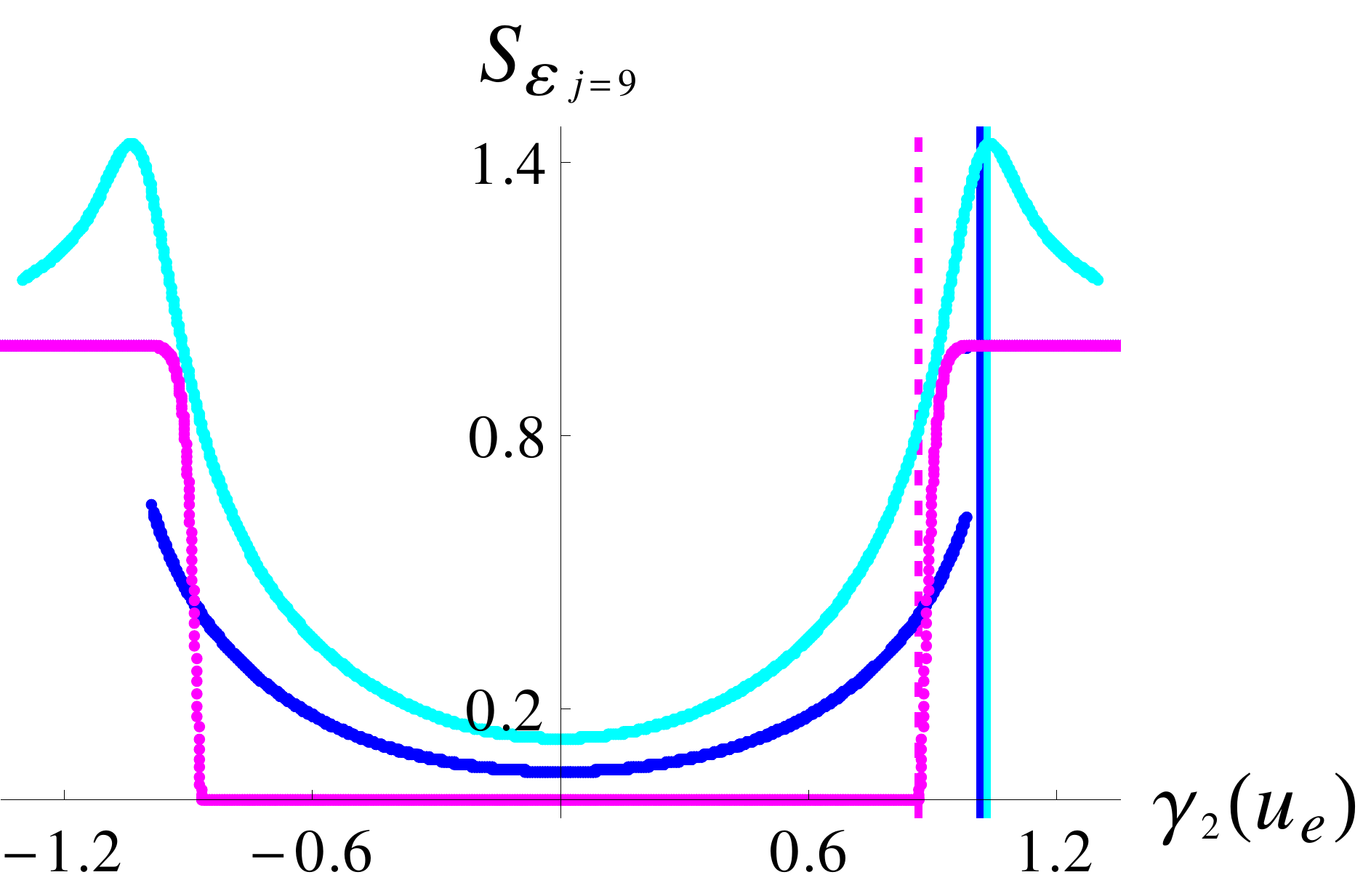}
\par\end{centering}

\caption{Entropy of entanglement as a function of $\gamma_{2}$ obtained SAS
with CS's minima (middle gray / magenta online), SAS minimized numerically
(dark gray / blue online) and quantum solution (light gray / cyan
online). Vertical lines show the transition according to the quantum
solution via fidelity's minimum (light gray / cyan online), SAS minimized
numerically (dark gray / blue online) and SAS with CS's minima (middle
gray dashed / magenta online). Left: j=1, center: j=5, right: j=9.
Assuming $k=2$ and using $N=18$, $\omega_{A}=\Omega_{1}=\Omega_{2}=2u_{e}$,
$\gamma_{1}=\frac{1}{2}u_{e}$, where $u_{e}$ stands for any energy
unit ($\hbar=1$).\label{fig:8}}
\end{figure*}

\begin{figure*}[t]
\begin{centering}
\includegraphics[scale=0.29]{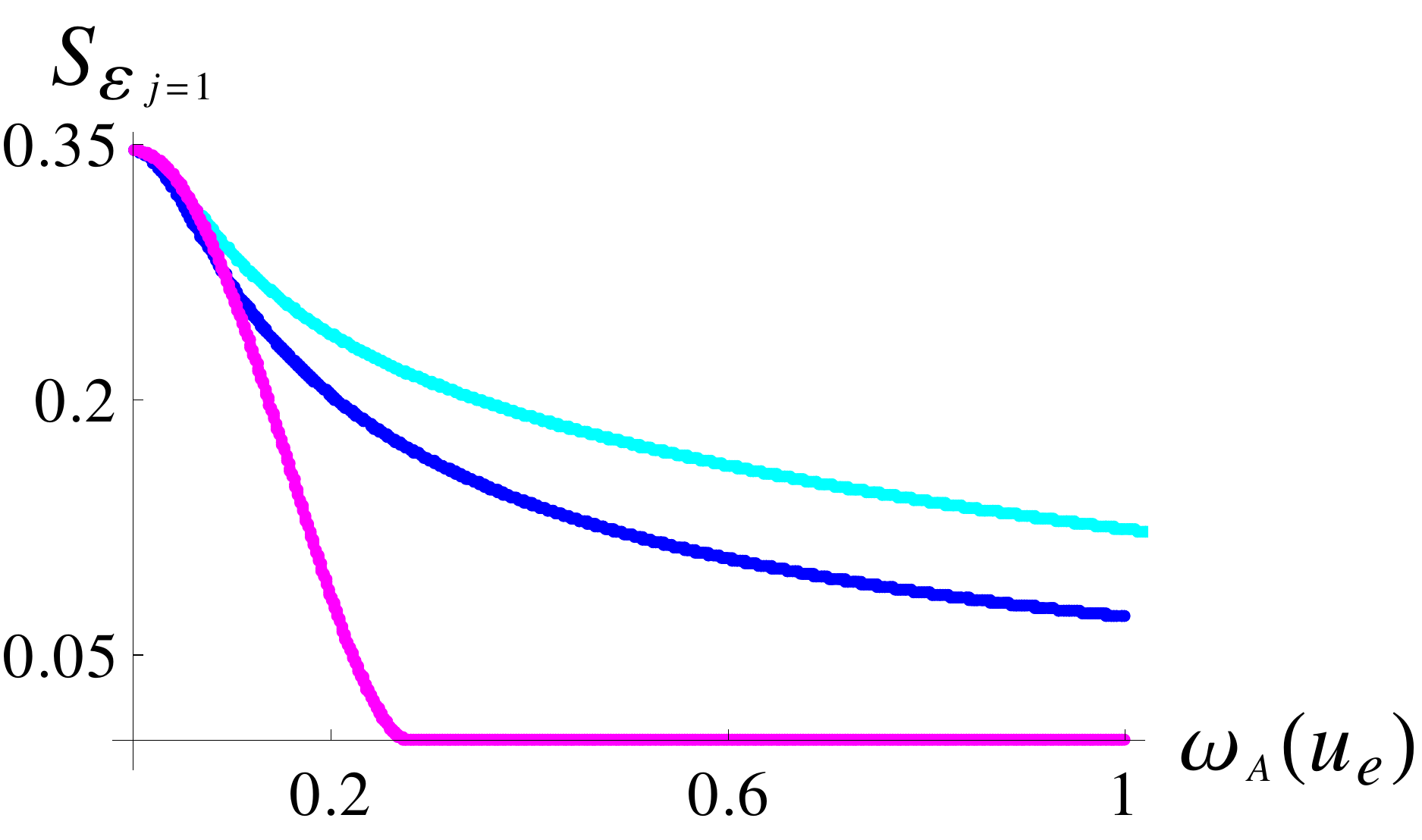}\qquad{}\includegraphics[scale=0.29]{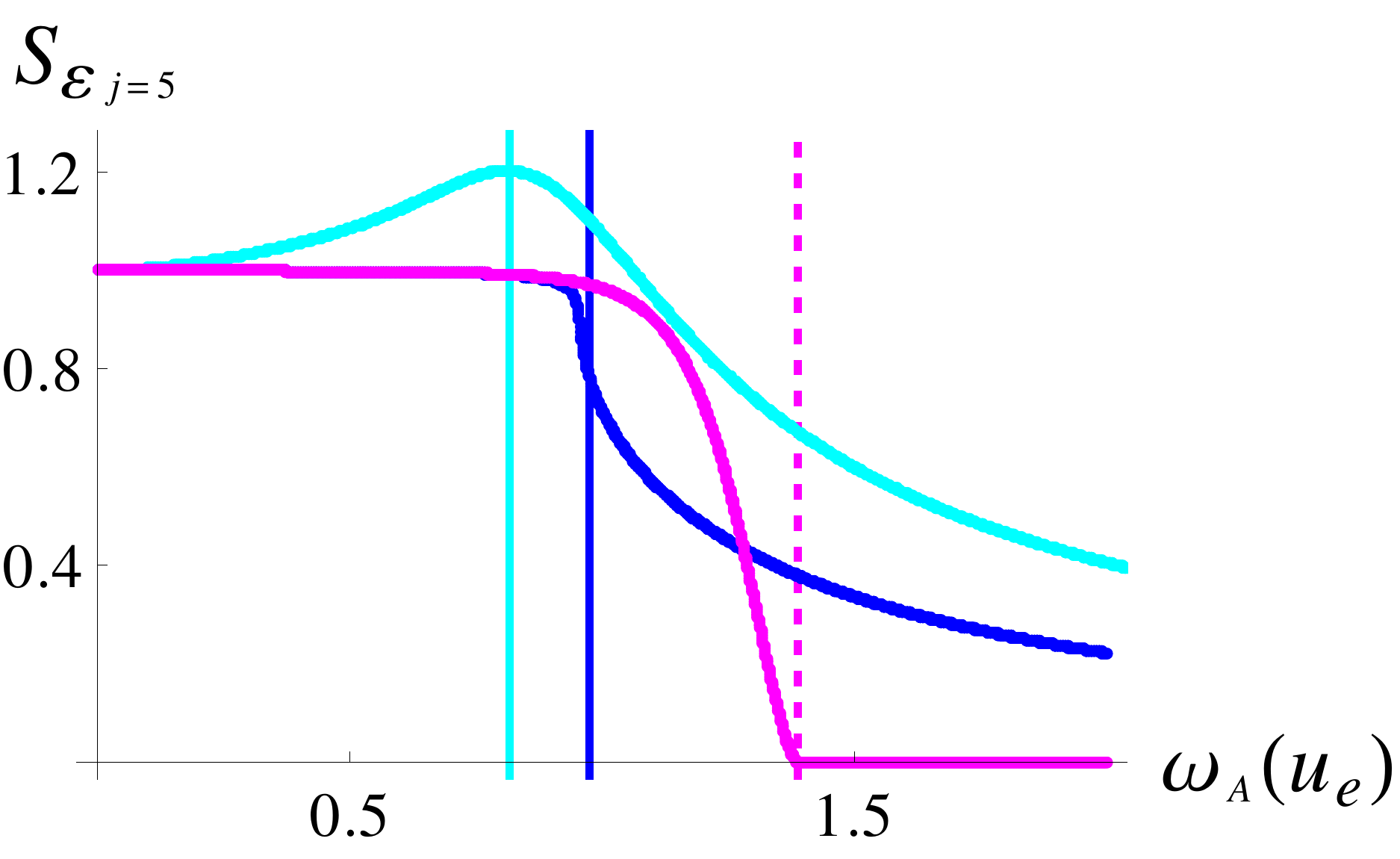}\qquad{}\includegraphics[scale=0.29]{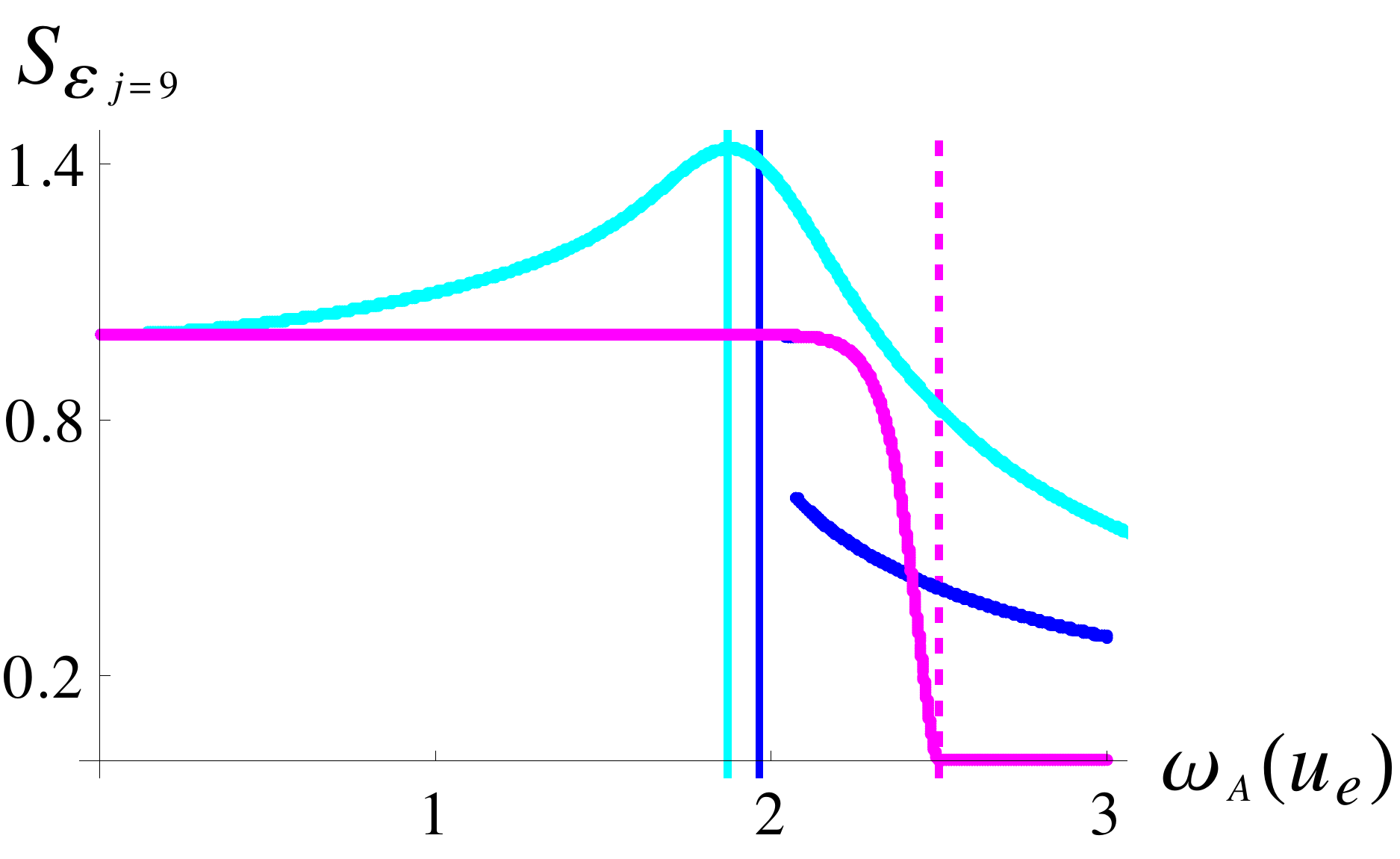}
\par\end{centering}

\caption{Entropy of entanglement as a function of $\omega_{A}$ obtained SAS
with CS's minima (middle gray / magenta online), SAS minimized numerically
(dark gray / blue online) and quantum solution (light gray / cyan
online). Vertical lines show the transition according to the quantum
solution via fidelity's minimum (light gray / cyan online), SAS minimized
numerically (dark gray / blue online) and SAS with CS's minima (middle
gray dashed / magenta online). Left: j=1, center: j=5, right: j=9.
Assuming $k=2$ and using $N=18$, $\Omega_{1}=\Omega_{2}=2u_{e}$,
$\gamma_{1}=\frac{1}{2}u_{e}$, $\gamma_{2}=1u_{e}$, where $u_{e}$
stands for any energy unit ($\hbar=1$).\label{fig:9}}
\end{figure*}

\section*{Results}

Writing the complex labels $\alpha_{\imath}$ and $\xi$ as $\alpha_{\imath}=q_{\imath}+ip_{\iota}$
and $\xi=\tan\left(\frac{\theta}{2}\right)e^{i\phi}$, with $q_{\imath},p_{\imath}\in\mathbb{R}$,
$\theta\in\left[0,\pi\right)$, $\phi\in\left[0,2\pi\right)$, the
CS's energy surface is obtained by taking the expectation value of
the modeling Hamiltonian $H$ with respect to the state $\left|\bar{\alpha}\right\rangle \otimes\left|\xi\right\rangle _{j}$,
and has the form

\begin{multline}   \label{eq:13}   \mathcal{H}_{j,CS}\left(q_{\imath},p_{\imath},\theta,\phi\right):=\left\langle \bar{\alpha}\right|\otimes\left\langle \xi\right|_{j}H\left|\bar{\alpha}\right\rangle \otimes\left|\xi\right\rangle _{j} \\=-j\omega_{A}\cos\theta+{\displaystyle \sum_{\imath=1}^{k}}\Omega_{\imath}\left(q_{\imath}^{2}+p_{\imath}^{2}\right)-\frac{4j}{\sqrt{N}}\sin\theta\cos\phi{\displaystyle \sum_{\imath=1}^{k}}\gamma_{\imath}q_{\imath}. \end{multline}

The critical points which minimize it are then found to be

\[
\ \ \ \ \theta_{c}=q_{\imath_{c}}=p_{\imath_{c}}=0,\ \ \ \ \mbox{for}\;{\displaystyle \omega_{A}\geq\frac{8j}{N}{\displaystyle \sum_{\imath=1}^{k}}\frac{\gamma_{\imath}^{2}}{\Omega_{\imath}}},
\]

\[
\left.\begin{array}{c}
\cos\theta_{c}=\frac{N\omega_{A}}{8j}\left({\displaystyle \sum_{\imath=1}^{k}}\frac{\gamma_{\imath}^{2}}{\Omega_{\imath}}\right)^{-1},\\
\\
\phi_{c}=0,\pi,\\
\\
q_{\imath_{c}}=\frac{2j\gamma_{\imath}}{\Omega_{\imath}\sqrt{N}}\cos\phi_{c}\sin\theta_{c},\\
\\
p_{\imath_{c}}=0
\end{array}\right\} \ \ \ \mbox{for}{\displaystyle \;\omega_{A}<\frac{8j}{N}{\displaystyle \sum_{\imath=1}^{k}}\frac{\gamma_{\imath}^{2}}{\Omega_{\imath}}}.
\]

\begin{figure*}[t]
\begin{centering}
\includegraphics[scale=0.29]{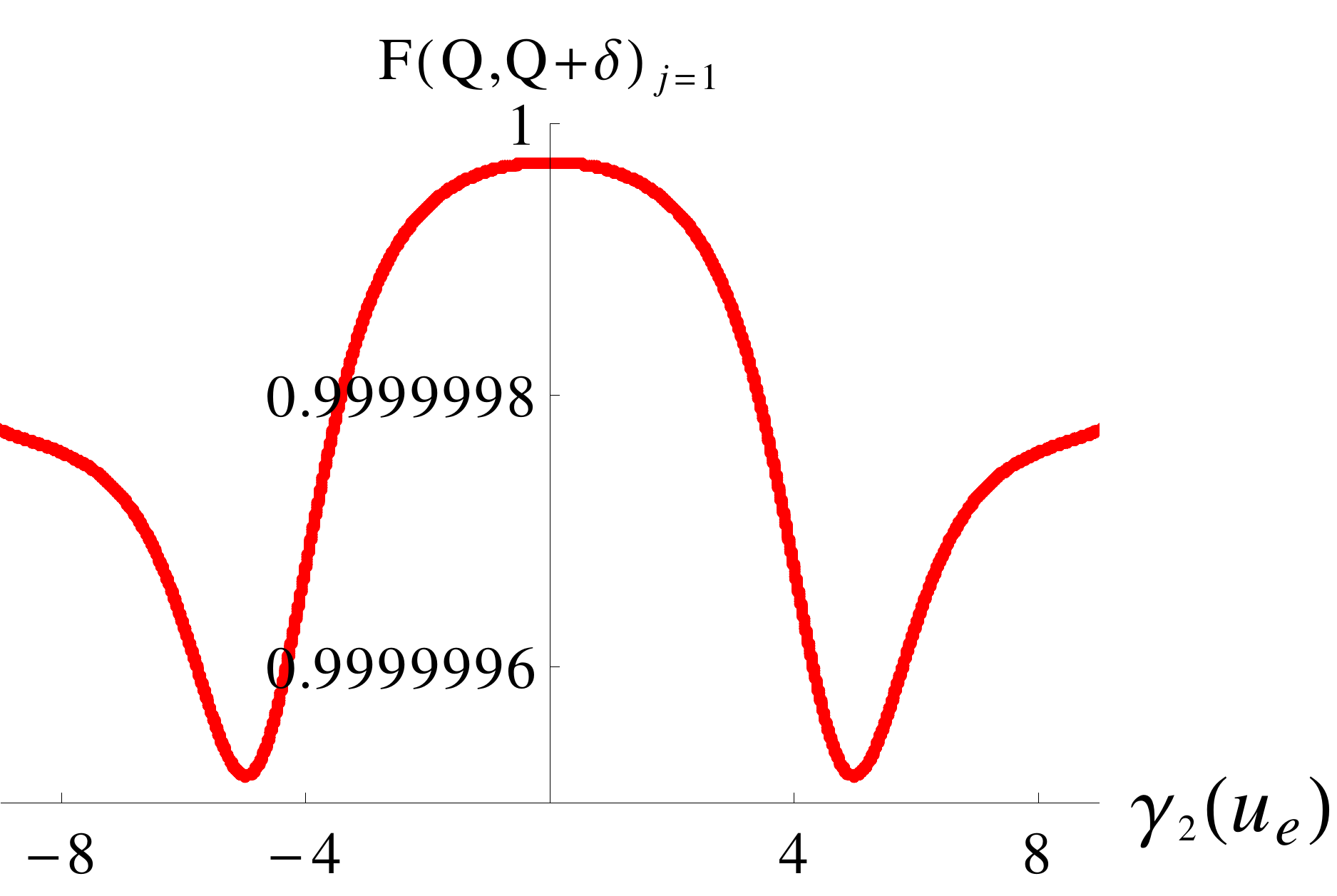}\qquad{}\includegraphics[scale=0.29]{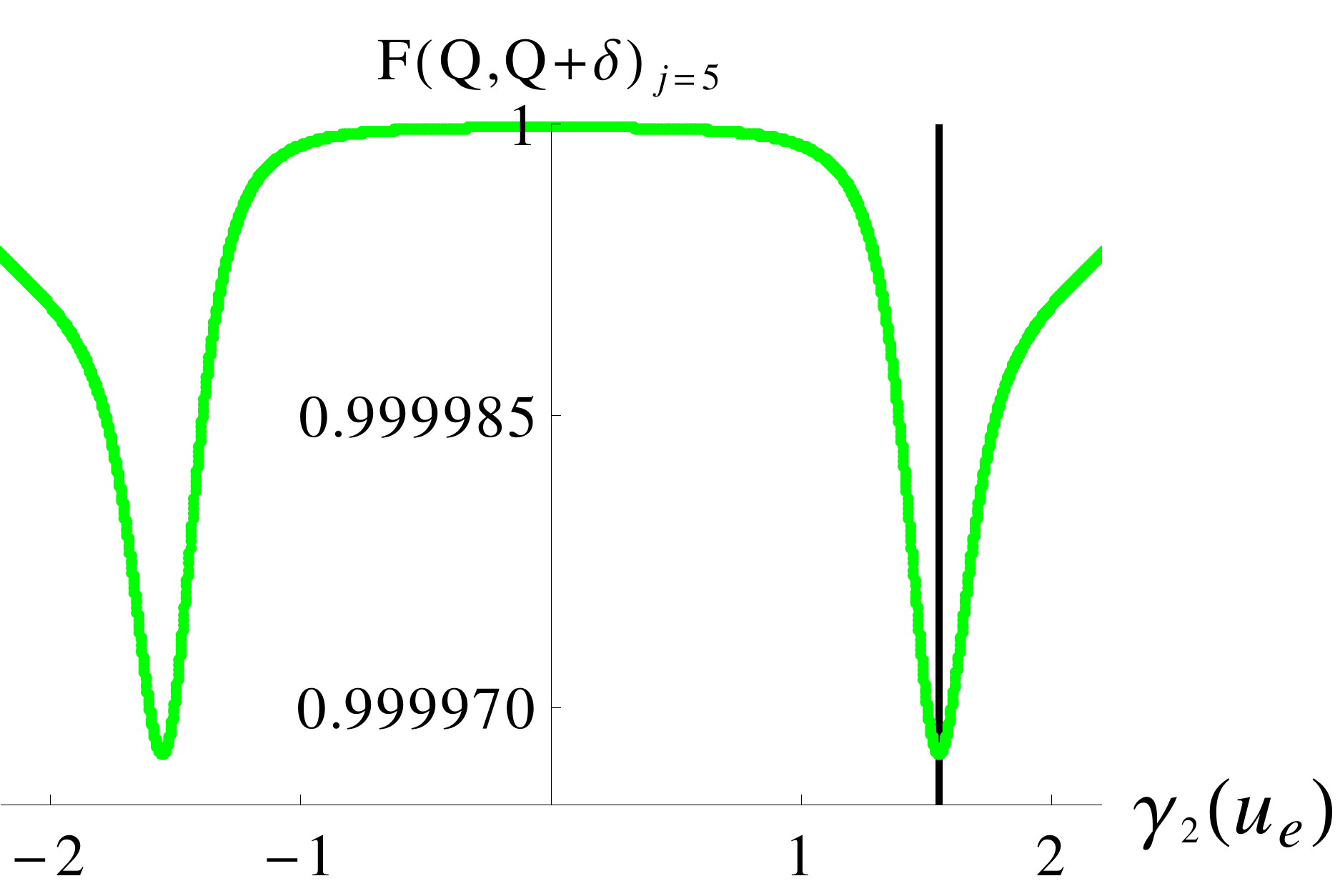}\qquad{}\includegraphics[scale=0.29]{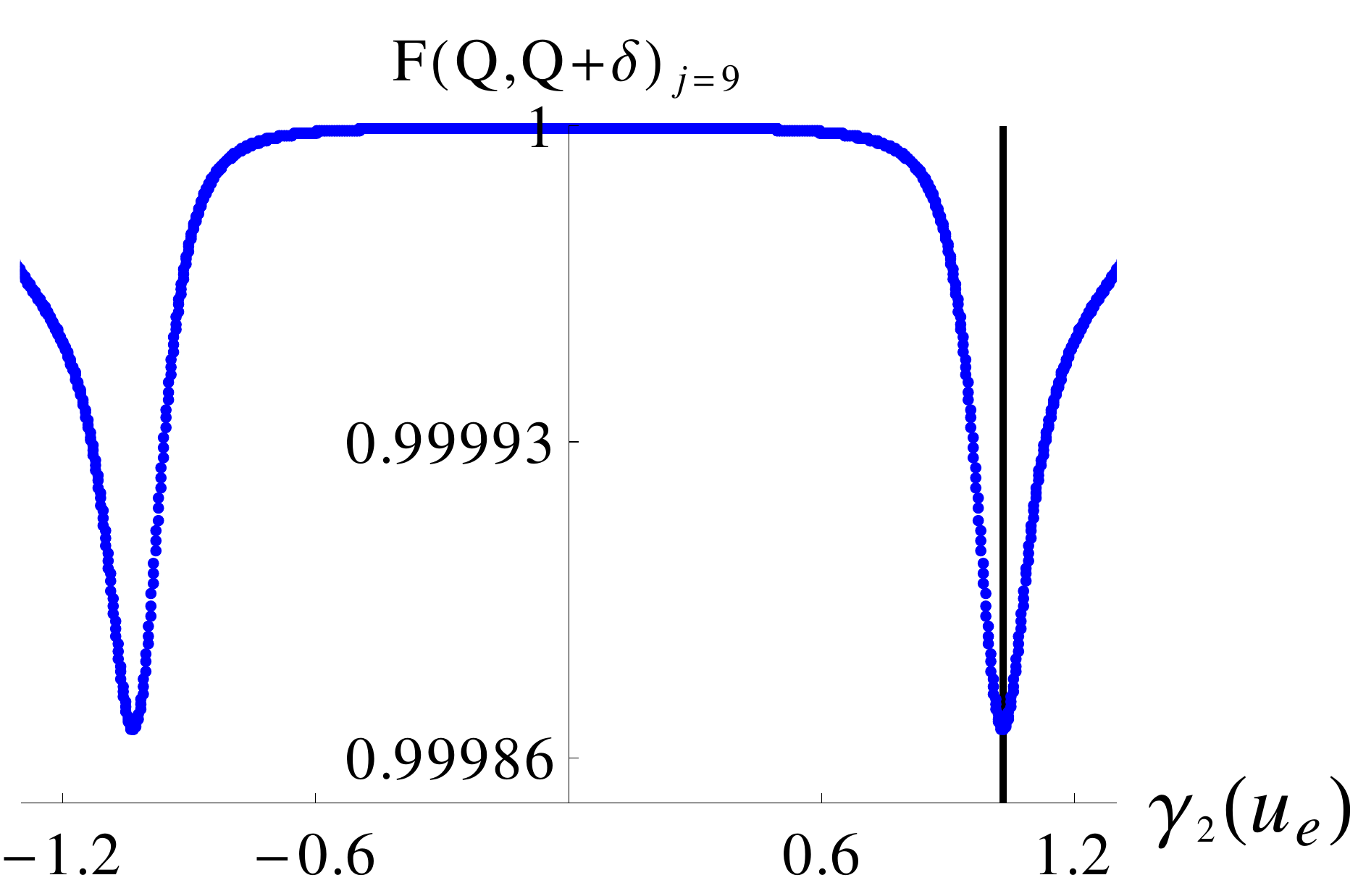}
\par\end{centering}

\caption{Fidelity between neighboring quantum states as a function of $\gamma_{2}$.
Left: j=1, center: j=5, right: j=9. Assuming $k=2$ and using $N=18$,
$\omega_{A}=\Omega_{1}=\Omega_{2}=2u_{e}$, $\gamma_{1}=\frac{1}{2}u_{e}$,
where $u_{e}$ stands for any energy unit ($\hbar=1$). Vertical black
line shows the fidelity's minimum (i.e. the quantum phase transition)
for $j=5$ and $j=9$.\label{fig:14}}
\end{figure*}

\begin{figure*}[t]
\begin{centering}
\includegraphics[scale=0.29]{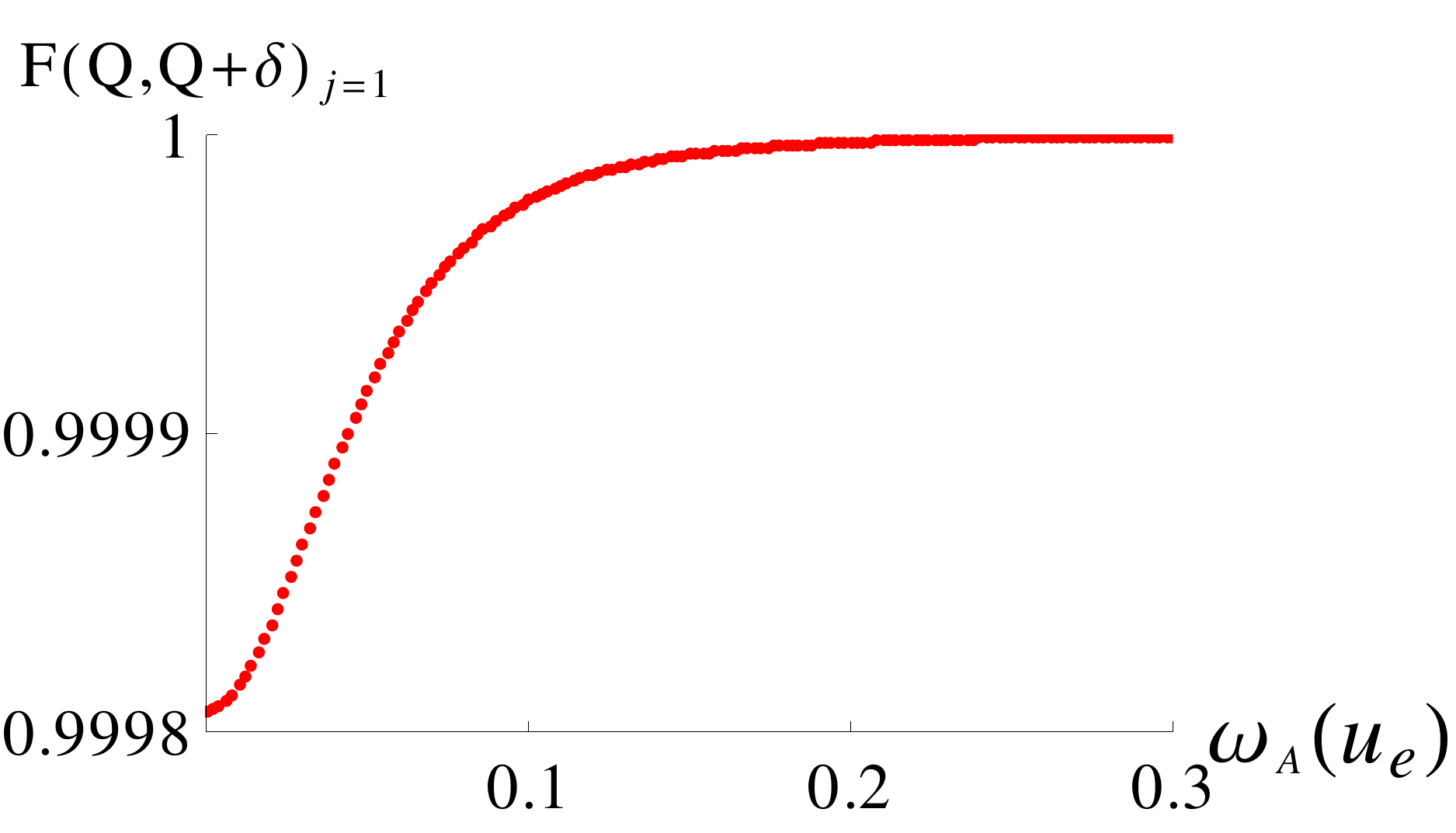}\qquad{}\includegraphics[scale=0.29]{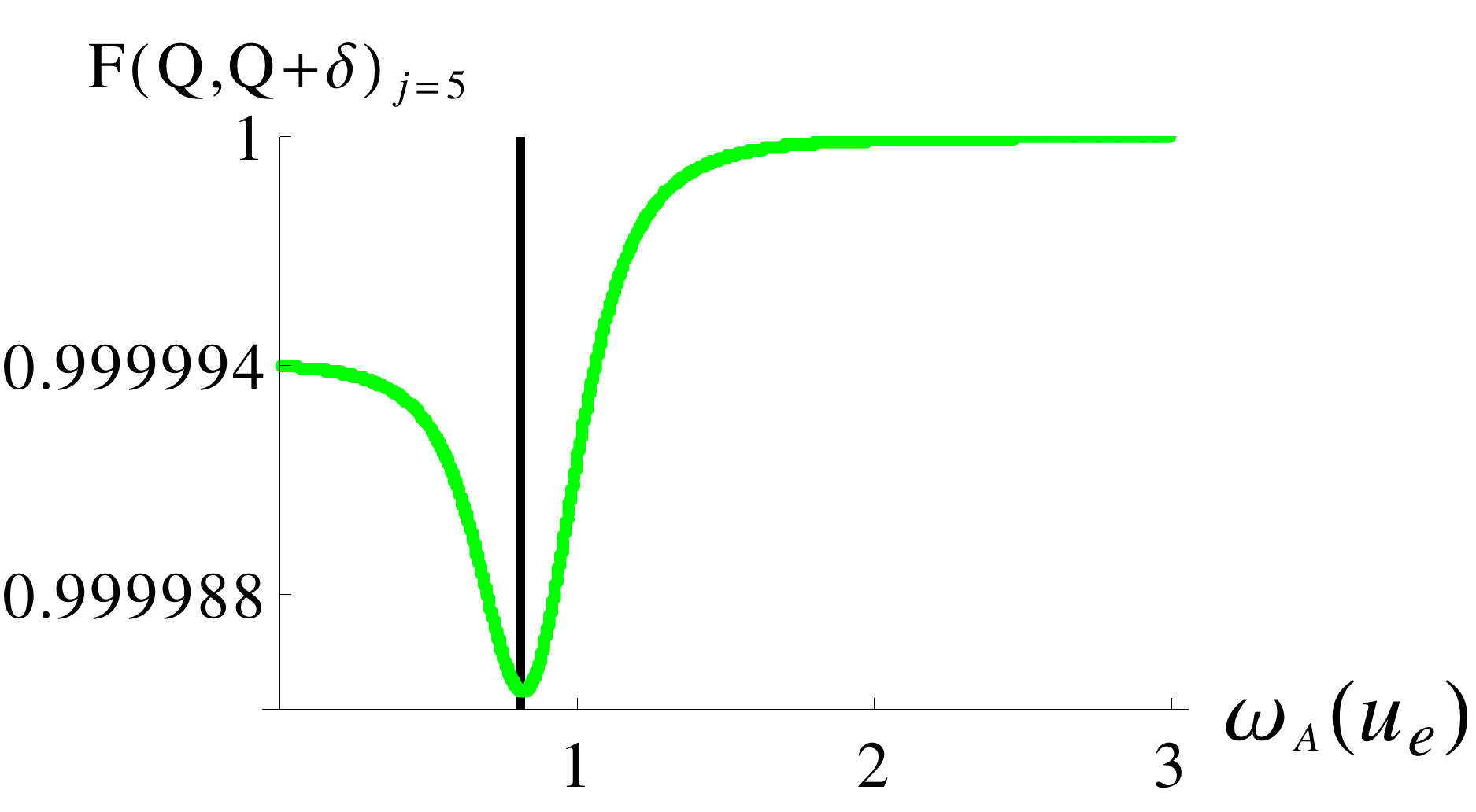}\qquad{}\includegraphics[scale=0.29]{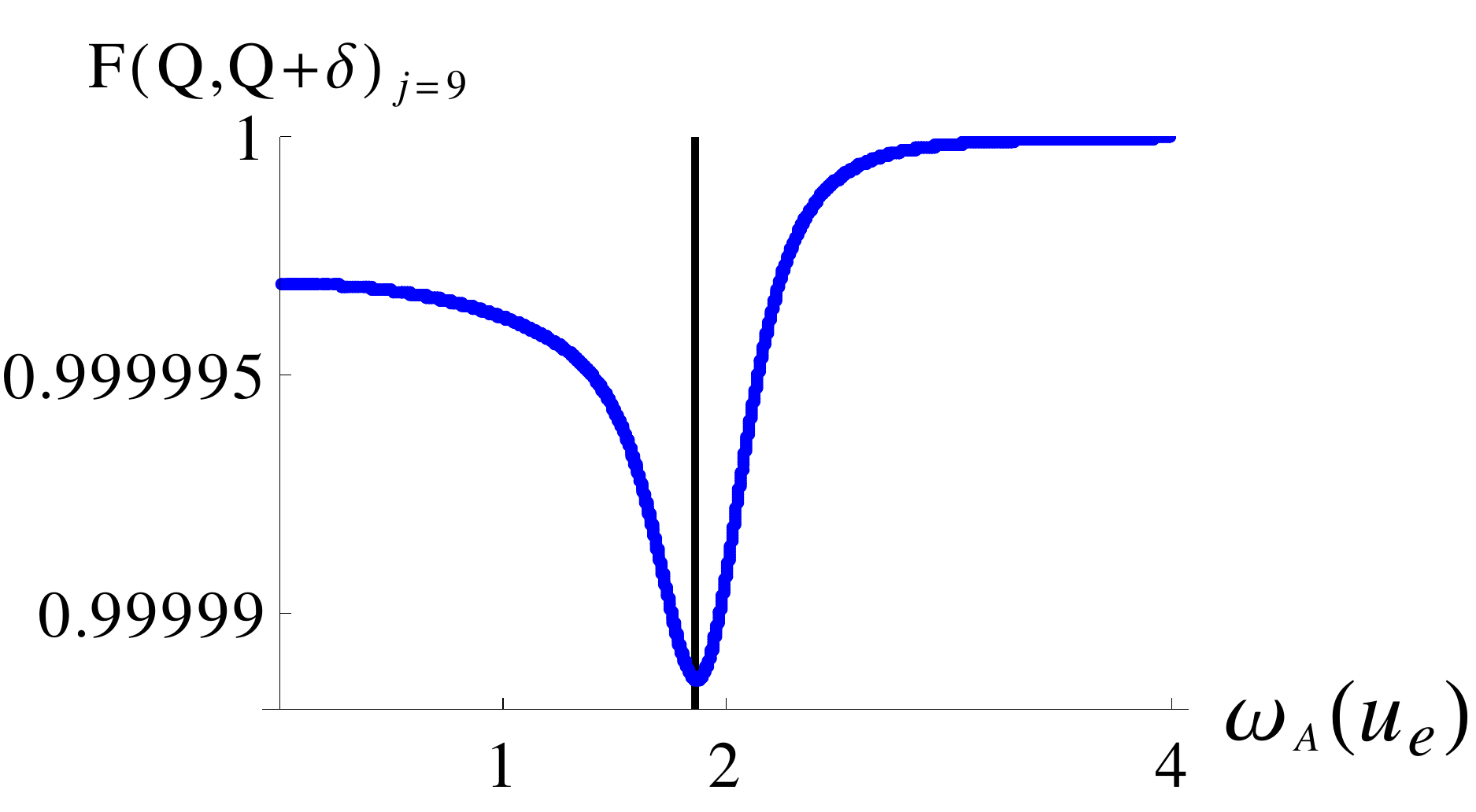}
\par\end{centering}

\caption{Fidelity between neighboring quantum states as a function of $\omega_{A}$.
Left: j=1, center: j=5, right: j=9. Assuming $k=2$ and using $N=18$,
$\Omega_{1}=\Omega_{2}=2u_{e}$, $\gamma_{1}=\frac{1}{2}u_{e}$, $\gamma_{2}=1u_{e}$,
where $u_{e}$ stands for any energy unit ($\hbar=1$). Vertical black
line shows the fidelity's minimum (i.e. the quantum phase transition)
for $j=5$ and $j=9$.\label{fig:15}}
\end{figure*}

Substituting these values into (\ref{eq:13}) we obtain the energy
of the coherent ground state as a function of the Hamiltonian parameters,

\begin{equation}
\mathcal{E}_{CS}\left(\omega_{A},\gamma_{\imath}\right)=\left\{ \begin{matrix}{\displaystyle -j\omega_{A}} & , & \mbox{for}\;{\displaystyle \delta\geq1}\\
{\displaystyle -\frac{j\omega_{A}}{2}\left(\frac{1}{\delta}+\delta\right)} & , & {\displaystyle \mbox{for}\;\delta<1},
\end{matrix}\right.\label{eq:14}
\end{equation}
where we have defined ${\displaystyle {\displaystyle \delta=\frac{N\omega_{A}}{8j\varsigma}}}$
with ${\displaystyle \varsigma={\displaystyle \sum_{\imath=1}^{k}}\frac{\gamma_{\imath}^{2}}{\Omega_{\imath}}}$.
Using the information of this coherent ground state we also obtain
the expectation values of the atomic relative population operator
$J_{z}$ and of the number of photons of mode $\imath$ operator $\nu_{\imath}:=a_{\imath}^{\dagger}a_{\imath}$:

\begin{equation}
\left\langle J_{z}\right\rangle _{CS}\left(\omega_{A},\gamma_{\imath}\right)=\left\{ \begin{matrix}{\displaystyle -j} & , & \mbox{for}\;{\displaystyle \delta\geq1}\\
{\displaystyle -j\delta} & , & {\displaystyle \mbox{for}\;\delta<1},
\end{matrix}\right.
\end{equation}

\begin{equation}
\left\langle \nu_{\imath}\right\rangle _{CS}\left(\omega_{A},\gamma_{\imath},\gamma_{\jmath}\right)=\left\{ \begin{matrix}{\displaystyle 0} & , & \mbox{for}\;{\displaystyle \delta\geq1}\\
{\displaystyle \frac{\gamma_{\imath}^{2}}{\Omega_{\imath}^{2}}\frac{j\omega_{A}}{2\varsigma}\left(\frac{1}{\delta}-\delta\right)} & , & {\displaystyle \mbox{for}\;\delta<1}.
\end{matrix}\right.
\end{equation}

Analogously as for the CS's energy surface, the SAS's energy surface
is obtained by taking the expectation value of the modeling Hamiltonian
$H$ with respect to the state $\left|\bar{\alpha},\xi_{j}\right\rangle _{+}$,
and has the more complicated form

\begin{multline}\label{eq:17}\mathcal{H}_{j,SAS}\left(q_{\imath},p_{\imath},\theta,\phi\right):=\left\langle \bar{\alpha},\xi_{j}\right|_{+}H\left|\bar{\alpha},\xi_{j}\right\rangle _{+}\\=\left(\frac{1+E\left(-\cos\theta\right)^{2j-2}}{1+E\left(-\cos\theta\right)^{2j}}\right)\left(-j\omega_{A}\cos\theta\right)\\+\left(\frac{1-E\left(-\cos\theta\right)^{2j}}{1+E\left(-\cos\theta\right)^{2j}}\right){\displaystyle \sum_{\ell=1}^{k}}\Omega_{\ell}\left(q_{\ell}^{2}+p_{\ell}^{2}\right)\\-\frac{4j}{\sqrt{N}}\sin\theta{\displaystyle \sum_{\ell=1}^{k}\left\{ \frac{\cos\phi\gamma_{\ell}q_{\ell}+E\left(-\cos\theta\right)^{2j-1}\sin\phi\gamma_{\ell}p_{\ell}}{1+E\left(-\cos\theta\right)^{2j}}\right\} }.\end{multline}

As a first approximation, we may substitute the critical values obtained
for the CS's energy surface into (\ref{eq:17}), we obtain the trial
state which approximates the lowest symmetry-adapted energy state,
and with respect to which we evaluate the expectation values of the
observables $H$, $J_{z}$ and $\nu_{\imath}$:

\begin{multline}\label{eq:18}\mathcal{E}_{SAS}\left(\omega_{A},\gamma_{\imath}\right)\\=\left\{ \begin{array}{cc}{\displaystyle -j\omega_{A}} & ,\,\mbox{for}\;{\displaystyle \delta\geq1}\\\\{\displaystyle -j\omega_{A}\left[\delta\left(\frac{1+\varepsilon\left(-\delta\right)^{2j-2}}{1+\varepsilon\left(-\delta\right)^{2j}}\right)\right.}\\{\displaystyle \left.+\frac{1}{2}\left(\frac{1}{\delta}-\delta\right)\right]} & ,\,\mbox{for}\;{\displaystyle \delta<1,}\end{array}\right.\end{multline}

\begin{multline}\left\langle J_{z}\right\rangle _{SAS}\left(\omega_{A},\gamma_{\imath}\right)\\=\left\{ \begin{matrix}{\displaystyle -j} & , & \mbox{for}\;{\displaystyle \delta\geq1}\\{\displaystyle {\displaystyle -j\delta}\left(\frac{1+\varepsilon\left(-\delta\right)^{2j-2}}{1+\varepsilon\left(-\delta\right)^{2j}}\right)} & , & {\displaystyle \mbox{for}\;\delta<1,}\end{matrix}\right.\end{multline}

\begin{multline}\left\langle \nu_{\imath}\right\rangle _{SAS}\left(\omega_{A},\gamma_{\imath},\gamma_{\jmath}\right)\\=\left\{ \begin{matrix}{\displaystyle 0} & , & \mbox{for}\;{\displaystyle \delta\geq1}\\{\displaystyle \frac{\gamma_{\imath}^{2}}{\Omega_{\imath}^{2}}\frac{j\omega_{A}}{2\varsigma}\left(\frac{1}{\delta}-\delta\right)\left(\frac{1-\varepsilon\left(-\delta\right)^{2j}}{1+\varepsilon\left(-\delta\right)^{2j}}\right)} & , & {\displaystyle \mbox{for}\;\delta<1,}\end{matrix}\right.\end{multline}where
${\displaystyle \varepsilon=\exp\left\{ {\displaystyle \frac{-j\omega_{A}\sigma}{\varsigma}}\right\} }$
with ${\displaystyle \sigma={\displaystyle \sum_{\imath=1}^{k}}\frac{\gamma_{\imath}^{2}}{\Omega_{\imath}^{2}}}$.

Of course, we can minimize eq. (\ref{eq:17}) numerically for the
SAS and obtain the expectation value of the relevant matter and field
observables.

In our numerical analysis we study the case with two modes of the
radiation field, as it is the maximum number of orthogonal modes that
can be present in a 3D cavity with the restrictions that the modes
interact with the electric dipole moment of the atoms and to be in
resonance with the frequency associated with the energy difference
between the two levels of the atoms. This latter restriction is just
considered to have the maximum transition probability between states.

For the exact quantum solution we must resort to numerical diagonalization
of the Hamiltonian and use the lowest eigenstate to compute the expectation
values of the relevant observables.

\begin{figure}[H]
\begin{centering}
\includegraphics[scale=0.209]{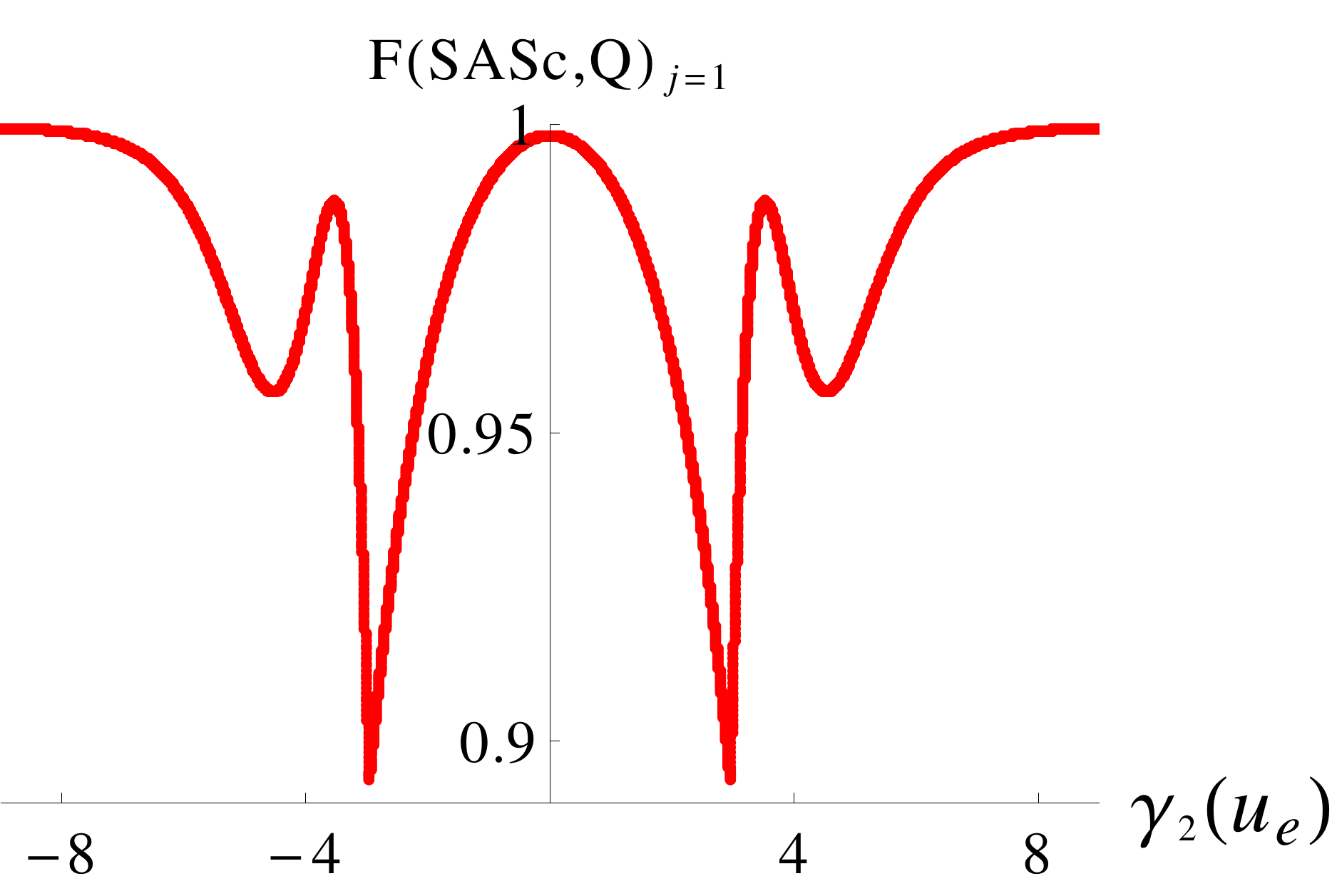}\qquad{}\includegraphics[scale=0.209]{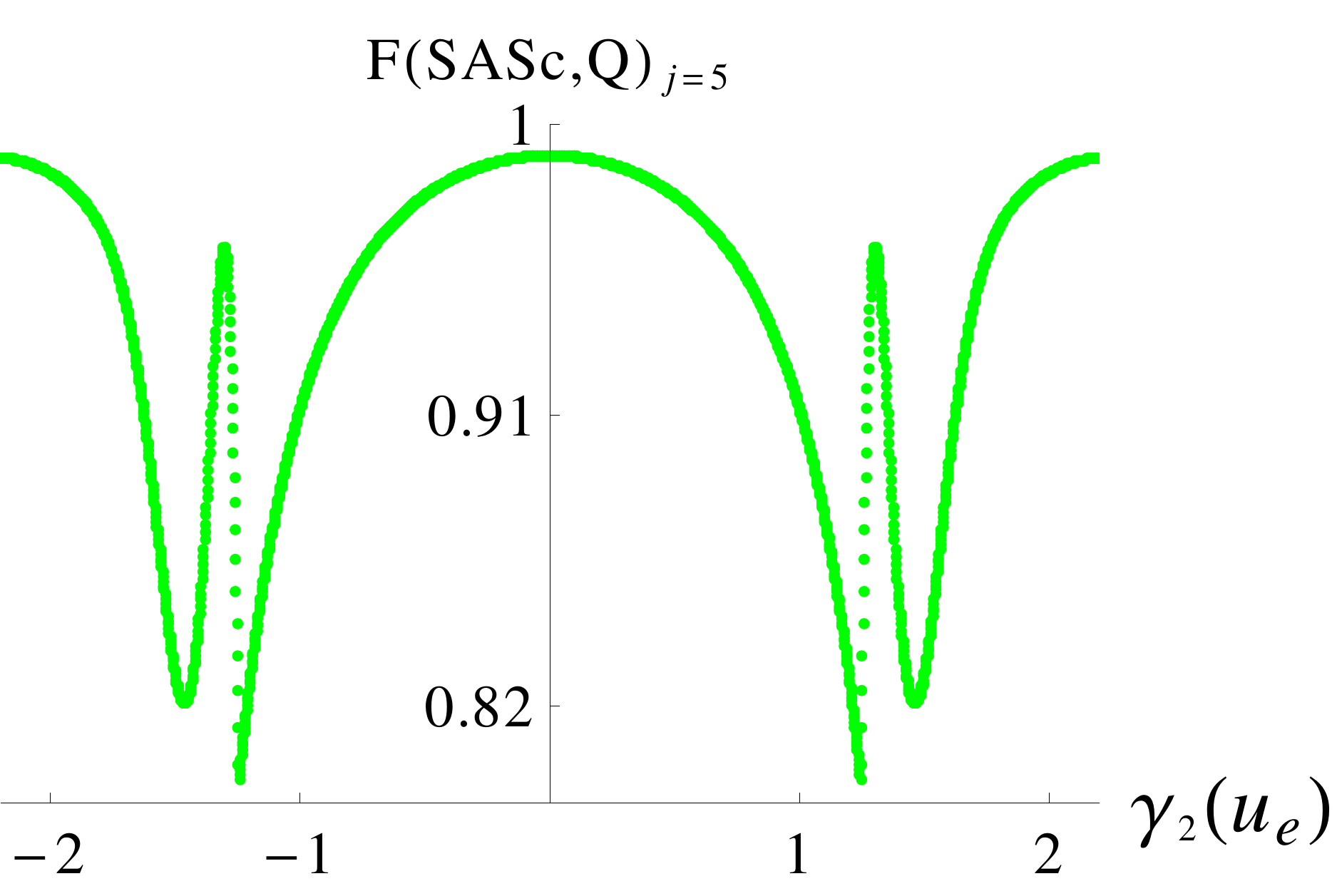}\qquad{}\includegraphics[scale=0.209]{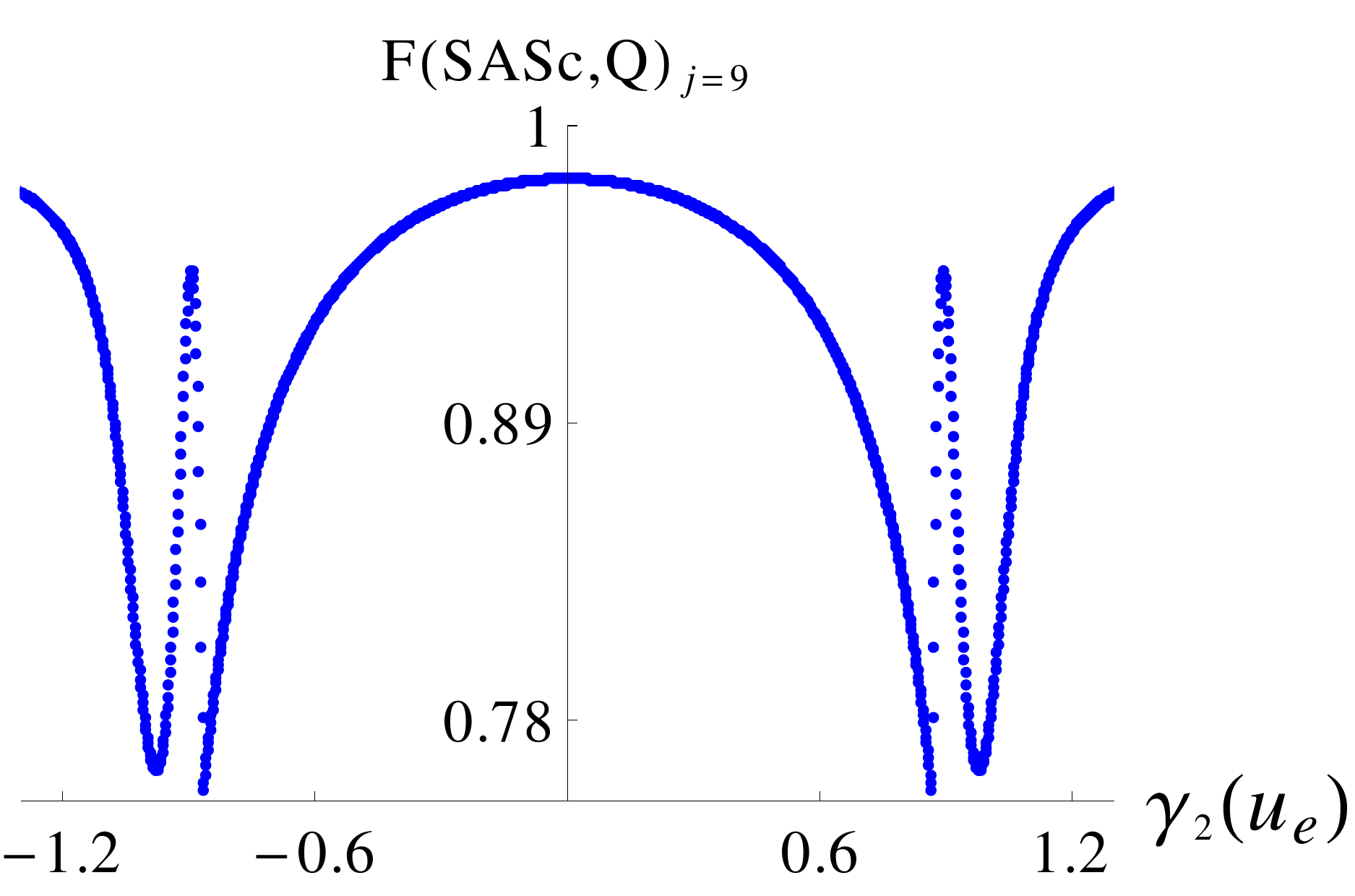}
\par\end{centering}

\caption{Fidelity between SAS with CS's minima and quantum solution as a function
of $\gamma_{2}$. Up (left): j=1, up (right): j=5, down: j=9. Assuming
$k=2$ and using $N=18$, $\omega_{A}=\Omega_{1}=\Omega_{2}=2u_{e}$,
$\gamma_{1}=\frac{1}{2}u_{e}$, where $u_{e}$ stands for any energy
unit ($\hbar=1$).\label{fig:10}}
\end{figure}

\begin{figure}[H]
\begin{centering}
\includegraphics[scale=0.209]{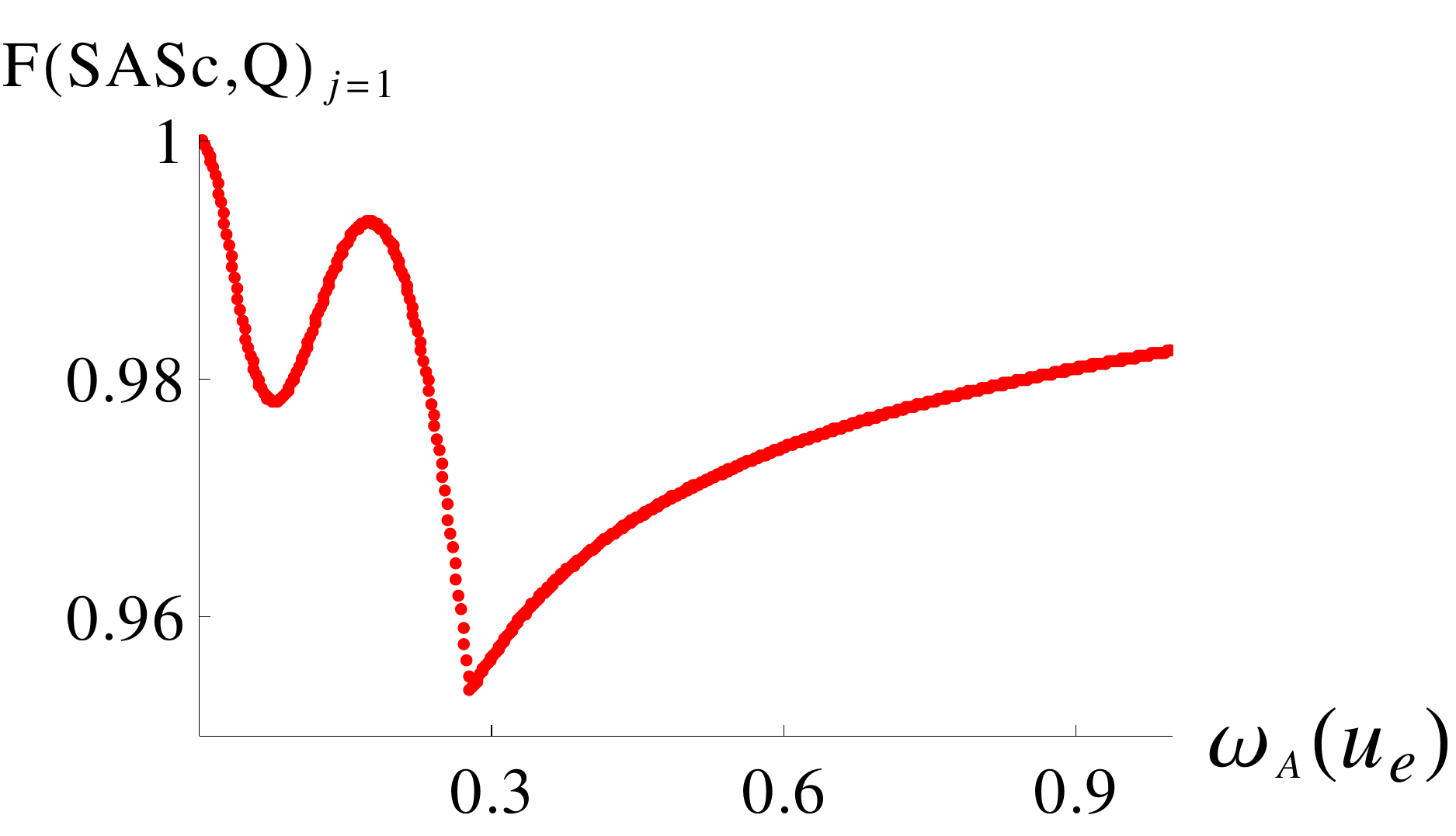}\qquad{}\includegraphics[scale=0.209]{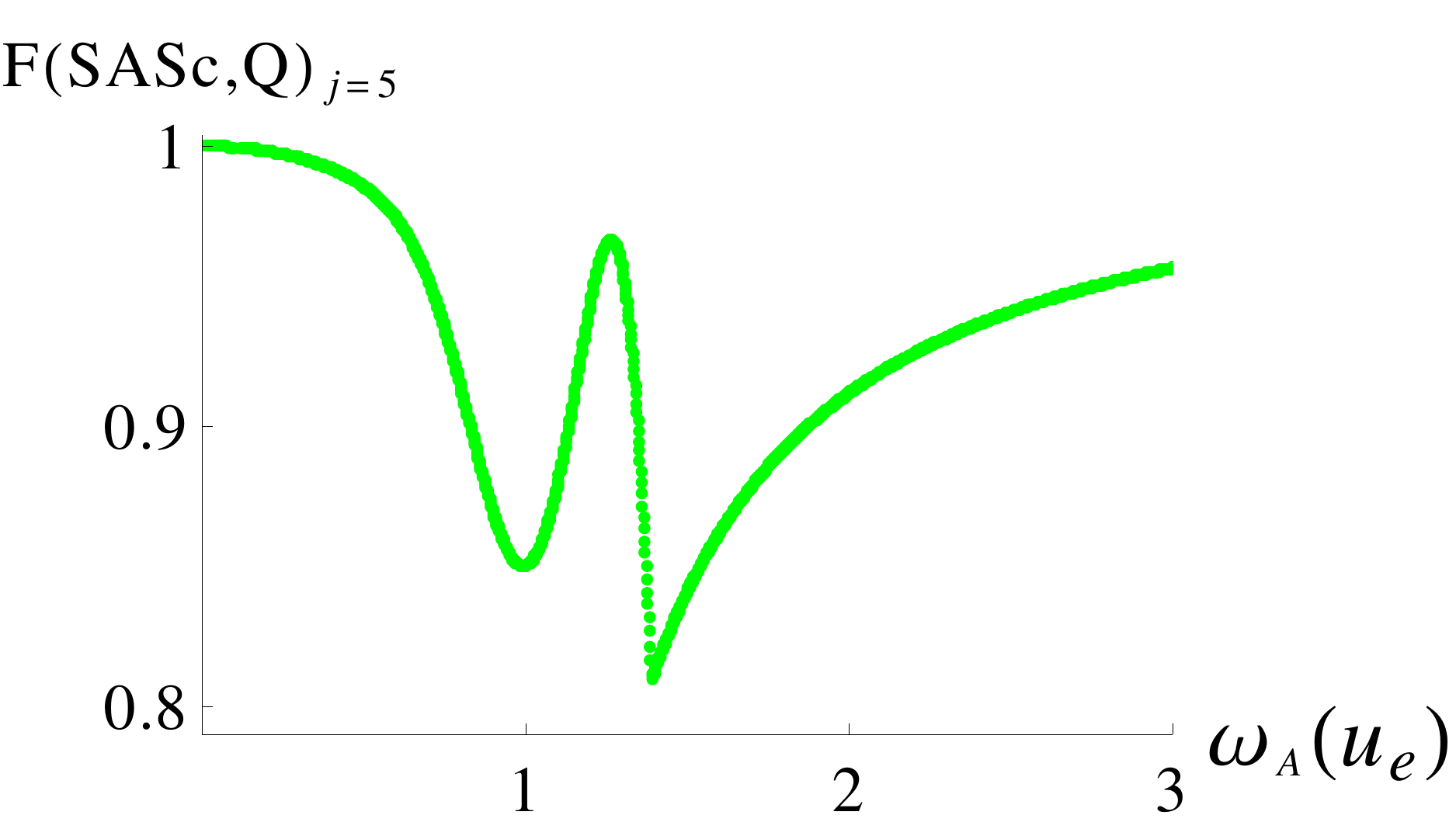}\qquad{}\includegraphics[scale=0.209]{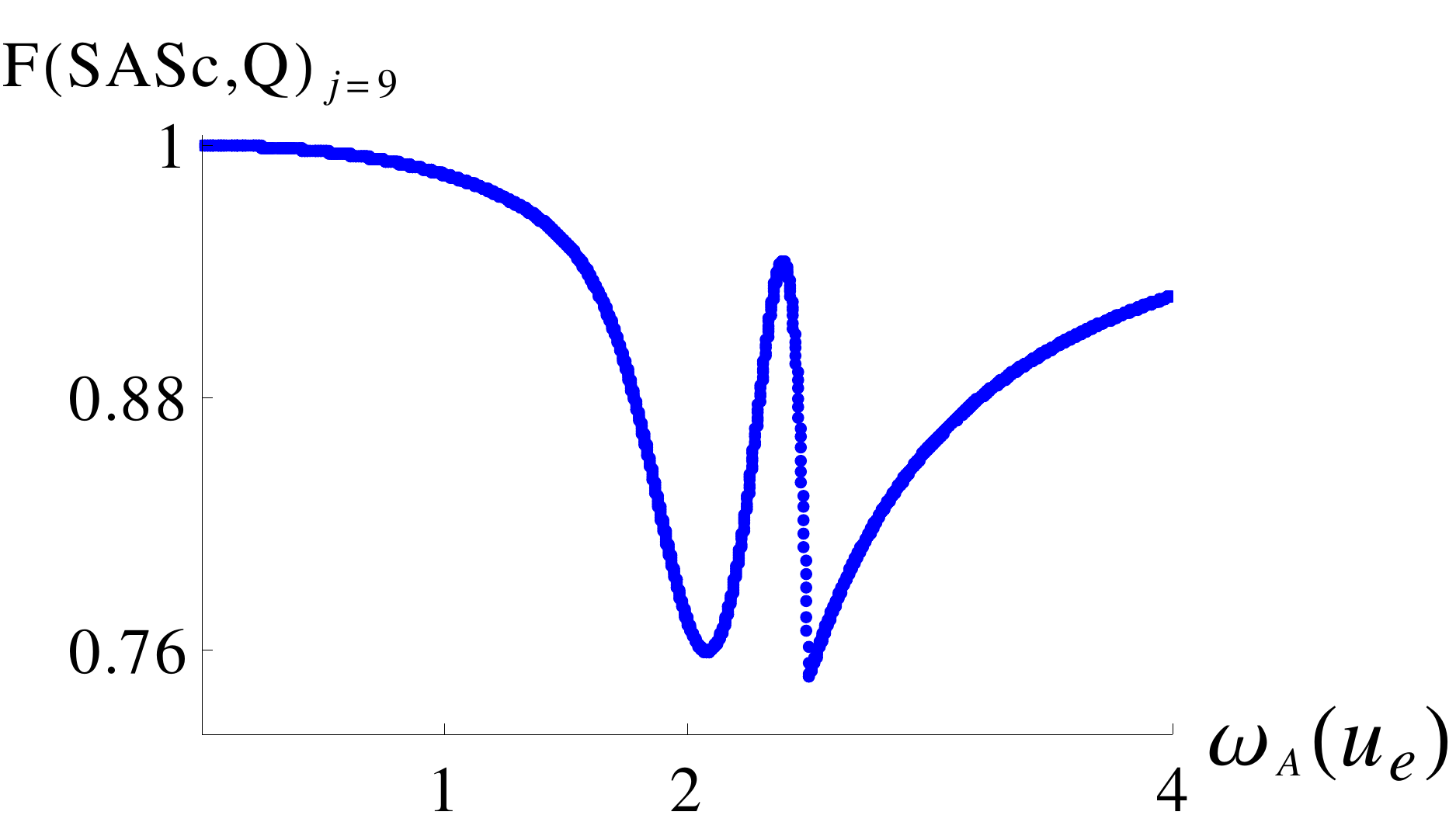}
\par\end{centering}

\caption{Fidelity between SAS with CS's minima and quantum solution as a function
of $\omega_{A}$. Up (left): j=1, up (right): j=5, down: j=9. Assuming
$k=2$ and using $N=18$, $\Omega_{1}=\Omega_{2}=2u_{e}$, $\gamma_{1}=\frac{1}{2}u_{e}$,
$\gamma_{2}=1u_{e}$, where $u_{e}$ stands for any energy unit ($\hbar=1$).\label{fig:11}}
\end{figure}

\begin{figure}[H]
\begin{centering}
\includegraphics[scale=0.25]{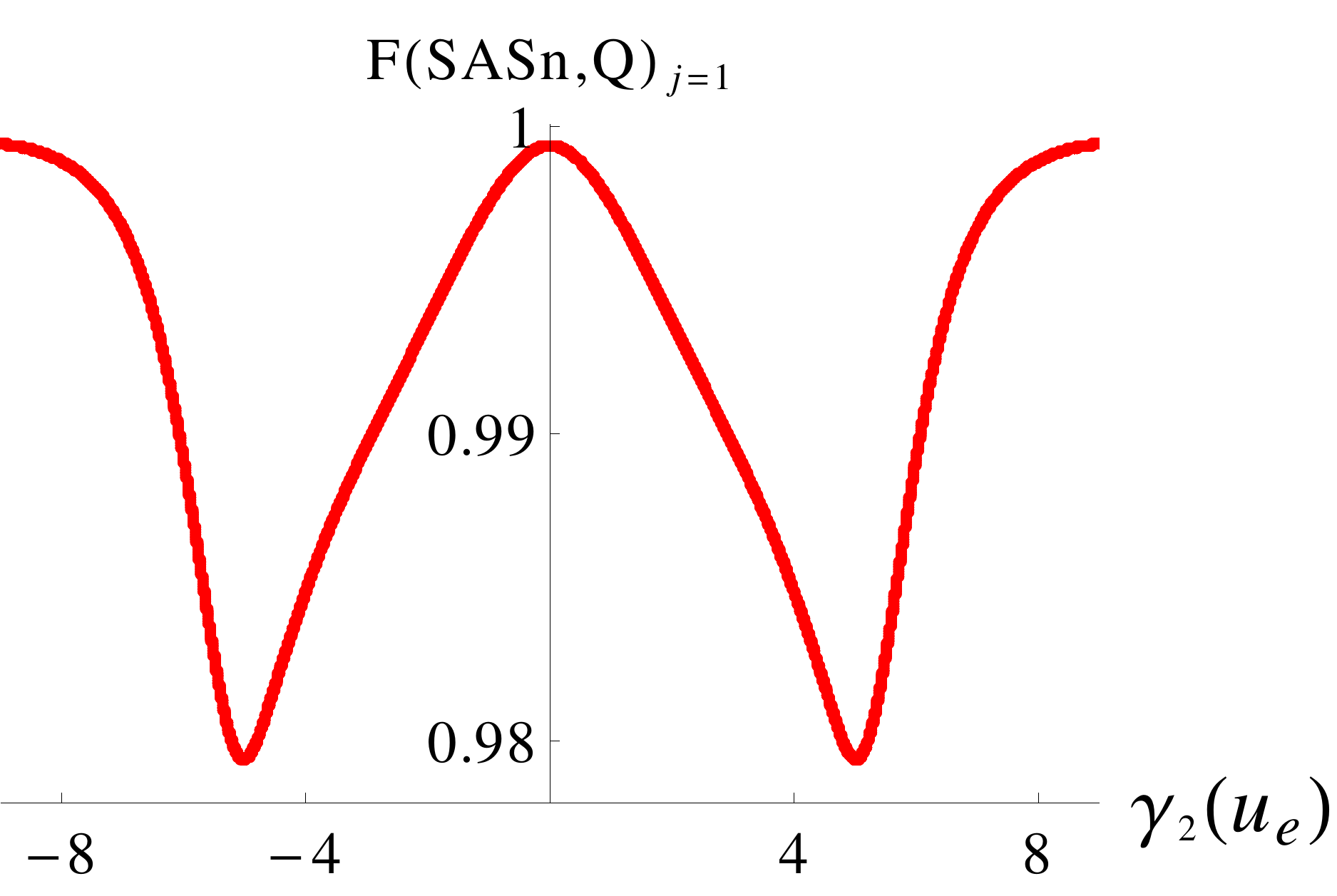}\qquad{}\includegraphics[scale=0.25]{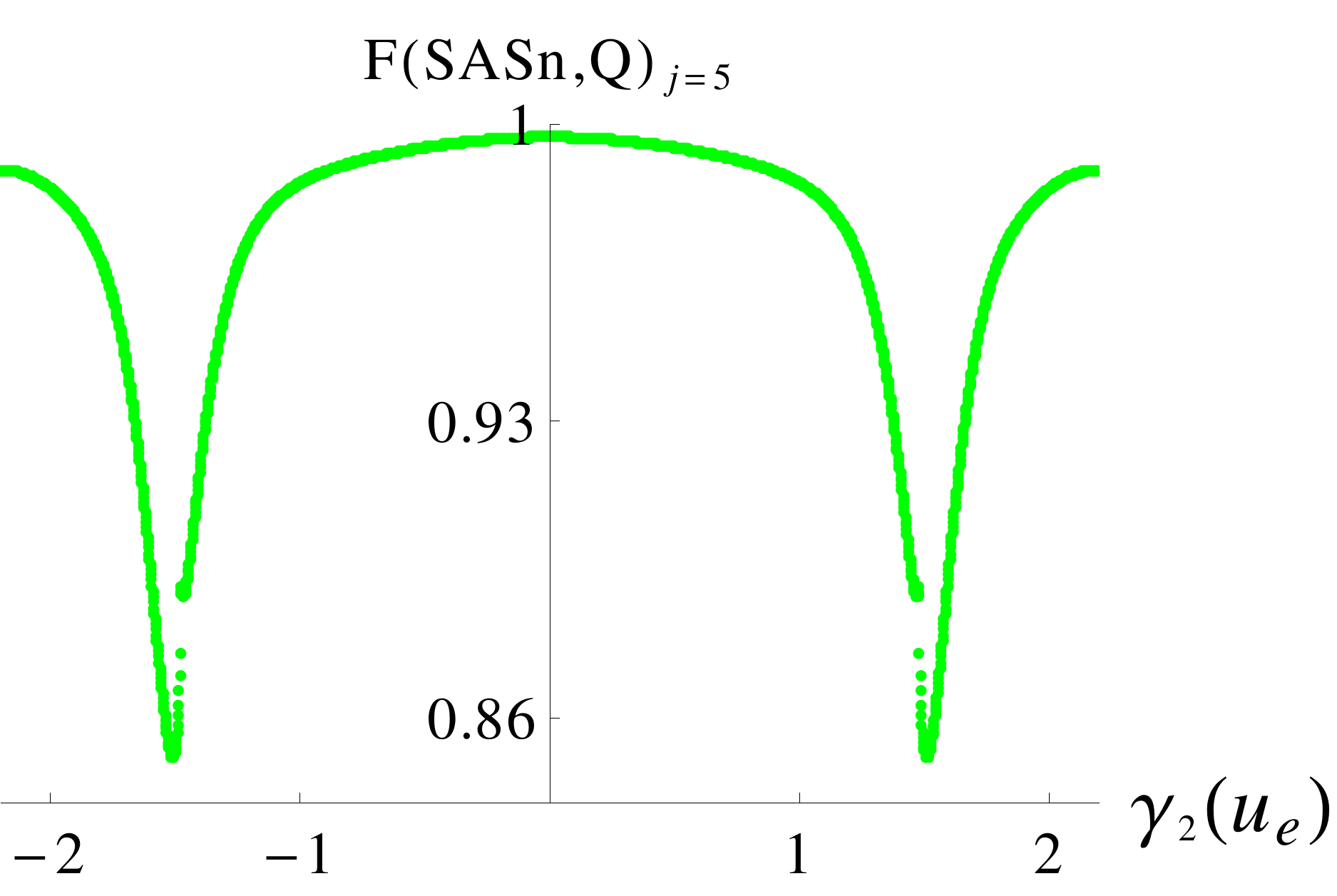}\qquad{}\includegraphics[scale=0.25]{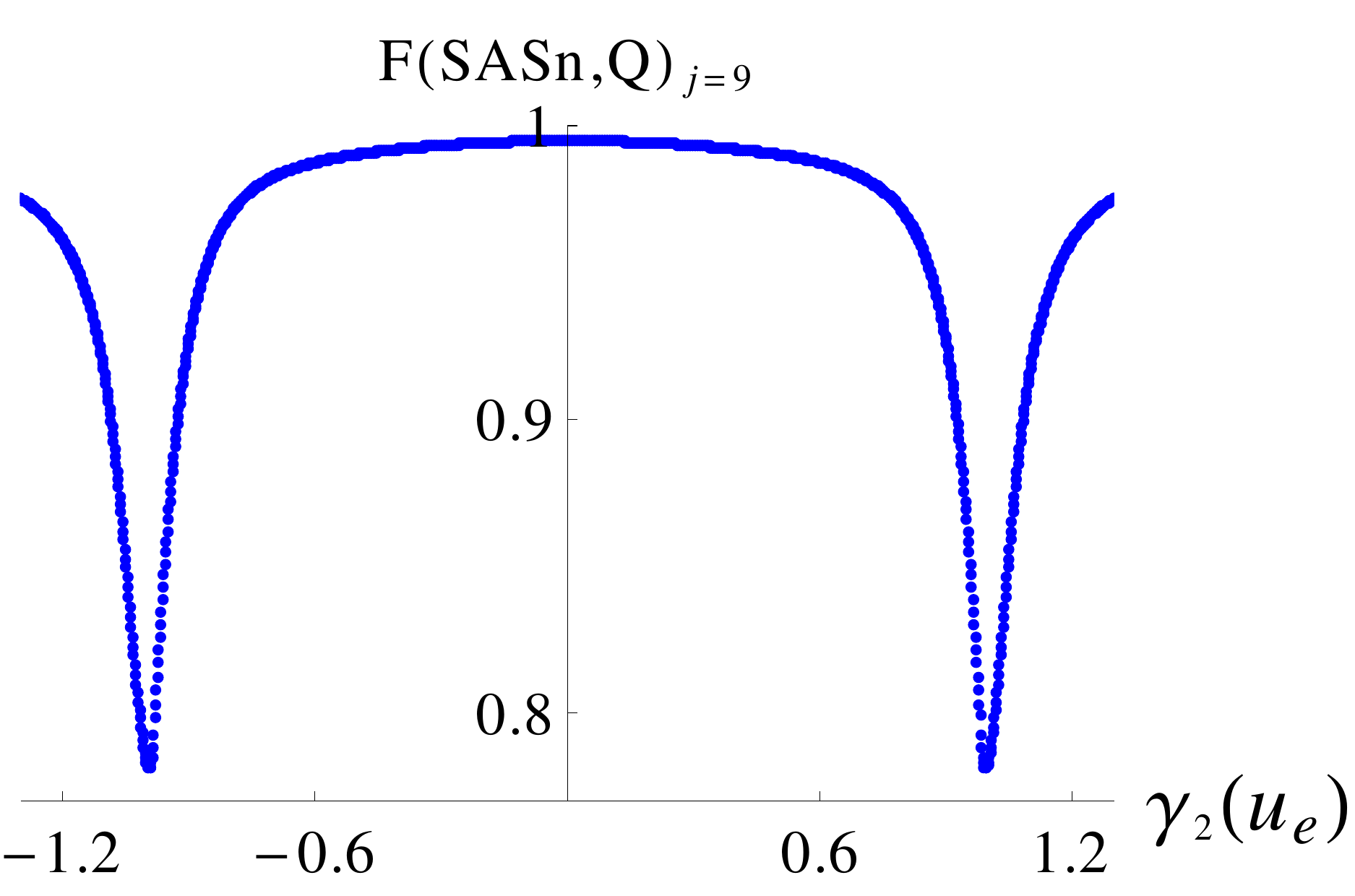}
\par\end{centering}

\caption{Fidelity between SAS minimized numerically and quantum solution as
a function of $\gamma_{2}$. Up: j=1, center: j=5, down: j=9. Assuming
$k=2$ and using $N=18$, $\omega_{A}=\Omega_{1}=\Omega_{2}=2u_{e}$,
$\gamma_{1}=\frac{1}{2}u_{e}$, where $u_{e}$ stands for any energy
unit ($\hbar=1$).\label{fig:12}}
\end{figure}

\begin{figure}[H]
\begin{centering}
\includegraphics[scale=0.25]{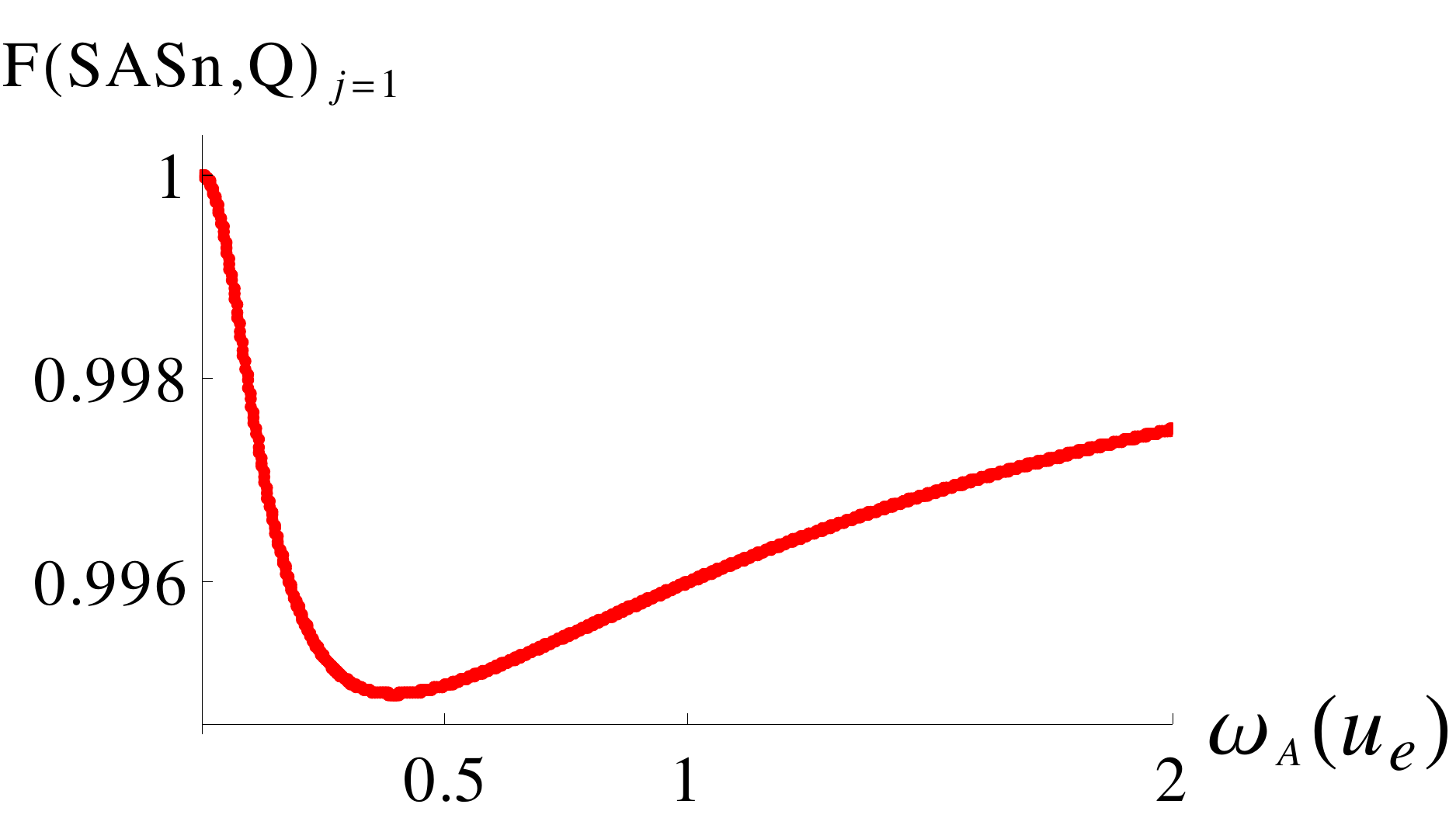}\qquad{}\includegraphics[scale=0.25]{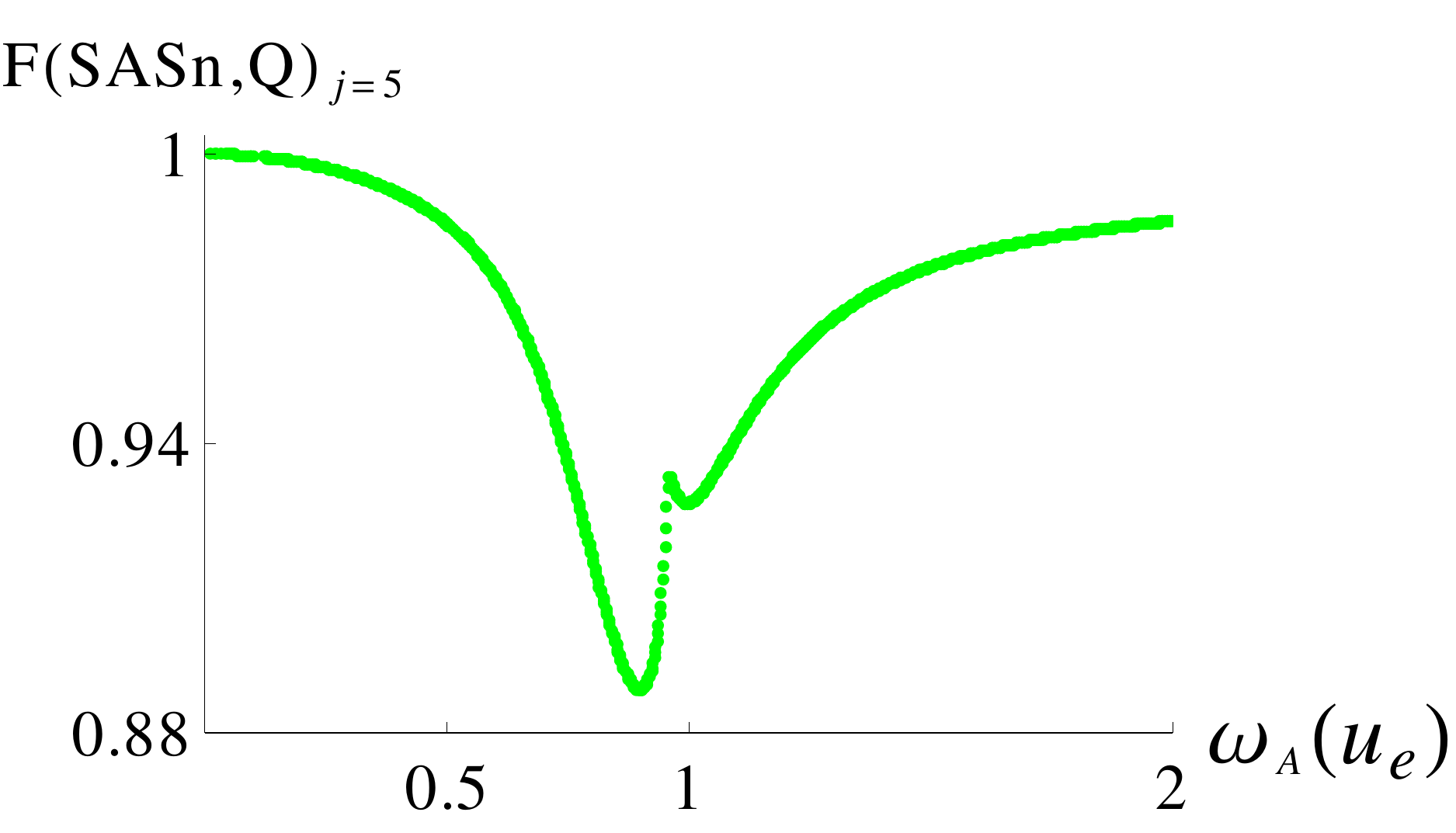}\qquad{}\includegraphics[scale=0.25]{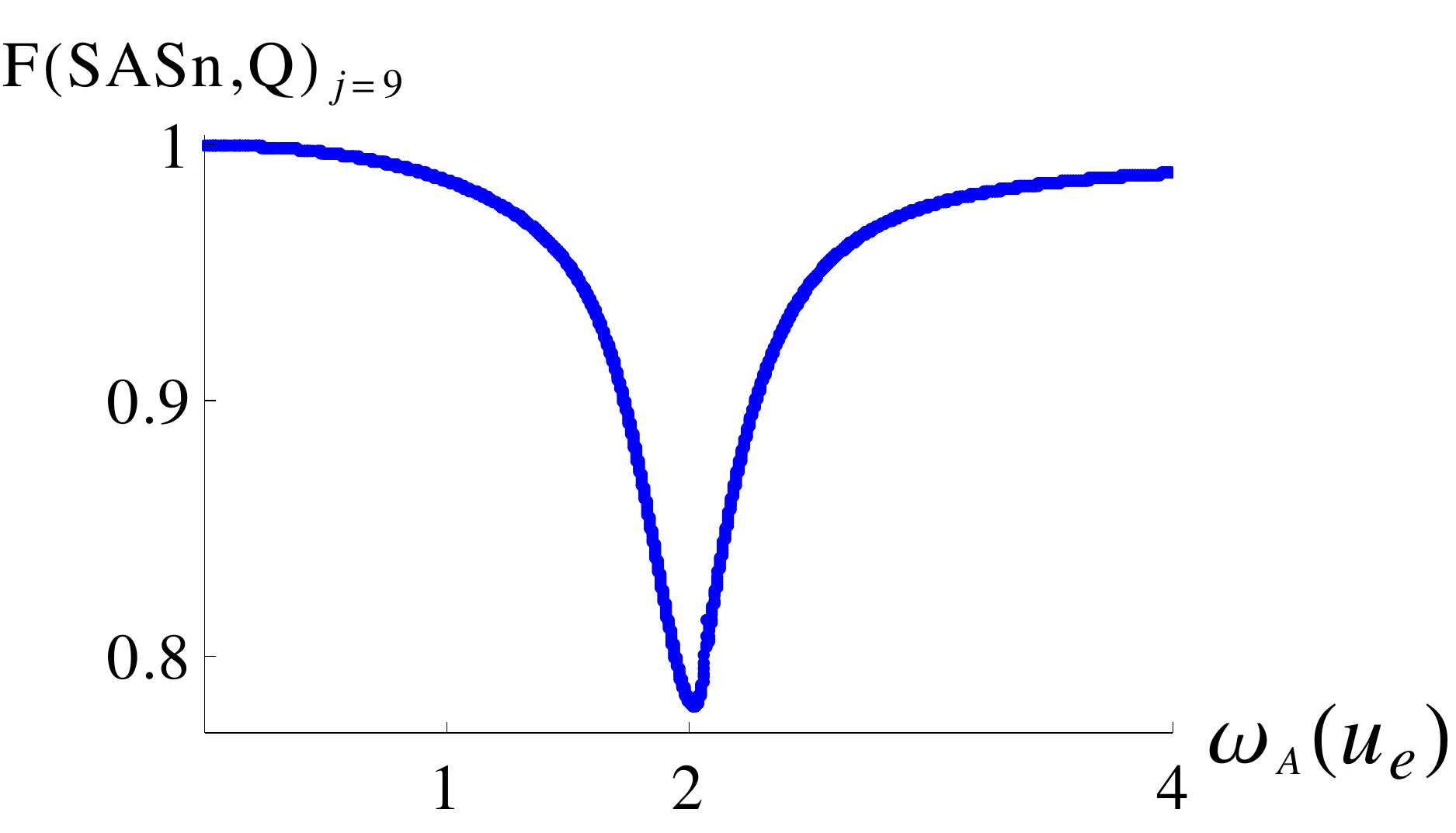}
\par\end{centering}

\caption{Fidelity between SAS minimized numerically and quantum solution as
a function of $\omega_{A}$. Up: j=1, center: j=5, down: j=9. Assuming
$k=2$ and using $N=18$, $\Omega_{1}=\Omega_{2}=2u_{e}$, $\gamma_{1}=\frac{1}{2}u_{e}$,
$\gamma_{2}=1u_{e}$, where $u_{e}$ stands for any energy unit ($\hbar=1$).\label{fig:13}}
\end{figure}

The results, properties of the ground state related to the CS, those
related to the SAS using the critical points of the CS (which have
the advantage of also providing analytical solutions), those of the
SAS minimized numerically and those of the quantum solution through
numerical diagonalization, are shown in figures \ref{fig:1} - \ref{fig:13}
and are discussed below.

One advantage of having analytical solutions is, of course, that the
order of the transition may be easily found. Equations (\ref{eq:14})
and (\ref{eq:18}) show a second-order QPT at $\delta=\frac{N\omega_{A}}{8j\varsigma}=1$
with the CS and SAS using CS's minima (SASc) approximations. In figure
\ref{fig:1} it can be seen that the data of the SAS using numerical
minimization (SASn) has a small discontinuity (the QPT) at $\gamma_{2}\approx1.485$
for $j=5$ and $\gamma_{2}\approx1.015$ for $j=9$, while in figure
\ref{fig:2} this discontinuity is at $\omega_{A}\approx0.975$ for
$j=5$ and $\omega_{A}\approx1.965$ for $j=9$. Note that the SASn
solution always approximate better the exact quantum result, as the
cooperation number increases this approximation gets better, in fact,
for $2j=18=N$ the loci of the separatrix between the normal and collective
regions for the quantum and SASn solutions are indistinguishable (except
in the zoomed inset). The true loci of the QPT may be found through
the fidelity: figures \ref{fig:14} and \ref{fig:15} show the fidelity
between neighboring states of the quantum solution, where the exact
QPT is characterized by the minimum, which is localized at $\gamma_{2}\approx1.550$
for $j=5$, $\gamma_{2}\approx1.031$ for $j=9$ in figure \ref{fig:14};
and $\omega_{A}\approx0.817$ for $j=5$, $\omega_{A}\approx1.870$
for $j=9$ in figure \ref{fig:15}.

The discrepancies between the transition values of the SASc approximation
and the exact quantum solution become obvious when looking at figures
\ref{fig:10} and \ref{fig:11}, where the fidelity between SASc and
the quantum solution drops (and oscillates) in a vicinity of the separatrix.
Therefore, we conclude that SASc offer a good approximation (with
an analytic expression) to the exact quantum solution far from the
QPT for low cooperation numbers, but as $j\rightarrow\infty$, the
interval where the SASc fail to reproduce the correct behavior, becomes
smaller.

Figures \ref{fig:12} and \ref{fig:13} show the fidelity drop at
the separatrix for the SASn. The resemblance to figures \ref{fig:14}
and \ref{fig:15} is uncanny, showing the benefits of restoring the
Hamiltonian symmetry into the trial variational states. This improvement
comes with the disadvantage of losing the analytic expression, but
still has an advantage over the quantum solution: the computational
time. SASn are obtained by numerically minimizing a real function,
which is far easier to do (computationally speaking) than numerically
diagonalizing the Hamiltonian matrix.

Figures \ref{fig:1} - \ref{fig:7} show the comparison between the
different approximations to the ground state: CS, SASc, SASn and quantum
solution. We show the behavior of $\mathcal{E}:=\left\langle H\right\rangle $,
$\left\langle J_{z}\right\rangle $ and $\left\langle \nu_{\imath}\right\rangle $
as functions of the atomic frequency $\omega_{A}$ and one of the
coupling constants $\gamma_{2}$, for different cooperation numbers.
It can be noticed that the discontinuity in the second derivative
of the energy (as modeled with CS and SASc) translates into a discontinuity
in the first derivative of $\left\langle J_{z}\right\rangle $ and
$\left\langle \nu_{\imath}\right\rangle $, thus characterizing the
QPT by means of an abrupt change in the expectation values of the
observables. In general, it can be observed that the four methods
(CS, SASc, SASn and quantum solution) converge in the limit $\delta\rightarrow0$,
where the case $j\rightarrow\infty$ is particularly interesting as
the interval around the QPT, where all the approximations fail to
reproduce the correct behavior, becomes smaller.

\begin{figure}[H]
\begin{centering}
\includegraphics[scale=0.26]{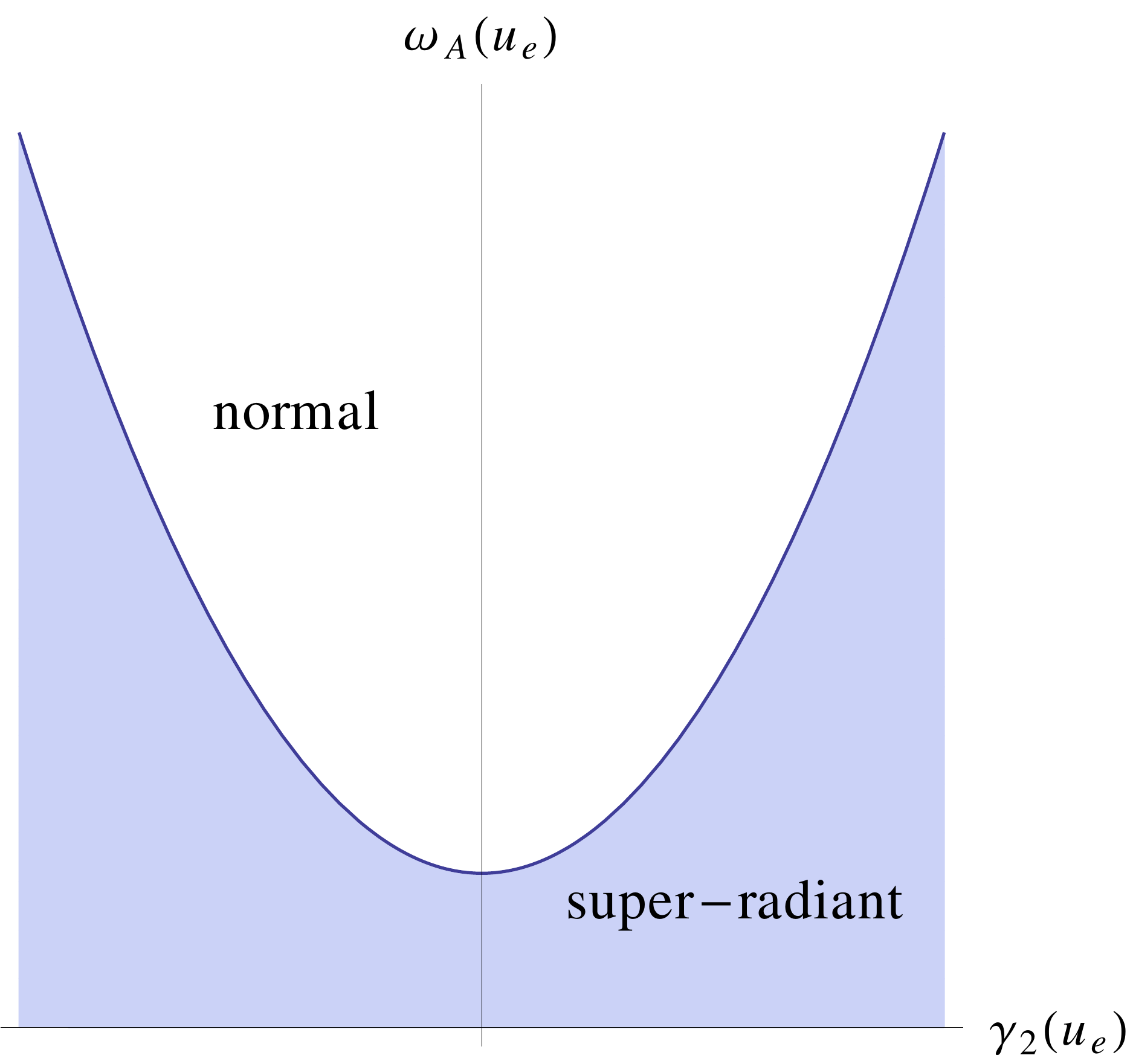}\qquad{}\includegraphics[scale=0.26]{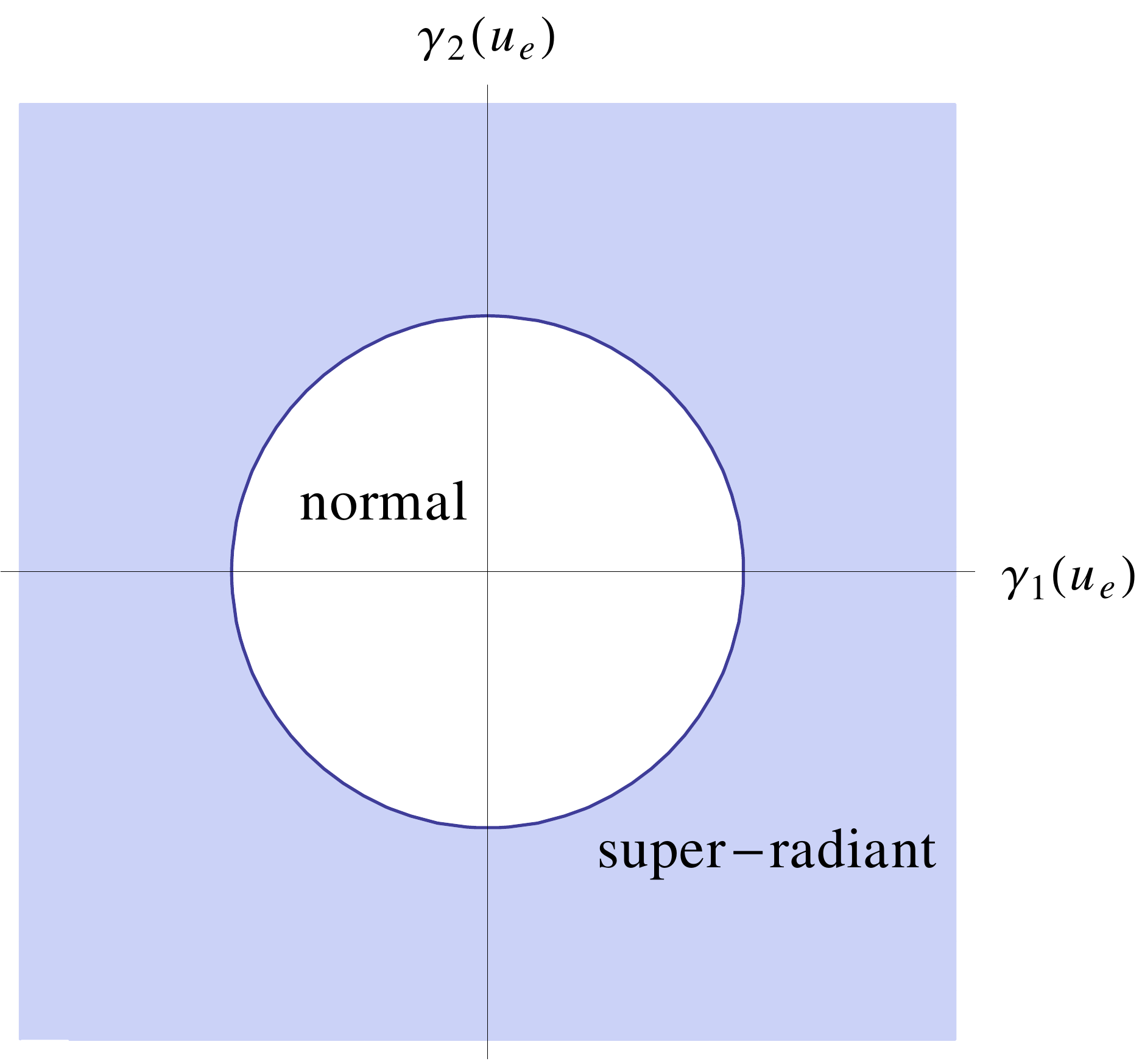}
\par\end{centering}

\caption{Pictographic representation of the phase diagrams in the plane $(\gamma_{2},\omega_{A})$
(up) and $(\gamma_{1},\gamma_{2})$ (down) obtained using CS. Normal
region (white) is defined as the region where $\delta\geq1$ and super-radiant
region (gray / light blue online) is defined as the region where $\delta<1$.
$u_{e}$ stands for any energy unit ($\hbar=1$).\label{fig:16}}
\end{figure}

It is worth mentioning the significance and importance of figures
\ref{fig:7} and \ref{fig:16} as they show aspects of the multi-mode
Dicke model which are not present in the single-mode case. In figure
\ref{fig:7} it is shown how the different modes of radiation (orthogonal
in principle) interact through the matter field, analogously as it
occurs with different atoms interacting through the radiation field.
On the other hand, figure \ref{fig:16} shows (pictorically) the phase
diagrams of the two-mode system, in which it can be observed that
any two points in the super-radiant region can be joined by a trajectory
that does not cross the normal region, a characteristic that the single-mode
system does not have. 

Figures \ref{fig:8} and \ref{fig:9} show the comparison between
SASc, SASn and the quantum solution for the entropy of entanglement
$S_{\varepsilon}$ as a function of the atomic frequency $\omega_{A}$
and one of the coupling constants $\gamma_{2}$, using different cooperation
numbers. A characterization of the QPT can be made by observing that
the entropy of entanglement obtained using the quantum solution shows
a maximum at the transition, an attribute that SASc and SASn approximations
fail to reproduce.

\FloatBarrier

\section*{Discussion and Conclusions}

From figures presented we conclude that SASc offer a good approximation
(with an analytic expression) to the exact quantum solution far from
the QPT for low cooperation numbers, but as $j\rightarrow\infty$,
the interval where the SASc fail to reproduce the correct behavior,
becomes smaller.

On the other hand, the SASn provide a better approximation to the
quantum solution. This improvement comes with the disadvantage of
losing the analytic expression, but still has the advantage over the
quantum solution of the computational time. SASn are obtained by numerically
minimizing a real function, which is far easier to do (computationally
speaking) than numerically diagonalizing the Hamiltonian matrix.

A characterization of the QPT can be made by looking at the entropy
of entanglement; that obtained using the quantum solution shows a
maximum at the transition, an attribute that SASc and SASn approximations
fail to reproduce.

The behavior of the expectation values of the relevant observables
of the system $\left\langle H\right\rangle $, $\left\langle J_{z}\right\rangle $
and $\left\langle \nu_{\imath}\right\rangle $, is also affected by
the QPT (figures \ref{fig:1} - \ref{fig:7}), thus allowing us to
characterize the QPT by means of its influence over the observables.
In general, it can be observed that the four methods (CS, SASc, SASn
and quantum solution) converge in the limit $\delta\rightarrow0$,
where the case $j\rightarrow\infty$ is particularly interesting as
the interval around the QPT, where all the approximations are weaker,
becomes smaller.

In conclusion, we have shown how the use of variational states to
approximate the ground state of a system can be useful to characterize
the QPT in a multi-mode Dicke model using the expectation value of
the observables relevant to the system and the entropy of entanglement
between matter and radiation. We have also introduced a not very commonly
used dependence: the cooperation number, showing its influence over
the behavior of the system, paying particular attention to the QPT
and the accuracy of the used approximations. Some aspects of the multi-mode
Dicke model which are not present in the single-mode case were also
briefly discussed.

\begin{acknowledgments}
We thank R. L\'opez-Pe\~na, O. Casta\~nos and S. Cordero for their comments
and discussion. This work was partially supported by DGAPA-UNAM under
project IN101217. L. F. Q. thanks CONACyT-M\'exico for financial support
(Grant \#379975).
\end{acknowledgments}

\end{document}